\documentclass[aps,prd,amsmath,amssymb,showpacs,floatfix,superscriptaddress,nofootinbib,twocolumn]{revtex4-1} 

\pdfoutput=1
\usepackage{natbib}
\usepackage[utf8]{inputenc} 
\usepackage{amsmath}
\usepackage{amssymb}
\usepackage{bm}
\usepackage{latexsym}
\usepackage{graphicx,epstopdf}
\usepackage{epstopdf}
\DeclareGraphicsExtensions{.ps, .png}
\usepackage{dcolumn} 
\usepackage{footnote}
\usepackage{tabularx,ragged2e,booktabs}
\usepackage[normalem]{ulem}
\usepackage{float}
\restylefloat{table}
\usepackage{color}
\usepackage{comment}

\usepackage{subcaption}
\usepackage{ragged2e} 
\DeclareCaptionJustification{justified}{\justifying}

\usepackage[font=small,labelfont=bf,
   justification=justified,
   format=plain]{caption}

\usepackage{tikz}
\usepackage{tikz-3dplot}
\usepackage[outline]{contour} 
\usepackage{xcolor}

\colorlet{veccol}{green!50!black}
\colorlet{projcol}{blue!70!black}
\colorlet{myblue}{blue!80!black}
\colorlet{myred}{red!90!black}
\colorlet{mydarkblue}{blue!50!black}
\tikzset{>=latex} 
\tikzstyle{proj}=[projcol!80,line width=0.08] 
\tikzstyle{area}=[draw=veccol,fill=veccol!80,fill opacity=0.6]
\tikzstyle{vector}=[-stealth,myblue,thick,line cap=round]
\tikzstyle{unit vector}=[->,veccol,thick,line cap=round]
\tikzstyle{dark unit vector}=[unit vector,veccol!70!black]
\usetikzlibrary{angles,quotes} 
\contourlength{1.3pt}

\def\lsim{\mathrel{\raise.3ex\hbox{$$<$$\kern-.75em\lower1ex\hbox{$\sim$}}}}
\def\gsim{\mathrel{\raise.3ex\hbox{$$>$$\kern-.75em\lower1ex\hbox{$\sim$}}}}

\newcommand{\fnl}{f_{\mathrm{NL}}}
\newcommand{\sigfnl}{\sigma_{f_{\mathrm{NL}}}}
\newcommand{\lmax}{l_{\mathrm{max}}}
\renewcommand{\vec}{\bm}

\newcommand{\dqc}{\frac{\derivd^3q}{(2\pi)^3}}

\newcommand{\ihMpc}{\ h\text{Mpc}^{-1}}

\newcommand{\derivd}{\,\mathrm{d}} 

\newcommand{\vk}{\vec k}
\newcommand{\vq}{\vec q}

\newcommand{\beq}{\begin{equation}}
\newcommand{\eeq}{\end{equation}}

\newcommand{\bea}{\begin{eqnarray}}
\newcommand{\eea}{\end{eqnarray}}

\def\mnras{Mon.\ Not.\ R.\ Astron.\ Soc.\ }
\definecolor{darkgreen}{cmyk}{0.85,0.2,1.00,0.2} 
\definecolor{purple}{cmyk}{0.5,1.0,0,0}

\definecolor{ultramarine}{rgb}{0.07, 0.04, 0.56}
\definecolor{cadmiumgreen}{rgb}{0.0, 0.42, 0.24}
\definecolor{indigo(dye)}{rgb}{0.0, 0.25, 0.42}
\usepackage[linktocpage=true]{hyperref}
\hypersetup{
colorlinks=true,
citecolor=ultramarine,
linkcolor=cadmiumgreen,
urlcolor=indigo(dye),
pdfauthor={},
pdftitle={},
pdfsubject={}
}

\begin{document}
	
\title{Measuring $\fnl$ with the SPHEREx Multi-tracer Redshift Space Bispectrum} 

\author{Chen Heinrich}\email{chenhe@caltech.edu}
\affiliation{California Institute of Technology, Pasadena, California 91125,USA}

\author{Olivier Dor\'e}
\affiliation{Jet Propulsion Laboratory, California Institute of Technology, Pasadena, California 91109, USA}
\affiliation{California Institute of Technology, Pasadena, California 91125,USA}

\author{Elisabeth Krause}
\affiliation{Department of Astronomy and Steward Observatory, University of Arizona, Tucson, Arizona 85721, USA}

\begin{abstract}

The bispectrum is an important statistics helpful for measuring the primordial non-Gaussianity parameter $f_{\mathrm{NL}}$ to less than order unity in error, which would allow us to distinguish between single and multi-field inflation models. The Spectro-Photometer for the History of the Universe, Epoch of Reionization and Ices Explorer (SPHEREx) mission is particularly well-suited for making this measurement with its $\sim$100-band all-sky observations in the near-infrared. Consequently, the SPHEREx data will contain galaxies with spectroscopic-like redshift measurements as well as those with much larger errors. In this paper, we evaluate the impact of photometric redshift errors on $f_{\mathrm{NL}}$ constraints in the context of an updated multi-tracer forecast for SPHEREx, finding that the azimuthal averages of the first three even bispectrum multipoles are no longer sufficient for capturing most of the information (as opposed to the case of spectroscopic surveys shown in the literature). The final SPHEREx result with all five galaxy samples and six redshift bins is however not severely impacted because the total result is dominated by the samples with the best redshift errors, while the worse samples serve to reduce cosmic variance. Our fiducial result of $\sigma_{f_{\mathrm{NL}}} = 0.7$ from bispectrum alone is increased by $18\%$ and $3\%$ when using $l_{\rm max}=0$ and $2$ respectively. We also explore the impact on parameter constraints when varying the fiducial redshift errors, as well as using subsets of multi-tracer combinations or triangles with different squeezing factors. Note that the fiducial result here is not the final SPHEREx capability, which is still on target for being $\sigma_{f_{\mathrm{NL}}} = 0.5$ once the power spectrum will be included.

\end{abstract}

\maketitle

\section{Introduction}
\label{sec:intro}

Single-field slow-roll inflation models predict that the primordial perturbations in the Universe are mostly Gaussian, with non-Gaussian deviations on the order of $O(10^{-2})$~\cite{Maldacena:2002vr, Creminelli:2004yq, Acquaviva:2002ud}, while multi-field inflation models can yield local non-Gaussianities of order $\fnl^{\rm loc} \gtrsim 1$~\cite{Bartolo:2004if}. Detecting or constraining the amount of primordial non-Gaussianity (PNG) at this level can help us distinguish between single and multi-field inflation models, and shed light into the process by which inflation proceeded. 

The best current limits on PNG come from observations of the cosmic microwave background (CMB) temperature and polarization by the Planck satellite: $\fnl = -0.9 \pm 5.1$~(68\% CL)~\cite{Planck:2019kim} (We will now drop the superscript ``loc" for the rest of our paper as it pertains to local non-Gaussianities only.). While the CMB is a two-dimensional map at the surface of last scattering, the large-scale-structure (LSS) of the Universe provides a three-dimensional map that gives access to more measurable modes, and therefore the ability to improve upon $\fnl$ constraints from the CMB.

The best measurements from LSS so far have $\sigma(\fnl) \sim$ $20-30$: The most robust measurements are coming from the three-dimensional power spectrum of quasars or galaxies~\cite{2022MNRAS.514.3396M, Castorina:2019wmr, Cabass:2022ymb, DAmico:2022gki, Cagliari:2023mkq}, while the photometric galaxy clustering observations achieve similar precisions although with more challenging systematics errors (e.g.~\cite{Rezaie:2023lvi}). Future spectroscopic surveys with increasing sky coverage such as Euclid~\cite{Amendola:2016saw}, DESI~\cite{DESI:2016}, and SPHEREx (Spectro-Photometer for the History of the Universe, Epoch of Reionization and Ices Explorer)~\cite{Dore:2014cca} would improve on Planck constraints to $\sigma(\fnl)$ of a few. Various techniques may also be employed to tighten constraints, for example cross-correlating with the CMB lensing, using higher-order statistics such as the bispectrum and the trispectrum, cross-correlating with the kinetic Sunyaev-Zel'dovich signal or even using a field-level inference (see e.g.~\cite{DESI:2023duv, Giri:2023, Gualdi:2021yvq, Andrews:2022nvv}).

Among the upcoming surveys, SPHEREx~\cite{Dore:2014cca} is a unique survey specifically designed to measure $\fnl$ to $\sigma(\fnl) \sim 0.5$ in just its nominal mission. Being a spectral survey without a spectrometer, it uses the Linear Variable Filter (LVF) technology to capture 102 spectral channels as it steps across the entire sky. This results in an all-sky spectral survey in the near-infrared (NIR) that enables us to measure galaxy redshifts in a large volume and infer the impacts of PNG on the distribution of matter. The 102-band observation lands itself somewhere between a traditional photometric and spectroscopic observation, thereby inheriting advantages as well as some challenges from both sides.

Recent studies suggest that for spectroscopic surveys, most of the constraining power on cosmological parameters can be captured with just the even $\ell$ and $m = 0$ modes of the bispectrum spherical harmonics decomposition~\cite{Gagrani:2016rfy, Byun:2022rvn}. Here, we re-evaluate this claim for the photometric redshift surveys in the context of the SPHEREx bispectrum forecast, and find that this claim does not hold in general in the presence of large enough photometric redshift errors. We show however that the impact is minimal on the final SPHEREx forecast because of the multi-tracer approach in SPHEREx where samples with small redshift errors dominate the results. 

Our forecast represents an improved update to the original SPHEREx forecast from Ref.~\cite{Dore:2014cca}: We include redshift space distortions (RSD) in the linear regime as well as a more complete bias modeling to second-order, and perform a full multi-tracer analysis, while binning not only in triangle shapes but also in triangle orientations, allowing for a more precise modeling of the photometric redshift errors as Gaussian damping rather than a hard cut-off in $k_{\parallel}$ for modes along the line-of-sight. We also show how results could be impacted if the photometric redshift errors were to vary from their current fiducial values.

Finally, we study the trade-off between the data vector size reduction and the $\fnl$ constraining power from selecting subsets of the galaxy samples or triangle shapes. Compared to the original forecast, our results were conducted with a more conservative $k_{\rm max} = 0.2 (1+z) \ihMpc$ to keep the modeling to the linear regime only, while using a similar galaxy sample specification with 5 samples and 11 redshift bins (although in practice we use the first six redshift bins where most of the bispectrum constraining power comes from). Our result does not yet include the modeling of the window function effects, as well as the wide-angle and general relativity (GR) effects which are important on the large scales we are probing. We leave these studies for future work.

The paper is structured as follows. In Section~\ref{sec:background}, we describe the background related to PNG and the multi-tracer galaxy bispectrum in redshift space. In Section~\ref{sec:fisher}, we describe the Fisher formalism used to forecast parameter errors for both the Fourier bispectrum and the bispectrum multipoles. We specify the SPHEREx forecast setup in Section~\ref{sec:forecast_setup} and show results in Section~\ref{sec:results}. Finally, we conclude in Section~\ref{sec:conclusion}.

\section{Background}
\label{sec:background}

We now present the background on the bispectrum signal modeling. We start by describing the modeling of the galaxy density in the presence of primordial non-Gaussianity, then the multi-tracer galaxy bispectrum in redshift space, and finally the definition and parametrization for the bispectrum multipoles.

\subsection{Galaxy density in the presence of primordial non-Gaussianty}

We consider here the local type primordial non-Gaussianity parameterized by $\fnl$
\beq
\Phi(\vec x)=\varphi(\vec x)+\fnl\left(\varphi^2(\vec
x)-\left\langle\varphi^2\right\rangle\right),
\label{eq:ngpot}
\eeq
where $\varphi$ is an auxiliary primordial Gaussian potential.
Using the Poisson equation, we can relate the primordial potential to the linearly evolved primordial matter density perturbation $\delta_{\rm m,p}$ as  (valid on subhorizon scales in the Newtonian limit)
\beq
\Phi(\vec k)=\frac{\delta_\text{m,p}(\vec k,z)}{\alpha(k,z)},
\label{eq:linpoiss}
\eeq
where
\beq
\alpha(k,z)=\frac{2 k^2 c^2 D(z) T(k)}{3H_0^2
\Omega_\text{m}},
\label{eq:alphaparam}
\eeq
and where $D(z)$ is the linear growth factor normalized at $z =0$, $T(k)$ the transfer function, $\Omega_m$ the matter density and $H_0$ the Hubble constant. On large scales where $T(k) = 1$, $\alpha$ scales as $k^2/H^2$. 

Working in perturbation theory, we expand the linear matter density contrast $\delta_\text{m}(\vec k)$ as
\beq
\delta_\text{m}(\vec k)=\delta_\text{m}^{(1)}(\vec k)+\delta_\text{m}^{(2)}(\vec
k)+\delta_\text{m}^{(3)}(\vec k)+\ldots\;.
\label{eq:matterexpansion}
\eeq
Given Eq.~\ref{eq:linpoiss} the linearly evolved primordial matter density field up to second order is then given by
\bea
\delta_\text{m,p}(\vec k,z) &=& \alpha(k,z)\, \Phi(\vec k) = \delta_\text{m,p}^{(1)}(\vec k,z) + \fnl\, \delta_\text{m,p}^{(2)}(\vec k,z), \notag \\
\label{eq:ngexp}
\eea
where
\beq
\delta_\text{m,p}^{(1)}(\vec k,z) = \alpha(k,z)\varphi(\vec k),
\eeq
and
\beq
\delta_\text{m,p}^{(2)}(\vec k,z) = \alpha(k,z)\, \int \dqc
\varphi(\vec q)\varphi(\vec k-\vec q).
\eeq

Gravitational evolution also contributes to nonlinear coupling starting at second order, so that the non-linearly evolved matter density field receives an additional term proportional to $F_2$:
\bea
\delta^{(1)}_\text{m}(\vec k,z)&=&\delta_\text{m,p}^{(1)}(\vec k,z),
\label{eq:delta_m_(1)}\\
\delta^{(2)}_\text{m}(\vec k,z)&=&\int \dqc \delta^{(1)}_\text{m,p}(\vec
q)\delta^{(1)}_\text{m,p}(\vec k-\vec q)
F_2(\vec q,\vec k-\vec q) \notag \\
&+&\fnl\, \delta_\text{m,p}^{(2)}(\vec k,z), \label{eq:delta_m_(2)}
\eea
where $F_2$ is the second-order mode coupling kernel
\beq
F_2(\vk_1, \vk_2) = \frac{5}{7} + \frac{1}{2} \frac{\vk_1 \cdot \vk_2} {k_1 k_2} \left( \frac{k_1}{k_2} + \frac{k_2}{k_1} \right) 
+ \frac{2}{7} \frac{(\vk_1 \cdot \vk_2)^2} {k_1^2 k_2^2}.
\eeq

Galaxy surveys observe the density of galaxies, which are biased tracers of the underlying dark matter density. We follow Ref.~\cite{Tellarini:2016sgp} in our modeling for the galaxy density and the galaxy bispectrum in the section to follow. We consider the following bivariate bias model for the Eulerian galaxy density constrast, where the dependence on $\varphi$ arises in the presence of non-Gaussianities:
\bea
\delta_\mathrm{g}(\vec x) 
&=& 
b_{10} \delta_{\rm m}(\vec x) 
+ b_{01} \varphi(\vec x) \notag\\
&+& 
\frac{1}{2} b_{20} \left( \delta_{\rm m}(\vec x) \right)^2 + b_{11} \delta_{\rm m}(\vec x)  \varphi(\vec x)  \notag\\
&+& 
\frac{1}{2} b_{02} \left( \varphi(\vec x)  \right)^2 
+ \frac{1}{2} b_{s_2} (s^2 - \langle s^2 \rangle) - b_{01} n^2. 
\label{eq:delta_g}
\eea
Here we have the tidal term~\cite{Catelan:2000vn, 2012PhRvD..86h3540B}
\beq
s^2(\vk) = \int \frac{d\vq}{(2\pi)^3} \mathcal{S}_2( \vq, \vk-\vq) \delta_{\rm m}^{(1)}(\vq)\delta_{\rm m}^{(1)}(\vk-\vq),
\eeq
and the non-Gaussian shift term due to the displacement of galaxies with respect to their initial positions $\vq$ in Lagrangian coordinates 
\beq
n^2(\vk) = 2 \int \frac{d\vq}{(2\pi)^3} \mathcal{N}_2( \vq, \vk-\vq) \frac{\delta_{\rm m}^{(1)}(\vq)\delta_{\rm m}^{(1)}(\vk-\vq)}
{\alpha(|\vk-\vq|)},
\eeq
where
\beq
\mathcal{S}(\vk_1, \vk_2) = \frac{(\vk_1 \cdot \vk_2)^2} {k_1^2 k_2^2} - \frac{1}{3},
\eeq
and
\beq
\mathcal{N}(\vk_1, \vk_2) = \frac{\vk_1 \cdot \vk_2} {2 k_1^2}.
\eeq
We note that we have ignored the stochastic contributions to $\delta_{\mathrm{g}}$ in Eq.~\ref{eq:delta_g}, and also do not marginalize over any potential deviations of the shot noise from Poisson predictions (see Ref.~\cite{Rizzo:2022lmh} for example for the full stochastic contributions to the tree-level bispectrum).

Taking only the first order terms in Eq.~\ref{eq:delta_g} we have
\beq
\delta_\mathrm{g}(\vec k)=\left(b_{10}+\frac{b_{01}}{\alpha(k)}\right)\delta_\text{m,p}(\vec k). 
\eeq
We will expound on the specific modeling and values we take for each bias parameter in Section~\ref{sec:forecast_setup}, where we will assume an universal mass function, for which the relationship between $b_{10}$ and $b_{01}$ becomes
\beq
    b_{01} = 2 \fnl \delta_c (b_{10}-1),
\label{eq:umf}
\eeq
and we recover the well-known linear order result~\cite{dalal2008}
\beq
\delta_\mathrm{g}(\vec k) = \left(b_{10}+\frac{2\fnl\delta_\text{c}
(b_{10}-1)}{\alpha(k)}\right)\delta_\text{m,p}(\vec k),
\label{eq:dalalfield}
\eeq
where $\delta_c = 1.686$ is the threshold for spherical collapse. 

We note that recent studies have shown how the universal mass function assumption for obtaining the relation Eq.~\ref{eq:umf} for $b_{01}$ (also known as $b_{\phi}\fnl$) could be inaccurate and could bias constraints on $\fnl$ (e.g.~\cite{Barreira:2020ekm}). So in a realistic analysis, one could marginalize over theoretically-informed priors for $b_{\phi}$ or choose to constrain the combination $b_{\phi}\fnl$ instead~\cite{Barreira:2020ekm, Barreira:2022sey}. Multi-tracer analysis can also help to improve the $\fnl$ constraints for suitably chosen galaxy samples (e.g. by maximizing the combination $|b_{10}^{B} b_{01}^A - b_{10}^{A} b_{01}^B|$ in a two-tracer analysis)~\cite{Barreira:2023rxn}. 

\subsection{Multi-tracer galaxy bispectrum in redshift space}
\label{sec:multitracer}

Let us define the multi-tracer bispectrum $B_{\rm ggg}^{ABC}$ as 
\beq
\langle \delta_g^{A}(\vk_1) \delta_{g}^{B}(\vk_2) \delta_{g}^{C}(\vk_3) \rangle = (2\pi)^3 \, \delta_D(\vk_1 + \vk_2  + \vk_3) B_{ggg}^{ABC}(\vk_1, \vk_2),
\eeq
where $A, B, C$ denote the galaxy samples. Now the wavevectors $\vec k_1, \vec k_2, \vec k_3$ will be associated with samples $A, B, C$ respectively. We only use the galaxy bispectrum with samples from the same redshift bin $i$ centered at $z_i$. Note that $A, B, C = 1..n_b$, so that there are $n_b = n_{\rm tracer}^3$ multitracer combinations for each redshift bin. 

To model the bispectrum in redshift space, we include the linear Kaiser effects on large scales, and the damping of small scales due to redshift errors, ignoring for now the Alcock-Pazynski effects. 

The redshift errors $\sigma_{z,i}^A = \tilde{\sigma}_{z}^A (1+z_i)$ decreases our ability to measure modes parallel to the line-of-sight. We choose to model this effect with a Gaussian suppression using
\beq
F^{ABC}_{i}(\vec{k}_1, \vec{k}_2) = e^{-\frac{1}{2} \left[k_1^2 \mu_1^2 (\sigma_{p,i}^{A})^2 + k_2^2 \mu_2^2 (\sigma_{p,i}^{B})^2 + k_3^2 \mu_3^2 (\sigma_{p,i}^{C})^2\right]}, 
\label{eq:FoG}
\eeq
and 
\beq
\sigma_{p,i}^A(z) = \tilde{\sigma}_{z}^A (1+z_i) \frac{2\pi c}{H(z)}.
\eeq
Note that here $\sigma_{p,i}(z)$ can no longer be factored out for the multi-tracer bispectrum for which each galaxy sample may have a different redshift error. 

\begin{widetext}
The multi-tracer bispectrum in redshift space is then modeled as
\beq
B^{ABC}_{ggg}(\vk_1, \vk_2| z_i) = F^{ABC}_i(\vec{k}_1, \vec{k}_2) \, \left[ 2 Z_1^{A}(\vk_1) Z_1^{B}(\vk_2) Z_2^{C}(\vk_1, \vk_2)\, P(k_1) P(k_2) + 2\mathrm{\ cycl.\ perm.} \right], 
\eeq 
where
\bea
Z_1^{X}(\vk_1) =\; &b_{10}^{X}& (1 + \beta \mu_1^2)  + \frac{b_{01}^{X}}{\alpha(k_1)} \\
Z_2^{X}(\vk_1, \vk_2) =\; &b_{10}^{X}& \left[ F_2(\vk_1, \vk_2) + \fnl \frac{\alpha(k)}{\alpha(k_1) \alpha(k_2)} \right] 
+ \frac{b_{20}^{X}}{2} + \frac{1}{2} b_{s_2}^{X} S_2(\vk_1, \vk_2) 
\notag \\
&+& \frac{b_{11}^{X}}{2} \left[ \frac{1}{\alpha(k_1)} + \frac{1}{\alpha(k_2)} \right] 
+ \frac{b_{02}^{X}}{2}\frac{1}{\alpha(k_1) \alpha(k_2)} 
- b_{01}^{X} \left[ \frac{ N_2(\vk_1, \vk_2 )}{ \alpha(k_2) } + \frac{ N_2(\vk_2, \vk_1) }{ \alpha(k_1) } \right]
\notag \\
&+& f\mu^2 \left[ G_2(\vk_1, \vk_2) + \fnl \frac{\alpha(k)}{\alpha(k_1)\alpha(k_2)} \right] 
+ \frac{1}{2} f \mu k \left( \frac{\mu_1}{k_1} Z_1^{X}(\vk_2) + \frac{\mu_2}{k_2} Z_1^{X}(\vk_1) \right),
\label{eq:Z2_kernel}
\eea

where $\mu_i = \hat{\vk}_i \cdot \hat{\vec{n}}$, where $\hat{\vec{n}}$ is the line-of-sight, $k = k_3$ and $\mu = -\mu_3$, and $G_2$ is the second-order velocity kernel
\beq
G_2(\vk_1, \vk_2) = \frac{3}{7} + \frac{1}{2} \frac{\vk_1 \cdot \vk_2} {k_1 k_2} \left( \frac{k_1}{k_2} + \frac{k_2}{k_1} \right) 
+ \frac{4}{7} \frac{(\vk_1 \cdot \vk_2)^2} {k_1^2 k_2^2}.
\eeq

The last term in Eq.~\ref{eq:Z2_kernel} is often also written as 
\bea
\frac{1}{2} f \mu k \left( \frac{\mu_1}{k_1} Z_1^{X}(\vk_2) + \frac{\mu_2}{k_2} Z_1^{X}(\vk_1) \right)
= f^2 \mu^2 k^2 \frac{\mu_1 \mu_2}{2 k_1 k_2} + b_{10}^{X} \frac{f \mu k }{2} \left(\frac{\mu_1}{k_1}  + \frac{\mu_2}{k_2} \right)
+
b_{01}^{X} \frac{f\mu k}{2} \left[ \frac{\mu_1}{k_1 \alpha(k_2)} + \frac{\mu_2}{k_2 \alpha(k_1)}\right].
\eea

\end{widetext}

\subsection{Fourier space and multipole space bispectrum}

\subsubsection{Fourier space bispectrum}

Given the shape of a triangle specified by ($k_1, k_2, k_3$), only two more parameters are needed to describe the orientation of the triangle due to the symmetry of the problem: Three degrees of freedom are taken away from the 9 coordinates that describe $\vk_1, \vk_2, \vk_3$, by virtue of the triangle condition $\vk_1 + \vk_2 + \vk_3 = 0$; one more is taken away because of the azimuthal symmetry of the signal around the line-of-sight vector. 

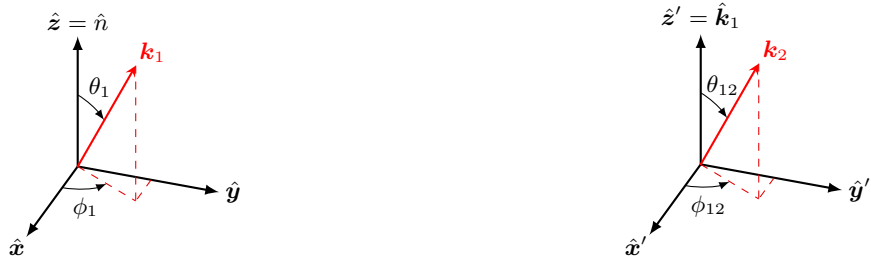
\begin{figure*}[t!]
    \centering 
    \begin{subfigure}[t]{0.45\textwidth}
      
        \tdplotsetmaincoords{60}{110}
        \begin{tikzpicture}[scale=2,tdplot_main_coords]
          
          \def\rvec{1.2}
          \def\thetavec{30}
          \def\phivec{60}
          
          \coordinate (O) at (0,0,0);
          \draw[thick,->] (0,0,0) -- (1,0,0) node[below left=-3]{$\hat{\vec{x}}$};
          \draw[thick,->] (0,0,0) -- (0,1,0) node[right=-1]{$\hat{\vec{y}}$};
          \draw[thick,->] (0,0,0) -- (0,0,1) node[above=-1]{$\hat{\vec{z}} = \hat{n}$};
          
          \tdplotsetcoord{P}{\rvec}{\thetavec}{\phivec}
          \draw[vector,red] (O)  -- (P) node[above right=-2] {$\vk_1$};
          \draw[dashed,myred]   (O)  -- (Pxy);
          \draw[dashed,myred]   (P)  -- (Pxy);
          \draw[dashed,myred]   (Py) -- (Pxy);
          
          \tdplotdrawarc[->]{(O)}{0.3}{0}{\phivec}
            {anchor=north}{$\phi_1$}
          \tdplotsetthetaplanecoords{\phivec}
          \tdplotdrawarc[->,tdplot_rotated_coords]{(0,0,0)}{0.55}{0}{\thetavec}
            {anchor=south west}{\hspace{-1.5mm}$\theta_1$}
        
        \end{tikzpicture}

    \end{subfigure}%
    ~ 
    \centering
    \begin{subfigure}[t]{0.45\textwidth}
        \tdplotsetmaincoords{60}{110}
        \begin{tikzpicture}[scale=2,tdplot_main_coords]
      
          \def\rvec{1.2}
          \def\thetavec{30}
          \def\phivec{60}
          
          \coordinate (O) at (0,0,0);
          \draw[thick,->] (0,0,0) -- (1,0,0) node[below left=-3]{$\hat{\vec{x}}'$};
          \draw[thick,->] (0,0,0) -- (0,1,0) node[right=-1]{$\hat{\vec{y}}'$};
          \draw[thick,->] (0,0,0) -- (0,0,1) node[above=-1]{$\hat{\vec{z}}' = \hat{\vk}_1$};
          
          \tdplotsetcoord{P}{\rvec}{\thetavec}{\phivec}
          \draw[vector,red] (O)  -- (P) node[above right=-2] {$\vk_2$};
          \draw[dashed,myred]   (O)  -- (Pxy);
          \draw[dashed,myred]   (P)  -- (Pxy);
          \draw[dashed,myred]   (Py) -- (Pxy);
          
          \tdplotdrawarc[->]{(O)}{0.3}{0}{\phivec}
            {anchor=north}{$\phi_{12}$}
          \tdplotsetthetaplanecoords{\phivec}
          \tdplotdrawarc[->,tdplot_rotated_coords]{(0,0,0)}{0.55}{0}{\thetavec}
            {anchor=south west}{\hspace{-2.2mm}$\theta_{12}$}
    
        \end{tikzpicture}
    \end{subfigure}%
    \caption{The coordinate systems we choose for parametrizing the triangle orientation for the Fourier space bispectrum. The two parameters that are used for determining the triangle orientation uniquely are $\theta_1$ and $\phi_{12}$. Note that $\phi_1$ is averaged over, as the bispectrum signal is invariant under changes in $\phi_1$. 
    }
    \label{fig:coordinates}
\end{figure*}

While many choices exist for parametrizing the orientation of a triangle, we follow Ref.~\cite{Scoccimarro:2015bla} using the following two angles: (1) $\theta_1$, the polar angle of $\vec{k}_1$ where $\hat{\vec{z}} = \hat{\vec n}$, and (2) $\phi_{12}$, the azimuthal angle of $\vk_2$ in a coordinate system $(\hat{\vec{x}}', \hat{\vec{y}}', \hat{\vec{z}}')$ where $\hat{\vec{z}}' = \hat{\vk}_1$. See Fig.~\ref{fig:coordinates} for an illustration.

In this parametrization we have
\beq
\mu_2 = \mathrm{cos}(\theta_{1})\, \mathrm{cos} (\theta_{12}) - \mathrm{sin} (\theta_1)\, \mathrm{sin} (\theta_{12})\, \mathrm{cos} (\phi_{12}), 
\label{eq:mu2_scoccimarro}
\eeq
and
\beq
\mu_3 = -\frac{k_1 \mu_1 + k_2 \mu_2}{k_3},
\label{eq:mu3_scoccimarro}
\eeq
where 
\beq
\theta_{12} = \mathrm{arccos}\left( \frac{-k_1^2 - k_2^2 + k_3^2}{2 k_1 k_2} \right).
\label{eq:theta_12}
\eeq
is the polar angle of $\vk_2$ in the primed coordinates, which is the angle between $\vk_1$ and $\vk_2$ and is always restricted to be between $[0, \pi]$.

When using this parametrization, we use orientation bins that are the linearly spaced in the variables $\mu_1 = \mathrm{cos}( \theta_{1})$ and $\phi_{12}$ since these angles are uniformly distributed.

\subsubsection{Bispectrum Multipoles}
\label{sec:multipoles}

Instead of parametrizing the orientation of the triangle, we can also expand the angle dependence in spherical harmonics:

\begin{widetext}

\bea
B_{lm}(k_1, k_2, k_3) &=& \frac{1}{4\pi} \,
\left(\frac{1}{2\pi} \int_{0}^{2\pi} \mathrm{d\, \phi_{1}}\right)
\left(\int_{-1}^{1} \mathrm{d\, cos}(\theta_1) \right)
\left( \int_{0}^{2\pi} \mathrm{d\, \phi_{12}} \right) \, 
%
B(k_1, k_2, k_3, \theta_1, \phi_{12}, \phi_1)\,
Y_{lm}^{*}(\theta_1, \phi_{12}) \notag \\
&=&  \frac{1}{4\pi} 
\left(\int_{-1}^{1} \mathrm{d\, cos} (\theta_1) \right)
\left( \int_{0}^{2\pi} \mathrm{d\, \phi_{12}} \right) \, 
%
B(k_1, k_2, k_3, \theta_1, \phi_{12})\,
Y_{lm}^{*}(\theta_1, \phi_{12}), \notag \\
\label{eq:blm_definition}
\eea
where we have averaged the signal over $\phi_1$, the azimuthal angle of $\vk_1$ around the line-of-sight $\hat{n}$ for which the signal is symmetric. The factor of $1/(4\pi)$ is a normalization convention. The inverse relation is: 
\beq
B(k_1, k_2, k_3, \theta_1, \phi_{12}, \phi_1) \notag\\
= \sum_l \sum_{m=-l}^{l} B_{lm}(k_1, k_2, k_3) Y_{lm}(\theta_1, \phi_{12}).
\eeq

\end{widetext}

Here the spherical harmonics are normalized such that 
\beq
Y_{00} = 1 \;\;\;\;\mathrm{and}\;\;\;
\int \frac{d^2 \hat{\vec{n}}}{4\pi}Y_{lm}(\hat{\vec{n}})Y_{l'm'}^{*}(\hat{\vec{n}}) = \delta_{ll'} \delta_{mm'}.
\eeq
The $B_{lm}$ with $m = 0$ are usually called the bispectrum multipoles. In our study, we will refer to all the $lm$ modes loosely as bispectrum multipoles. We investigate the effects effect of truncating the sum at various $\lmax$ and investigate the effect of omitting the odd $l$ and $m\neq0$ modes.

\section{Fisher Formalism}
\label{sec:fisher}

\subsection{Fisher matrix for the Fourier bispectrum}

Let $\tilde{B}^{ABC}(k_1, k_2, k_3)$ represent the binned bispectrum over triangle shapes and orientations with bin centers denoted by $(k_1, k_2, k_3)$ and $(\theta, \phi)$ respectively. If we approximate the binned bispectrum by the value at the bin center, then the Fisher matrix for the multi-tracer bispectrum in a single redshift bin can be written as
\begin{widetext}

\bea
F_{ij} &=& \sum_{(k_1, k_2, k_3)}  
\sum_{(\theta, \phi)}
\sum_{(ABC)} \sum_{(A'B'C')} 
\frac{\partial \tilde{B}^{ABC}(k_1, k_2, k_3, \theta, \phi)}{\partial p_i} 
\left[ \mathrm{\tilde{Cov}}\right] ^{-1} 
\frac{\partial \tilde{B}^{A'B'C'}(k_1, k_2, k_3, \theta, \phi)}{\partial p_j},
\eea
where the sum is over $n_{\rm shape}$ allowed triangle shape bins with centers denoted by $(k_1, k_2, k_3)$ and $n_{\rm ori}$ orientation bins with centers denoted by $(\theta, \phi)$, and there is no correlation between different triangle shapes and orientations in the Gaussian approximation for the covariance matrix. There is however a correlation between the different multi-tracer combinations, so the covariance $\tilde{\mathrm{Cov}}$ is a $n_b \times n_b$ matrix where $n_b = n_{\rm tracers}^3$ is the number of multi-tracer combinations. It can be written as
\beq
\mathrm{\tilde{Cov}} = \mathrm{Cov}\,\frac{V}{N_{\rm modes}},
\eeq
where in the Gaussian approximation we have for a single mode

\bea
&&\langle
\delta_g^A(\vk_1)\, \delta_g^B(\vk_2)\, \delta_g^C(\vk_3)\, 
\delta_g^{A'}(\vk'_1)\, \delta_g^{B'}(\vk'_2)\,  \delta_g^{C'}(\vk'_3)\,  
\rangle 
\label{eq:covariance_bispectrum_multitracer_contraction}
\\
&\approx&  \left( P_{gg}^{AA'}(\vk_1) + \frac{\delta_{AA'}}{\bar{n}_g^{A}}\right) 
\left(P_{gg}^{BB'}(\vk_2) + \frac{\delta_{BB'}}{\bar{n}_g^{B}}\right)
\left( P_{gg}^{CC'}(\vk_3) + \frac{\delta_{CC'}}{\bar{n}_g^{C}}\right) 
\delta_{D}(\vk_1 + \vk'_1) \delta_{D}(\vk_2 + \vk'_2) \delta_{D}(\vk_3 + \vk'_3)
 \\
&\equiv& \mathrm{Cov}\left[B_{ggg}^{ABC}(\vk_1, \vk_2), B_{ggg}^{A'B'C'}(\vk'_1, \vk'_2)\right] \; \delta_{D}(\vk_1 + \vk'_1) \delta_{D}(\vk_2 + \vk'_2) \delta_{D}(\vk_3 + \vk'_3),
\label{eq:covariance_bispectrum_multitracer}
\eea
and where
\beq
\frac{V}{N_{\rm modes}}\, =  \frac{(2\pi)^5  }{ V (\mathrm{d}k_1\, \mathrm{d}k_2\, \mathrm{d}k_3\, k_1 k_2 k_3)\, (d\mu d\phi)\, \beta}.
\eeq

\end{widetext}

Here the number of modes is the number of closed triangles $N_{\rm modes}$ within a triangle shape and orientation bin given a survey volume which sets the fundamental frequency $k_F\equiv (2\pi) V^{-1/3}$, where $V^{-1/3}$ is the volume of the survey in the given redshift bin. In the limit that the bin width $\Delta k_i \gg k_F$, the following expression is a good approximation
\bea
&&N_{\rm modes} = \frac{K_{\Delta}}{k_F^6} \notag\\
&=& \frac{V^2}{(2\pi)^6} \left[ 8\pi^2 k_1 k_2 k_3 \Delta k_1 \Delta k_2 \Delta k_3 \beta   
\left(\frac{1}{4\pi} \Delta \mu_1 \Delta \phi_{12} \right) \right], \notag\\
\eea
where $8\pi^2 k_1 k_2 k_3 \Delta k_1 \Delta k_2 \Delta  k_3 \beta$ is the number of closed triangles in a triangle shape bin denoted by the bin centers ($k_1, k_2, k_3$) with bin widths $\Delta k_1, \Delta k_2$ and $\Delta k_3$, and $\beta = 0.5$ for degenerate triangles and 1 otherwise.
The factor $\frac{1}{4\pi} \Delta \mu_1 \Delta \phi_{12}$ corresponds to the fraction of triangles in a fixed triangle shape bin that falls into the orientation bin with centers $(\mu, \phi_{12})$. 

Note that we do not include the factor $s_B$ = 6, 2, 1 for equilateral, isoceles and scalene triangles respectively, often used in the orientation-averaged case. Since here we have triangle orientation bins, we instead have that triangles of the same equilateral or isoceles shape but different orientations can be correlated. These would appear as off-diagonal elements of the $n_{\rm ori} \times n_{\rm ori}$ covariance matrix block for a fixed triangle shape. Furthermore, the $s_B$ factor is not valid for multi-tracers, since the other possible contractions in Eq.~\ref{eq:covariance_bispectrum_multitracer_contraction} are not necessarily equal to the first contraction we wrote down in Eq.~\ref{eq:covariance_bispectrum_multitracer} in the multi-tracer case.  For our calculations however, we ignore the effects from those off-diagonal terms because a bin with an equilateral or isoceles triangle as its representative center contains many triangles that are not equilateral or isoceles. 

Finally, the length of our multi-tracer bispectrum data vector for each redshift bin is $n = n_b \times n_{\rm shape} \times n_{\rm ori} = 125 \times 10^4 \times 25 \approx 10^7$ for $n_k = 50$, $n_{\theta} = 5$, $n_{\phi} = 5$.  The exact number of triangle shapes $n_{\rm shape}$ varies for different redshift bins, since the values of $k_{\rm min} = 2\pi V^{-1/3}$ and $k_{\rm max} = 0.2 \ihMpc (1+z)$ are dependent on the redshift bin. We note also that we have used an approximate $k_{\rm min}$ value  corresponding to a cubic volume, whereas in reality, the redshift bins are shaped as spherical shells in the full sky limit, or part of a spherical shell when the survey window function is applied. A precise mode counting treatment with the exact window function will be evaluated in a future work.

\subsection{Fisher matrix for the bispectrum multipoles}

For the bispectrum multipoles, we have the following Fisher matrix
\bea
&&F_{ij} = \int dk_1 dk_2 dk_3 k_1 k_2 k_3 
\sum_{(l,m)} \sum_{(l',m')} \sum_{(ABC)} \sum_{(A'B'C')}\notag\\
&&
\frac{\partial B_{lm}^{ABC*}(k_1, k_2, k_3)}{\partial p_i} 
\left[ \frac{(2\pi)^5}{V} \mathrm{Cov} \right]^{-1} 
\frac{\partial B_{l'm'}^{A'B'C'}(k_1, k_2, k_3)}{\partial p_j}.\notag\\
\eea
Let $\tilde{B}_{lm}^{ABC}(k_1, k_2, k_3)$ represent the bispectrum binned over triangle shapes that fall into the bin specified by the centers $(k_1, k_2, k_3)$. Let this binned value be approximated by the bispectrum value at this center configuration, and we have that the Fisher matrix can be approximated as
\bea
F_{ij} =&& \sum_{(k_1, k_2, k_3)} 
\sum_{(l,m)} \sum_{(l',m')} \sum_{(ABC)} \sum_{(A'B'C')} \notag \\
&&\frac{\partial \tilde{B}_{lm}^{ABC*}(k_1, k_2, k_3)}{\partial p_i}
%
\left[ \mathrm{\tilde{Cov}}\right] ^{-1} 
\frac{\partial \tilde{B}_{l'm'}^{A'B'C'}(k_1, k_2, k_3)}{\partial p_j}.\notag\\
\eea 
The outer sum is taken over $n_{\rm shape}$ unique triangle shapes $(k_1, k_2, k_3)$, where $k_1, k_2$ and $k_3$ are values of the $k$-bin centers that satisfy the triangle inequality ($0 < k_1 \leq k_2 \leq k_3 \leq k_1+k_2$). There is also a sum over $n_{lm}$ pairs of $(l, m)$ values, where $l$ and $m$ are integers satisfying $0 \leq l \leq l_{\rm max}$ and $-l \leq m \leq l$ (and similarly for the $(l', m')$ pairs), as well as a sum over $n_b = n_{\rm tracers}^3$ multi-tracer combinations ($ABC$) where $A, B, C = 1 .. n_{\rm tracers}$ (and similarly for ($A'B'C'$)). 

Because there is no correlation between different triangle shapes (in the Gaussian covariance), whereas there is a correlation between different $(l,m)$ pairs and multi-tracer combinations, the covariance $\tilde{\mathrm{Cov}}$ of the binned bispectrum is a $N\times N$ matrix, where $N = n_{b} n_{lm}$. It can be calculated using
\beq
\mathrm{\tilde{Cov}} = \mathrm{Cov}\, \frac{(2\pi)^5  s_{B}}{ V \mathrm{d}k_1 \mathrm{d}k_2 \mathrm{d}k_3 k_1 k_2 k_3 \beta},
\eeq
where
\bea
\mathrm{Cov}[B_{lm}^{ABC*} &,& B_{l'm'}^{A'B'C'}] = \frac{1}{(4\pi)^2}\int \mathrm{d(cos}\theta) \mathrm{d}\phi \, \notag\\
&\times &Y_{lm}^{*}(\theta, \phi) Y_{l'm'}(\theta, \phi) 
\left( P^{AA'}(\vk_1) + \frac{1}{\bar{n}^A_g} \right)  \notag\\
&\times &\left( P^{BB'}(\vk_2) + \frac{1}{\bar{n}^B_g} \right)
\left( P^{CC'}(\vk_3) + \frac{1}{\bar{n}^C_g} \right),\notag\\
\eea
and $\beta = 0.5$ for degenerate triangles and 1 otherwise. The factor of $1/(4\pi)^2$ comes from the normalization of the bispectrum multipoles in Eq.~\ref{eq:blm_definition}.

The marginalized error on parameter $p_i$ is obtained using
\beq
\sigma_{p_i} = \left[F^{-1/2}\right]_{ii}.
\eeq

The length of the data vector is slightly reduced here, with $n = n_b \times n_{\rm shape} \times n_{\rm lm}$ for a single redshift bin, where $n_{\rm lm} = 25$ when all $(l,m)$ pairs are used up to $l_{\rm max} = 4$, which can be reduced $n_{\rm lm} = 2$ without much loss of information if only the even $l$ and $m = 0$ modes are used up to $l_{\rm max} = 2$, as we will show later in Section~\ref{sec:results}.

\section{Forecast Setup}
\label{sec:forecast_setup}

We now present the setup for our Fisher forecast: The galaxy bias modeling choices and the survey parameters. The fiducial cosmology here is consistent the Planck 2018 cosmology~\cite{Planck:2018vyg}: The primordial spectral amplitude $A_s = 2.100\times10^{-9}$ with the tilt $n_s = 0.9659$ and the running of the tilt $n_{\rm run} = 0$, the baryon density $\Omega_b h^2 = 0.02238$, the dark matter density $\Omega_c h^2 = 0.1201$, and the acoustic scale $100\theta_{\mathrm{MC}} = 1.0409$. For a given redshift bin, we marginalize over the following set of parameters: \{$A_s, n_s, n_{\rm run}, \fnl, \Omega_b h^2, \Omega_c h^2, 100\theta_{\rm MC}, b_{10}^X | X = 1..5\}$. For the joint Fisher forecast from all redshift bins, the constraining power on the cosmological parameters from each redshift bin is combined, while the individual $n_z \times n_{\rm sample} = 55$ linear galaxy bias parameters are constrained by their corresponding redshift bin only.

\subsection{Bias modeling}
\label{sec:bias_modeling}

We briefly summarize the galaxy bias modeling we chose and refer the readers to Ref.~\cite{Tellarini:2016sgp} for more details. Recall that the galaxy density field up to second order is described by
\bea
\delta_\mathrm{g}(\vec x) 
&=& 
b_{10} \delta_{\rm m}(\vec x) 
+ b_{01} \varphi(\vec x) \notag\\
&+& 
b_{20} \left( \delta_{\rm m}(\vec x) \right)^2 + b_{11} \delta_{\rm m}(\vec x)  \varphi(\vec x) \notag\\
&+& b_{02} \left( \varphi(\vec x)  \right)^2 
+ b_{s_2} (s^2 - \langle s^2 \rangle) - b_{01} n^2, 
\label{eq:delta_g_copy}
\eea
where we have now absorbed the factors of 1/2 into $b_{20}$, $b_{02}$ and $b_{s_2}$. There are a total of 6 different kind of bias parameters to model: $b_{10}$, $b_{01}$, $b_{20}$, $b_{11}$, $b_{02}$ and $b_{s_2}$, each having 55 distinct values for the 5 galaxy samples and 11 redshift bins. 

As previously noted, we assume an universal mass function and use the following relation for $b_{01}$:
\beq
    b_{01} = 2 \fnl \delta_c (b_{10}-1).
\eeq
For $b_{20}$, we treat it as a function of $b_{10}$ using a fit from simulations in Ref.~\cite{Lazeyras:2015lgp},
\beq
b_{20}^{\rm Lezeyras}(b_{10}) = \frac{1}{2} (0.412 - 2.143\, b_{10} + 0.929\, b_{10}^2 + 0.008\, b_{10}^3).
\label{eq:b20_lezeyras}
\eeq    
Similarly for $b_{s_2}$, we use a relation to $b_{10}$ obtained from a fit to simulation results from Ref.~\cite{2012PhRvD..86h3540B},
\beq
b_{s_2}^{\rm Saito}(b_{10}) = - \frac{2}{7} \, (b_{10} - 1).
\label{eq:bs2_saito}
\eeq
For $b_{11}$ and $b_{02}$, we can relate them to any given values of $b_{10}, b_{20}$ and $\fnl$ using, 
\beq
b_{11} = b_{11}^L + b_{01}^L,
\eeq
and
\beq
b_{02} = \fnl^2 4 \delta_c (\delta_c  b_{20}^L - 2 b_{10}^L),
\eeq
where the $L$ superscript denotes Lagrangian bias, and we suppressed the $E$ superscript for the Eulerian biases, and the Lagrangian quantities may be related to the Eulerian ones as follows:
\beq
b_{11}^L = 2 \fnl (\delta_c b_{20}^L - b_{10}^L),
\eeq
\beq
 b_{20}^L = b_{20} - \frac{8}{21} (b_{10}-1),
\eeq
\beq
b_{10}^L = b_{10} - 1,
\eeq
and
\beq
b_{01}^L = b_{01},
\eeq
The above results were obtained under the assumption of the universal mass function, conservation of the galaxy number density in a given volume, spherical collapse and no velocity bias between galaxies and matter. 

In the Fisher forecast, we vary the parameters $b_{10}$ for each sample and redshift bin when taking the derivative with respect to $b_{10}$, and let all other bias values vary according to the relations described above.

\subsection{Survey parameters}

The survey parameters have not changed significantly since the original SPHEREx forecast. There are eleven redshift bins ranging from $z = 0$ to $4.6$ (see  definitions in Table~\ref{tab:redshift_bin_definition} in Appendix~\ref{sec:appendix}), with the galaxies in each redshift bin divided by their redshift uncertainties falling in the bins $\tilde{\sigma}_{z} = \sigma_z/(1+z) =  0 - 0.003 - 0.01 - 0.03 - 0.1 - 0.2$. We use the maximum value of the redshift bin in our forecast for more conservative results: $\tilde{\sigma}_{z}^{A} = 0.003, 0.01, 0.03, 0.1$ and $0.2$ for $A = 1$ to 5 respectively. So a lower sample number means better redshift uncertainties. Note that in the original bispectrum forecast, it was the mean value of $\tilde{\sigma}_{z}^{A}$ that was used, leading to slightly more optimistic but also more realistic results.

The details of the procedure for obtaining the fiducial galaxy number densities and biases are found in Ref.~\cite{Dore:2014cca}, which we briefly summarize here. First, the simulated SPHEREx galaxy catalog (based on COSMOS~\cite{Scoville:2006vq, Ilbert:2008hz}) was piped through a template-fitting based photometric redshift measurement pipeline in order to produce a redshift and a redshift error estimate for each galaxy. These estimates were used to derive the galaxy number density for the galaxy sample in each redshift bin and $\tilde{\sigma}_z$ bin. 

The method of abundance matching was then used to obtain an estimate of the linear galaxy bias for each galaxy sample: The galaxies with the best redshift errors are matched to the host halos with the largest total mass. More specifically, the mass function in Ref.~\cite{Tinker:2008ff} was used to find the minimum halo mass, for which the halo bias is found using the fitting formula in Ref.~\cite{2010ApJ...724..878T} and set to the linear galaxy bias. The fiducial values for the  linear biases and number densities are available at the SPHEREx public Github repository\footnote{SPHEREx public github repository: \url{https://github.com/SPHEREx/Public-products/blob/master/galaxy_density_v28_base_cbe.txt}}, and also reproduced in Appendix~\ref{sec:appendix} for convenience in Tables~\ref{tab:galaxy_bias} and~\ref{tab:number_density} respectively.

\section{Results}
\label{sec:results}

\begin{figure}
  \centering
    \includegraphics[width=0.5\textwidth]{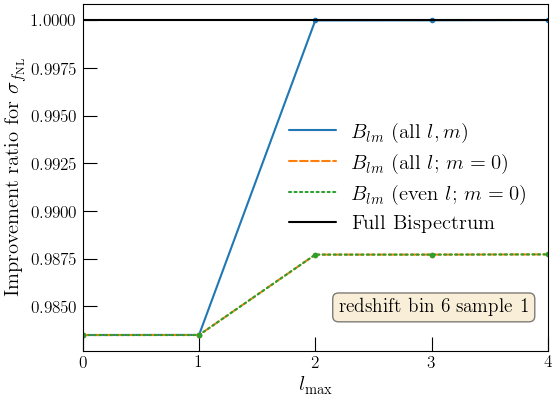}
  \caption{Plot of the improvement ratio for $\sigma_{\fnl}$ of bispectrum multipoles against the full bispectrum result, for a single tracer with good redshift error: redshift bin 6, sample 1, for which $z_{\rm mid}=1.3$ and $\sigma_z/(1+z)=0.003$. }
  \label{fig:plot_compare_multipoles_improvement_fnl_1}
\end{figure}

\begin{figure}
\centering
\begin{minipage}{.45\textwidth}
  \centering
    \includegraphics[width=1.0\textwidth]{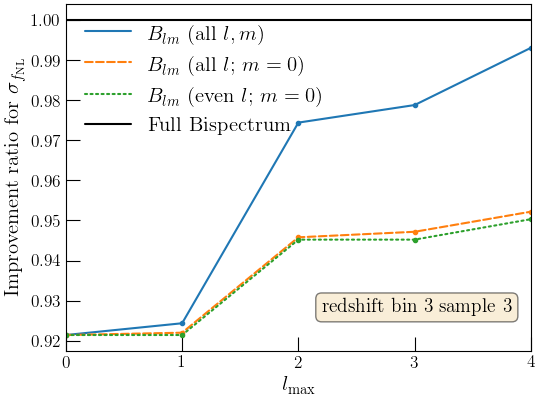}
\end{minipage}
\quad
\begin{minipage}{.45\textwidth}
  \centering
    \includegraphics[width=1.0\textwidth]{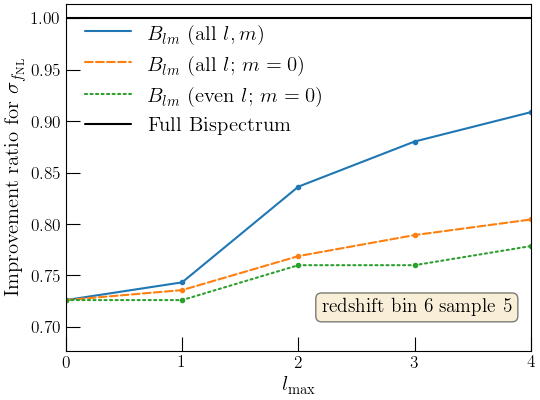}
\end{minipage}
\caption{Plot of the improvement ratio for $\sigma_{\fnl}$ of bispectrum multipoles against the full bispectrum result, for a single tracer with medium and worst redshift error. Left: A representative galaxy sample that contributes well to the total constraint: redshift bin 3, sample 3 ($z_{\rm mid}=0.5$, $\sigma_z/(1+z)=0.03$). Right: The galaxy sample with the worst redshift error: redshift bin 6, sample 5 ($z_{\rm mid}=1.3$, $\sigma_z/(1+z)=0.2$).}
\label{fig:plot_compare_multipoles_improvement_fnl_2}
\end{figure}

We now show the Fisher forecast results for the first six redshift bins ($0 < z < 1.6$) (as we verified that higher redshift bins contribute negligibly to the bispectrum constraint on $\fnl$ for SPHEREx).  We begin by looking at how photometric redshift errors affect the claim that the first three even $l$ and $m = 0$ modes are sufficient for capturing most of the constraining power. Then we investigate how constraints would be impacted if the redshift errors were to change from their fiducial values. Finally, we study how the constraints vary when using different subsets of multi-tracers, as well as different subsets of triangle shapes for various squeezing factors.

\subsection{The impact of photometric error on bispectrum multipoles}

To illustrate how much of the total constraint is captured by a set of bispectrum multipoles, we look at the improvement ratio defined as
\beq
\mathrm{Improvement\ ratio} = \frac{\sigma_{\fnl}(B_{\rm full})}{\sigma_{\fnl}(B_{lm})},
\label{eq:improvement_ratio}
\eeq
where $B_{\rm full}$ stands for the full constraint obtained by using the Fourier space bispectrum with all the triangle shape and orientation bins, whereas $B_{lm}$ stands for bispectrum multipoles with a specific set of $l,m$ values. In this notation, the improvement ratio will always be less than one for the bispectrum multipoles, and higher means better constraints.

We show this improvement ratio for the unmarginalized $\fnl$ uncertainty as a function of $\lmax$ in Figs.~\ref{fig:plot_compare_multipoles_improvement_fnl_1} and~\ref{fig:plot_compare_multipoles_improvement_fnl_2} for individual galaxy samples. Various subsets of multipoles are selected: ``all $l,m$" for all multipoles up to $\lmax$ (blue solid), ``all $l$; $m =0$" for all multipoles up to $\lmax$ but without the nonzero-$m$ modes (orange dashed); ``even $l$; $m = 0$" for all even $l$ up to $\lmax$ without the nonzero-$m$ modes (green dotted). For a spectroscopic survey, we expect that only $l = 0, 2$ and 4 and $m = 0$ will contribute to the signal, when the RSD modeling only includes the linear Kaiser effects and no window function effects are accounted for.

This behavior is indeed observed for tracers with the least redshift uncertainty. An example is shown in Fig.~\ref{fig:plot_compare_multipoles_improvement_fnl_1} for the tracer in redshift bin 6 sample 1 (with $z_{\rm mid} = 1.3$ and $\sigma_z/(1+z) = 0.003$, so giving sub-percent $\sigma_z$). This tracer behaves like a spectroscopic sample~\cite{Gagrani:2016rfy}: $\lmax = 4$ is enough to capture all constraining power; using only $m=0$ modes affects the error by less than 2\%; and removing odd multipoles have no impact on the constraints.

The story is however different when we look at tracers with larger redshift errors. In Fig.~\ref{fig:plot_compare_multipoles_improvement_fnl_2}, we show the same plot for two tracers: One with medium redshift uncertainty, redshift bin 3 sample 3 (with $z_{\rm mid}=0.5$, $\sigma_z/(1+z)=0.03$), which is also a representative sample that contributes well to the combined $\fnl$ constraint from all tracers and redshift bins, and the sample with the worst redshift error, redshift bin 6 sample 5 ($z_{\rm mid}=1.3$, $\sigma_z/(1+z)=0.2$).

The first observation is that $\lmax = 4$ no longer captures all the constraining power there is -- higher multipoles contain information too, because of the way the redshift error modeling is a Gaussian function in $k\mu$ which has non-zero multipole decomposition in all the multipoles. Another impact is that the odd multipoles also contribute slightly to the total constraints.
Finally, removing $m\neq0$ modes has a bigger impact on the $\fnl$ constraint than in the nearly-spectroscopic case: An additional 4\% (10\%) on top of the 1\% (10\%) increase in $\fnl$ uncertainty from truncating at $\lmax = 4$, for the representative (worst) galaxy sample. An overall consequence is that the monopole alone no longer contains most of the information: The uncertainty is degraded by roughly 8\% (23\%) for the representative (worst) sample (as opposed to 2\% in the case of the spectroscopic-like sample). 

While the findings of Ref.~\cite{Gagrani:2016rfy} no longer hold for a photometric redshift sample where errors are naturally bigger than in spectroscopic surveys, this is however not a problem for SPHEREx. When combining all five samples and six redshift bins in the multi-tracer analysis, we find that the marginalized constraint is $\sigfnl = 0.86$ (0.75) respectively for $\lmax = 0$ ($\lmax = 2$) where all the $l$ and $m$ modes have been included up to that $\lmax$. This represents a 18\% (3\%) increase from the full constraint $\sigfnl = 0.73$. 

This behavior closely parallels that of the representative sample with medium redshift uncertainty -- redshift bin 3 sample 3. The reason is that the samples with the worst redshift errors are also the ones that contribute the least to the total constraint in a multi-tracer analysis. Consequently, we can measure $\fnl$ with $\lmax = 2$ with marginal loss in the total constraining power. Note that we do not quote the $\lmax = 4$ case for the the full result because it is computationally expensive to compute in the multi-tracer case as it requires a larger sampling rate in the $\theta_1$ and $\phi_{12}$ parameters for accurate results.

We note also that there are compelling reasons for measuring the odd multipoles, even if they may not contribute significantly to the final $\fnl$ constraints. Because SPHEREx will reach larger scales than in previous surveys, it will start to probe a variety of effects that will become important on these scales. These include general relativistic effects which would be the main cosmological signal in the odd multipoles~\cite{Clarkson:2018dwn, Jeong:2019igb, Maartens:2019yhx, Jolicoeur:2020eup}, as well as wide-angle effects coming from the breakdown of the plane-parallel approximation on large angular separations~\cite{Noorikuhani:2022bwc, Foglieni:2023xca}. The wide-angle effects, together with the window function convolution~\cite{Pardede:2023ddq} would also induce odd multipoles on top of the GR signal. In this regard, measuring the odd $l$'s in the bispectrum as a unique signature for GR effects as well as for cross-checking wide-angle and window function effects in the even multipoles. Of course, the modeling of the bispectrum involving all these large scale effects could become quite involved, and one might choose to do the measurement in the power spectrum instead. We leave the more precise forecast involving all these large scale effects to future work. 

\subsection{Varying the redshift errors} 

Besides spreading the signal into more than just the $l = 0, 2, 4$ and $m = 0$ modes, the presence of photometric redshift error also poses the problem that the $\fnl$ constraint depends sensitively on the precise redshift error modeling (since the bispectrum signal is Gaussian damped where $\sigma_z^2$ is exponentiated).

To quantify how sensitive $\sigfnl$ is to the redshift error, we vary $\sigma_z$ by $\pm$20\% for all samples. This variation is at the level of the difference between choosing to use the maximum or the mean of the measured distribution for $\Delta z \equiv z_{\rm measured} - z_{\rm true}$ of a given sample. We find that the marginalized constraint $\sigfnl$ from the multi-tracer analysis of the first six redshift bins varied by $\pm8\%$ when the redshift error was changed by $\pm$20\%, giving $\sigfnl = 0.79$ and 0.67 respectively. 

Looking at individual redshift bins, we find that varying $\sigma_z$ by $\pm$20\% gives a $\pm 5\%$ change in $\sigfnl$ for redshift bin 3, whereas for the higher redshift bins the induced change in $\sigfnl$ was larger (about $12-13\%$ for example for redshift bin 6) because the larger variation in the redshift error $\sigma_z = \tilde{\sigma}_z (1+z)$.

\subsection{Varying multi-tracer combination}

\begin{figure*}[ht]
    \includegraphics[width=1.0\textwidth]{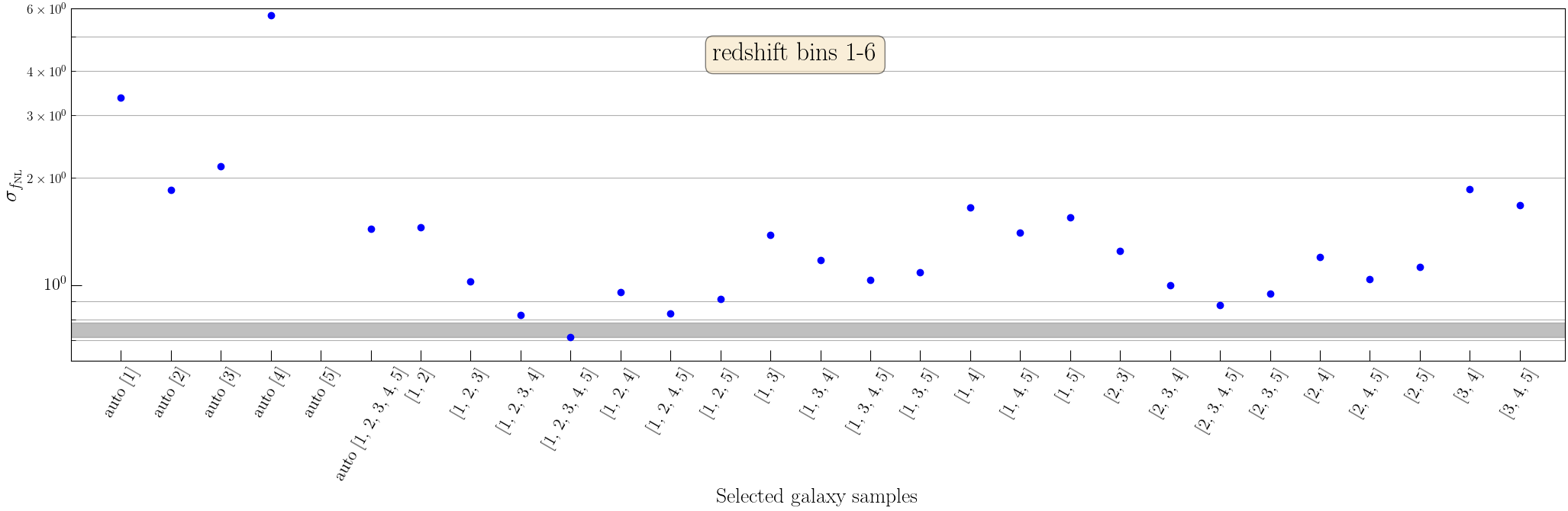}
    \caption{Plot of the marginalized $\sigma_{\fnl}$ for the full bispectrum in a single redshift bin (redhsift bin 3) vs various multi-tracer combinations. The shaded band contains constraints that are within 10\% of the best constraint from the 5-tracer combination. The worst constraints are outside the plot.}
    \label{fig:plot_error_fnl_vs_multitracer_combination}
\end{figure*}

Another interesting question to ask is whether we really need all five galaxy samples. The more galaxy samples we use, the higher the number of multi-tracer bispectrum combinations $n_{b} = n_{\rm samples}^3$. Because the number of mocks for accurately evaluating the covariance matrix using simulations needs to be much larger than the data vector size~\cite{Hartlap:2006kj, Taylor:2014ota}, a smaller data vector is more desirable. In Fig.~\ref{fig:plot_error_fnl_vs_multitracer_combination}, we plot the marginalized $\sigfnl$ from the Fourier bispectrum as a function of various multi-tracer combinations for the first six redshift bins combined. The plot is zoomed so that for some combinations the marginalized error is too high to be seen on the plot. 

We first notice the single-tracer results on the left of the plot, as well as the auto-bispectrum only result using all five tracers. We note that the samples 4 and 5 from all redshift bins combined are the least constraining by themselves, because of the large redshift error they have. The first samples are the most constraining ones, as expected. 

Next, we note that the following combinations are similar in $\fnl$ constraining power: The auto-only result from all five redshift bins (auto [1,2,3,4,5]), and the full multi-tracer result [1,2]. The latter includes the cross-bispectra (e.g. 112, 121, etc.) and already shows the power of the multi-tracer analysis to cancel cosmic variance. 

As we go further to the right side of the plot, we see that adding samples 3, 4 and 5 gradually gets down to the final result of $\sigfnl = 0.73$. Even if the samples 4 and 5 by themselves are not very constraining, they contribute to the cancellation of cosmic variance, with their large number densities -- it is easier to find a larger number of galaxies where the redshift error is measured with worse uncertainties. 

Finally, we show in a grey band 10\% degradation from the best $\fnl$ error from all five tracers. The closest combinations slightly above the 10\% line would be [1,2,3,4] and [1,2,4,5], both of which are four-tracer results, showing the importance of using the multi-tracer analysis to achieve the finest constraints. 

Because the redshift measurement pipeline is still under research and development, it is also interesting to ask ourselves how important it is for the best galaxy sample to reach the required redshift uncertainty of $\tilde{\sigma}_z \leq 0.003$ in order to measured $\sigfnl$ to its desired accuracy. We see from Fig.~\ref{fig:plot_error_fnl_vs_multitracer_combination} that removing the best sample completely ([2,3,4,5]) would lead to a $\sim$20\% degradation in the $\fnl$ error. 

Finally, we report that using the first 2, 3 or 4 best tracers, we obtain $\sigfnl = 1.4, 1.0$ and 0.8 respectively, representing a 91\%, 37\% and 12\% degradation respectively from the five-tracer version. Note however that the best set of multi-tracer combinations for the power spectrum may not be the same as that for the bispectrum, so in a combined analysis with both the power spectrum and the bispectrum, one may choose to select different subsets of tracers, as long as the covariance between the resulting multi-tracer combinations are properly accounted for.

\subsection{Squeezed triangles only result}

Finally, we explore how the $\fnl$ constraints are impacted if we choose to use only a subset of the available triangle shapes, namely triangles with squeezing factors $k_2/k_1$ and $k_3/k_1$ above a threshold $S_{\rm min}$, and we recall that we order $k_1 \leq k_2 \leq k_3$ for unique triangle shapes. 

As an example, we report results for the redshift bin 3 with all its five tracers in Table~\ref{tab:squeezed_triangles}. For the $k$-binning we've adopted, we find that the marginalized $\sigma(\fnl)$ only degrades by 17\% for $S_{\rm min} = 4$, while the number of triangles are dramatically reduced by a factor of 9 from 12000 to 1294. For a more intermediate case with $S_{\rm min} = 2$, the marginalized $\sigma(\fnl)$ only degrades by 11\% while the number of triangles is reduced by a factor of 3. 

The trade-off here is that the marginalized uncertainty for the linear galaxy biases takes a hit: About a factor of $\sim1.6$ ($\sim2.8$) worse for $S_{\rm min} = 2$ ($S_{\rm min} = 4$). This degradation in the galaxy bias uncertainties is roughly independent of the redshift error of the sample, since the linear galaxy bias for all five samples have a similar factor of degradation. It would be interesting to see whether the joint PS-Bis constraints on the galaxy biases would suffer less from excluding triangles, as the total constraint may rely more on the degeneracy breaking between the power spectrum and the bispectrum instead of the total number of triangles available.

\begin{table}[]
\begin{tabular}{|l|l|l|l|}
\hline
                  & $S_{\rm min}=2$   & $S_{\rm min}=3$   & $S_{\rm min}=4$   \\ \hline
$\sigma(\fnl)$                  & 1.06   & 1.11   & 1.17     \\ \hline
$\sigma(A_s)$                   & 1.4   & 1.8   & 2.2 \\ \hline
$\sigma(n_s)$                   & 1.4   & 1.8   & 2.4   \\ \hline
$\sigma(n_{\rm run})$           & 1.4   & 1.9   & 2.2   \\ \hline
$\sigma(b_{10}^1)$              & 1.6   & 2.2   & 2.8   \\ \hline
$N_{\rm triangles}$             & 3964  & 2072  & 1294 \\ \hline
\end{tabular}
\caption{Ratio of the marginalized parameter constraints for a subset of the triangle shapes with squeezing factor threshold $S_{\rm min}$ compared to all triangles. Constraints are for the 5-tracer analysis in the redshift bin 3, and are marginalized over the parameter set \{$A_s, n_s, \fnl, n_{\rm run}, \Omega_b h^2, \Omega_c h^2, 100\theta_{\rm MC}, b_{10}^X\,|\,X = 1..5$\}. We find that the number of triangles is reduced by a factor of 3, 6 and 9 respectively compared to using all triangles for the subsets with $S_{\rm min} = 2, 3$ and 4, while the fractional increase in the marginalized $\fnl$ error is only $6\%, 11\%$ and $17\%$ respectively. The linear bias parameters are however more strongly impacted (we show only for the bias for the first tracer as the other ones are similar).}
\label{tab:squeezed_triangles}
\end{table}

\section{Summary and discussion}
\label{sec:conclusion}

In this paper, we explored how the presence of photometric redshift errors alter the well-known claim that bispectrum multipoles with $l = 0, 2, 4$ and $m = 0$ capture most of the constraint on cosmological parameters, which was shown to be valid for spectroscopic surveys in Refs.~\cite{Gagrani:2016rfy} and~\cite{Byun:2022rvn}. We showed, in the context of our updated bispectrum forecast for SPHEREx, that individual galaxy samples with sufficient redshift errors suffer from information leaking into the higher multipoles, the $m \neq 0$ modes, and the odd multipoles. We expect this to also hold true for the power spectrum multipoles, though we did not demonstrate this explicitly.

Future photometric redshift surveys would need to account for this effect when measuring the bispectrum or power spectrum multipoles. In particular, we found that restricting to the monopole alone affect the unmarginalized $\fnl$ error by about 2\% for a nearly spectroscopic sample ($\sigma_z/(1+z) = 0.003$), whereas it is about 30\% for the SPHEREx sample with the worst redshift error ($\sigma_z/(1+z) = 0.2$). For reference, LSST has the requirement that $\sigma_z/(1+z) < 0.05$ with the goal being 0.02.~\cite{2009arXiv0912.0201L}, similar to the SPHEREx sample 3.

The behavior of the total result from combining all five samples and the first six redshift bins is dominated by the best redshift accuracy samples and does not suffer as severely. More precisely, using $l=0$ alone gave a marginalized error of $\sigma_{\fnl} = 0.86$, which is $18\%$ degradation compared to the full result using the Fourier bispectrum $\sigma_{\fnl} = 0.73$, whereas using $l_{\rm max} = 2$ with all $l,m$ modes gave $\sigma_{\fnl} = 0.75$, within 3\% of the full error.

Beside the above effects, the photo-$z$ error is also important to control carefully because of how sensitively it can affect the constraining power: Varying the redshift errors of all samples by $\pm 20\%$ led to a $\pm 8\%$ change on the final $\sigma_{\fnl}$ for the SPHEREx bispectrum . This variation is at the level of the difference between the mean and the maximum of the redshift error distribution for a given sample. This finding motivates future work in which we would need to characterize the precise shape of the photo-z error distribution from simulations and investigate the impact of its associated uncertainties on parameter constraints. 

Finally, we also explored the trade-off in constraining power that comes with reducing the data vector size by selecting subsets of the multi-tracer combinations and subsets of triangle shapes. We found that although samples 4 and 5 have large redshift errors and are not constraining by themselves, their larger number densities do help with reducing cosmic variance when used in combination with the other samples in a multi-tracer setting. We found that the first two, three and four tracers would raise the $\fnl$ error by $91\%, 37\%$ and $12\%$ respectively from the five-tracer result. Correspondingly, the data vector size reduction would be a factor of 16, 5, and 2 for using two, three and four tracers respectively, which does not seem to be compelling enough given the large amount of $\fnl$ constraint it would sacrifice.

For the triangle shapes, we looked at subsets of triangles with squeezing factor thresholds $S_{\rm min} = 2, 3$ and 4, and found in the case of a representative redshift bin (redshift bin 3), the marginalized $\fnl$ error went up by 6\%, 11\% and 17\% while the data vector size reduced by a factor of 3, 6 and 9  respectively. The bias parameters errors were the most impacted, going up by a factor 2 to 3 depending on the $S_{\rm min}$ chosen and in a fashion mostly independent of the redshift uncertainty of the sample. This is because the $\fnl$ constraint is dominated by squeezed triangles, whereas the galaxy bias parameter constraints receive contributions from all triangles. Since the power spectrum would also be constraining the bias parameters and the other cosmological parameters, singling out highly squeezed triangles might be an interesting option to reduce the data vector size in a combined analysis. 

Our fiducial analysis here can be extended in various ways. We have only explored a linear regime bispectrum forecast, staying within $k_{\rm max} = 0.2 (1+z) \ihMpc$. An accurate modeling of the bispectrum in the nonlinear regime would extend the $k$-range to higher values, and could increase the constraining power on $\fnl$ as the squeezing factor would also increase. There are a variety of redshift space effects that can be improved in our modeling, for example including AP effects and using a more precise redshift error distribution from simulations instead of assuming a Gaussian distribution. Moreover, we have omitted the modeling of wide-angle and GR effects which appear on large scales, as well as window function effects. The presence of wide-angle and window function effects would lead to the mixing the signal between different multipoles.

Additionally, it would also be highly desirable to explore compression methods for the bispectrum. Examples include the modal bispectrum (which has been recently extended to include RSD and $\fnl$ effects~\cite{Byun:2022rvn}), the Massively Optimized Parameter Estimation and Data compression technique (MOPED)~\cite{Heavens:1999am} (which uses 
the score function -- the gradient of the log likelihood -- to achieve a compression where the data vector size is the same as the number of parameters of interest) and its generalized versions leading to likelihood-free inference~\cite{Alsing:2017var} (for an application to the galaxy bispectrum, see e.g.~\cite{Gualdi:2017iey}).

In comparison with the previous SPHEREx forecast in Ref.~\cite{Dore:2014cca}, our assumptions and modeling differed in the following ways. We used a slightly more conservative $k_{\rm max} = 0.2 (1+z) \ihMpc$ and used only a linear modeling whereas the previous forecast had $k_{\rm max} = 0.25 (1+z) \ihMpc$ using a simple nonlinear modeling. We treated the signal dependence on the triangle orientations in order to model the impact of RSD and photometric redshift errors more accurately instead of using a cutoff in $k_{\parallel}$. We employed a slightly different bias prescription -- while we still only varied the linear galaxy biases, we modeled the signal with a few more second-order effects in addition to the $b_{20}$ term, namely the $b_{s_2}$, $b_{11}$, $b_{02}$ and $n^2$ terms which depended on the values of $b_{10}$.

Finally, we performed a bispectrum-only forecast for SPHEREx, whereas adding the power spectrum should be helpful in breaking various parameter degeneracies (e.g.~\cite{Gualdi:2021yvq}). In particular, we expect that the combined power spectrum and bispectrum forecast would improve significantly on the $n_{\rm run}$ constraint than that from the power spectrum or bispectrum alone~\cite{Dore:2014cca}. We also expect that there would be more degeneracy breaking for galaxy bias parameters. Adding the bispectrum should also help to mildly break the complete degeneracy between $b_{\phi} \fnl$ and $\fnl$ in the power spectrum, should one choose to use these parameter combinations~\cite{Barreira:2020ekm}. To do so accurately, we would need to take into account the covariance between the power spectrum and the bispectrum. Moreover, we have assumed a Gaussian covariance throughout our forecast, whereas including the non-Gaussian covariance could affect the constraints on $\fnl$~\cite{Biagetti:2021tua, Floss:2022wkq}. We leave these considerations for future work. 

In sum, the bispectrum multipoles are an important statistics that current and future surveys will measure in order to best constrain the primordial non-Gaussianity and study its implications for inflation. Being specifically optimized for this goal, the SPHEREx mission will achieve an all-sky observation in 102 NIR bands, making available multiple galaxy samples with redshift measurements ranging from spectrocopic-like to photometric-like. We presented a refined forecast that is able to account for the redshift space effects more accurately than before, and studied their impact on $\fnl$ constraints. Characterizing the constraining power of SPHEREx will be important as we seek to understand and make use of this dataset to shed light on the details of how inflation proceeded. 

\section*{Acknowledgement}

CH: I acknowledge my Maker for providing an amazing team of SPHEREx members who were supportive throughout the work, for the ability to do this work, and for the wisdom needed to navigate the project. I thank in particular Joyce Byun who provided a code comparison for the bispectrum multipoles, Yi-Kuan Chiang for useful feedback on the redshift error results, and Jamie Bock and the rest of the SPHEREx cosmology team who provided useful discussions and helped shaping the project. We thank Chris Hirata for careful feedback on the manuscript. We are grateful to the Texas Advanced Computing Center for the computing resources that enabled this work. All the authors thank the SPHEREx mission for providing the funding needed to accomplish this work.  Part of this work was done at Jet Propulsion
Laboratory, California Institute of Technology, under a contract with the National Aeronautics and Space Administration.

\begin{appendix}

\section{Survey Parameters}
\label{sec:appendix}

We now list the redshift bins and the fiducial values for the galaxy linear biases and number densities used in this work in Tables~\ref{tab:redshift_bin_definition}, ~\ref{tab:galaxy_bias} and~\ref{tab:number_density} respectively. 

\begin{table}[ht!]
\begin{tabular}{|c|c|c|}
\hline
Redshift bin & $z_{\rm min}$  & $z_{\rm min}$   \\ \hline
1  & 0 & 0.2  \\ \hline
2  & 0.2 & 0.4 \\ \hline
3  & 0.4 & 0.6  \\ \hline
4  & 0.6 & 0.8 \\ \hline
5  & 0.8 & 1.0  \\ \hline
6  & 1.0 & 1.6  \\ \hline
7  & 1.6 & 2.2  \\ \hline
8  & 2.2 & 2.8 \\ \hline
9  & 2.8 & 3.4  \\ \hline
10 & 3.4 & 4.0  \\ \hline
11 & 4.0 & 4.6  \\ \hline
\end{tabular}
\caption{Redshift bin definition for the 11 redshift bins for SPHEREx.}
\label{tab:redshift_bin_definition}
\end{table}
\begin{table}[ht!]
\begin{tabular}{|c|c|c|c|c|c|}
\hline
Redshift bin & Sample 1   & Sample 2   & Sample 3   & Sample 4    & Sample 5    \\ \hline
1  & 1.3 & 1.2 & 1.0 & 0.98 & 0.83 \\ \hline
2  & 1.5 & 1.4 & 1.3 & 1.3  & 1.2  \\ \hline
3  & 1.8 & 1.6 & 1.5 & 1.4  & 1.3  \\ \hline
4  & 2.3 & 1.9 & 1.7 & 1.5  & 1.4  \\ \hline
5  & 2.1 & 2.3 & 1.9 & 1.7  & 1.6  \\ \hline
6  & 2.7 & 2.6 & 2.6 & 2.2  & 2.1  \\ \hline
7  & 3.6 & 3.4 & 3.0 & 3.6  & 3.2  \\ \hline
8  & 2.3 & 4.2 & 3.2 & 3.7  & 4.2  \\ \hline
9  & 3.2 & 4.3 & 3.5 & 2.7  & 4.1  \\ \hline
10 & 2.7 & 3.7 & 4.1 & 2.9  & 4.5  \\ \hline
11 & 3.8 & 4.6 & 5.0 & 5.0  & 5.0  \\ \hline
\end{tabular}
\caption{Fiducial linear galaxy bias $b_{10}$ for SPHEREx's 5 galaxy samples and 11 redshift bins.}
\label{tab:galaxy_bias}
\end{table}
\begin{table*}[]
\begin{tabular}{|c|c|c|c|c|c|}
\hline
Redshift bin & Sample 1        & Sample 2        & Sample 3        & Sample 4        & Sample 5        \\ \hline
1  & $9.97\times10^{-3}$  & $1.23\times10^{-2}$   & $1.34\times10^{-2}$   & $2.29\times10^{-2}$   & $1.49\times10^{-2}$   \\ \hline
2  & $4.11\times10^{-3}$  & $8.56\times10^{-3}$  & $8.57\times10^{-3}$  & $1.29\times10^{-2}$   & $7.52\times10^{-3}$  \\ \hline
3  & $5.01\times10^{-4}$ & $2.82\times10^{-3}$  & $3.62\times10^{-3}$  & $5.35\times10^{-3}$  & $3.27\times10^{-3}$  \\ \hline
4  & $7.05\times10^{-5}$ & $9.37\times10^{-4}$ & $2.94\times10^{-3}$  & $4.95\times10^{-3}$  & $2.50\times10^{-3}$   \\ \hline
5  & $3.16\times10^{-5}$ & $4.30\times10^{-4}$  & $2.04\times10^{-3}$  & $4.15\times10^{-3}$  & $1.83\times10^{-3}$  \\ \hline
6  & $1.64\times10^{-5}$ & $5.00\times10^{-5}$    & $2.12\times10^{-4}$ & $7.96\times10^{-4}$ & $7.34\times10^{-4}$ \\ \hline
7  & $3.59\times10^{-6}$ & $8.03\times10^{-6}$ & $6.97\times10^{-6}$ & $7.75\times10^{-5}$ & $2.53\times10^{-4}$ \\ \hline
8  & $8.07\times10^{-7}$ & $3.83\times10^{-6}$ & $2.02\times10^{-6}$ & $7.87\times10^{-6}$ & $5.41\times10^{-5}$ \\ \hline
9  & $1.84\times10^{-6}$ & $3.28\times10^{-6}$ & $1.43\times10^{-6}$ & $2.46\times10^{-6}$ & $2.99\times10^{-5}$ \\ \hline
10 & $1.50\times10^{-6} $ & $1.07\times10^{-6}$ & $1.93\times10^{-6}$ & $1.93\times10^{-6}$ & $9.41\times10^{-6}$ \\ \hline
11 & $1.13\times10^{-6}$ & $6.79\times10^{-7}$ & $6.79\times10^{-7}$ & $1.36\times10^{-6}$ & $2.04\times10^{-6}$ \\ \hline
\end{tabular}
\caption{Galaxy number density in ($h$/Mpc)$^3$ for SPHEREx's 5 galaxy samples and 11 redshift bins. }
\label{tab:number_density}
\end{table*}

\end{appendix}

\bibliography{bismult.bib}

\begin{thebibliography}{50}%
\makeatletter
\providecommand \@ifxundefined [1]{%
 \@ifx{#1\undefined}
}%
\providecommand \@ifnum [1]{%
 \ifnum #1\expandafter \@firstoftwo
 \else \expandafter \@secondoftwo
 \fi
}%
\providecommand \@ifx [1]{%
 \ifx #1\expandafter \@firstoftwo
 \else \expandafter \@secondoftwo
 \fi
}%
\providecommand \natexlab [1]{#1}%
\providecommand \enquote  [1]{``#1''}%
\providecommand \bibnamefont  [1]{#1}%
\providecommand \bibfnamefont [1]{#1}%
\providecommand \citenamefont [1]{#1}%
\providecommand \href@noop [0]{\@secondoftwo}%
\providecommand \href [0]{\begingroup \@sanitize@url \@href}%
\providecommand \@href[1]{\@@startlink{#1}\@@href}%
\providecommand \@@href[1]{\endgroup#1\@@endlink}%
\providecommand \@sanitize@url [0]{\catcode `\\12\catcode `\$12\catcode
  `\&12\catcode `\#12\catcode `\^12\catcode `\_12\catcode `\%12\relax}%
\providecommand \@@startlink[1]{}%
\providecommand \@@endlink[0]{}%
\providecommand \url  [0]{\begingroup\@sanitize@url \@url }%
\providecommand \@url [1]{\endgroup\@href {#1}{\urlprefix }}%
\providecommand \urlprefix  [0]{URL }%
\providecommand \Eprint [0]{\href }%
\providecommand \doibase [0]{http://dx.doi.org/}%
\providecommand \selectlanguage [0]{\@gobble}%
\providecommand \bibinfo  [0]{\@secondoftwo}%
\providecommand \bibfield  [0]{\@secondoftwo}%
\providecommand \translation [1]{[#1]}%
\providecommand \BibitemOpen [0]{}%
\providecommand \bibitemStop [0]{}%
\providecommand \bibitemNoStop [0]{.\EOS\space}%
\providecommand \EOS [0]{\spacefactor3000\relax}%
\providecommand \BibitemShut  [1]{\csname bibitem#1\endcsname}%
\let\auto@bib@innerbib\@empty
\bibitem [{\citenamefont {Maldacena}(2003)}]{Maldacena:2002vr}%
  \BibitemOpen
  \bibfield  {author} {\bibinfo {author} {\bibfnamefont {J.~M.}\ \bibnamefont
  {Maldacena}},\ }\href {\doibase 10.1088/1126-6708/2003/05/013} {\bibfield
  {journal} {\bibinfo  {journal} {JHEP}\ }\textbf {\bibinfo {volume} {05}},\
  \bibinfo {pages} {013} (\bibinfo {year} {2003})},\ \Eprint
  {http://arxiv.org/abs/astro-ph/0210603} {arXiv:astro-ph/0210603} \BibitemShut
  {NoStop}%
\bibitem [{\citenamefont {Creminelli}\ and\ \citenamefont
  {Zaldarriaga}(2004)}]{Creminelli:2004yq}%
  \BibitemOpen
  \bibfield  {author} {\bibinfo {author} {\bibfnamefont {P.}~\bibnamefont
  {Creminelli}}\ and\ \bibinfo {author} {\bibfnamefont {M.}~\bibnamefont
  {Zaldarriaga}},\ }\href {\doibase 10.1088/1475-7516/2004/10/006} {\bibfield
  {journal} {\bibinfo  {journal} {JCAP}\ }\textbf {\bibinfo {volume} {10}},\
  \bibinfo {pages} {006} (\bibinfo {year} {2004})},\ \Eprint
  {http://arxiv.org/abs/astro-ph/0407059} {arXiv:astro-ph/0407059} \BibitemShut
  {NoStop}%
\bibitem [{\citenamefont {Acquaviva}\ \emph {et~al.}(2003)\citenamefont
  {Acquaviva}, \citenamefont {Bartolo}, \citenamefont {Matarrese},\ and\
  \citenamefont {Riotto}}]{Acquaviva:2002ud}%
  \BibitemOpen
  \bibfield  {author} {\bibinfo {author} {\bibfnamefont {V.}~\bibnamefont
  {Acquaviva}}, \bibinfo {author} {\bibfnamefont {N.}~\bibnamefont {Bartolo}},
  \bibinfo {author} {\bibfnamefont {S.}~\bibnamefont {Matarrese}}, \ and\
  \bibinfo {author} {\bibfnamefont {A.}~\bibnamefont {Riotto}},\ }\href
  {\doibase 10.1016/S0550-3213(03)00550-9} {\bibfield  {journal} {\bibinfo
  {journal} {Nucl. Phys. B}\ }\textbf {\bibinfo {volume} {667}},\ \bibinfo
  {pages} {119} (\bibinfo {year} {2003})},\ \Eprint
  {http://arxiv.org/abs/astro-ph/0209156} {arXiv:astro-ph/0209156} \BibitemShut
  {NoStop}%
\bibitem [{\citenamefont {Bartolo}\ \emph {et~al.}(2004)\citenamefont
  {Bartolo}, \citenamefont {Komatsu}, \citenamefont {Matarrese},\ and\
  \citenamefont {Riotto}}]{Bartolo:2004if}%
  \BibitemOpen
  \bibfield  {author} {\bibinfo {author} {\bibfnamefont {N.}~\bibnamefont
  {Bartolo}}, \bibinfo {author} {\bibfnamefont {E.}~\bibnamefont {Komatsu}},
  \bibinfo {author} {\bibfnamefont {S.}~\bibnamefont {Matarrese}}, \ and\
  \bibinfo {author} {\bibfnamefont {A.}~\bibnamefont {Riotto}},\ }\href
  {\doibase 10.1016/j.physrep.2004.08.022} {\bibfield  {journal} {\bibinfo
  {journal} {Phys. Rept.}\ }\textbf {\bibinfo {volume} {402}},\ \bibinfo
  {pages} {103} (\bibinfo {year} {2004})},\ \Eprint
  {http://arxiv.org/abs/astro-ph/0406398} {arXiv:astro-ph/0406398} \BibitemShut
  {NoStop}%
\bibitem [{\citenamefont {Akrami}\ \emph {et~al.}(2020)\citenamefont {Akrami}
  \emph {et~al.}}]{Planck:2019kim}%
  \BibitemOpen
  \bibfield  {author} {\bibinfo {author} {\bibfnamefont {Y.}~\bibnamefont
  {Akrami}} \emph {et~al.} (\bibinfo {collaboration} {Planck}),\ }\href
  {\doibase 10.1051/0004-6361/201935891} {\bibfield  {journal} {\bibinfo
  {journal} {Astron. Astrophys.}\ }\textbf {\bibinfo {volume} {641}},\ \bibinfo
  {pages} {A9} (\bibinfo {year} {2020})},\ \Eprint
  {http://arxiv.org/abs/1905.05697} {arXiv:1905.05697 [astro-ph.CO]}
  \BibitemShut {NoStop}%
\bibitem [{\citenamefont {{Mueller}}\ \emph {et~al.}(2022)\citenamefont
  {{Mueller}}, \citenamefont {{Rezaie}}, \citenamefont {{Percival}},
  \citenamefont {{Ross}}, \citenamefont {{Ruggeri}}, \citenamefont {{Seo}},
  \citenamefont {{Gil-Mar{\'\i}n}}, \citenamefont {{Bautista}}, \citenamefont
  {{Brownstein}}, \citenamefont {{Dawson}}, \citenamefont {{de la Macorra}},
  \citenamefont {{Palanque-Delabrouille}}, \citenamefont {{Rossi}},
  \citenamefont {{Schneider}},\ and\ \citenamefont
  {{Y{\'e}che}}}]{2022MNRAS.514.3396M}%
  \BibitemOpen
  \bibfield  {author} {\bibinfo {author} {\bibfnamefont {E.-M.}\ \bibnamefont
  {{Mueller}}}, \bibinfo {author} {\bibfnamefont {M.}~\bibnamefont {{Rezaie}}},
  \bibinfo {author} {\bibfnamefont {W.~J.}\ \bibnamefont {{Percival}}},
  \bibinfo {author} {\bibfnamefont {A.~J.}\ \bibnamefont {{Ross}}}, \bibinfo
  {author} {\bibfnamefont {R.}~\bibnamefont {{Ruggeri}}}, \bibinfo {author}
  {\bibfnamefont {H.-J.}\ \bibnamefont {{Seo}}}, \bibinfo {author}
  {\bibfnamefont {H.}~\bibnamefont {{Gil-Mar{\'\i}n}}}, \bibinfo {author}
  {\bibfnamefont {J.}~\bibnamefont {{Bautista}}}, \bibinfo {author}
  {\bibfnamefont {J.~R.}\ \bibnamefont {{Brownstein}}}, \bibinfo {author}
  {\bibfnamefont {K.}~\bibnamefont {{Dawson}}}, \bibinfo {author}
  {\bibfnamefont {A.}~\bibnamefont {{de la Macorra}}}, \bibinfo {author}
  {\bibfnamefont {N.}~\bibnamefont {{Palanque-Delabrouille}}}, \bibinfo
  {author} {\bibfnamefont {G.}~\bibnamefont {{Rossi}}}, \bibinfo {author}
  {\bibfnamefont {D.~P.}\ \bibnamefont {{Schneider}}}, \ and\ \bibinfo {author}
  {\bibfnamefont {C.}~\bibnamefont {{Y{\'e}che}}},\ }\href {\doibase
  10.1093/mnras/stac812} {\bibfield  {journal} {\bibinfo  {journal} {\mnras}\
  }\textbf {\bibinfo {volume} {514}},\ \bibinfo {pages} {3396} (\bibinfo {year}
  {2022})}\BibitemShut {NoStop}%
\bibitem [{\citenamefont {Castorina}\ \emph {et~al.}(2019)\citenamefont
  {Castorina} \emph {et~al.}}]{Castorina:2019wmr}%
  \BibitemOpen
  \bibfield  {author} {\bibinfo {author} {\bibfnamefont {E.}~\bibnamefont
  {Castorina}} \emph {et~al.},\ }\href {\doibase 10.1088/1475-7516/2019/09/010}
  {\bibfield  {journal} {\bibinfo  {journal} {JCAP}\ }\textbf {\bibinfo
  {volume} {09}},\ \bibinfo {pages} {010} (\bibinfo {year} {2019})},\ \Eprint
  {http://arxiv.org/abs/1904.08859} {arXiv:1904.08859 [astro-ph.CO]}
  \BibitemShut {NoStop}%
\bibitem [{\citenamefont {Cabass}\ \emph {et~al.}(2022)\citenamefont {Cabass},
  \citenamefont {Ivanov}, \citenamefont {Philcox}, \citenamefont
  {Simonovi\'c},\ and\ \citenamefont {Zaldarriaga}}]{Cabass:2022ymb}%
  \BibitemOpen
  \bibfield  {author} {\bibinfo {author} {\bibfnamefont {G.}~\bibnamefont
  {Cabass}}, \bibinfo {author} {\bibfnamefont {M.~M.}\ \bibnamefont {Ivanov}},
  \bibinfo {author} {\bibfnamefont {O.~H.~E.}\ \bibnamefont {Philcox}},
  \bibinfo {author} {\bibfnamefont {M.}~\bibnamefont {Simonovi\'c}}, \ and\
  \bibinfo {author} {\bibfnamefont {M.}~\bibnamefont {Zaldarriaga}},\ }\href
  {\doibase 10.1103/PhysRevD.106.043506} {\bibfield  {journal} {\bibinfo
  {journal} {Phys. Rev. D}\ }\textbf {\bibinfo {volume} {106}},\ \bibinfo
  {pages} {043506} (\bibinfo {year} {2022})},\ \Eprint
  {http://arxiv.org/abs/2204.01781} {arXiv:2204.01781 [astro-ph.CO]}
  \BibitemShut {NoStop}%
\bibitem [{\citenamefont {D'Amico}\ \emph {et~al.}(2022)\citenamefont
  {D'Amico}, \citenamefont {Lewandowski}, \citenamefont {Senatore},\ and\
  \citenamefont {Zhang}}]{DAmico:2022gki}%
  \BibitemOpen
  \bibfield  {author} {\bibinfo {author} {\bibfnamefont {G.}~\bibnamefont
  {D'Amico}}, \bibinfo {author} {\bibfnamefont {M.}~\bibnamefont
  {Lewandowski}}, \bibinfo {author} {\bibfnamefont {L.}~\bibnamefont
  {Senatore}}, \ and\ \bibinfo {author} {\bibfnamefont {P.}~\bibnamefont
  {Zhang}},\ }\href@noop {} {\  (\bibinfo {year} {2022})},\ \Eprint
  {http://arxiv.org/abs/2201.11518} {arXiv:2201.11518 [astro-ph.CO]}
  \BibitemShut {NoStop}%
\bibitem [{\citenamefont {Cagliari}\ \emph {et~al.}(2023)\citenamefont
  {Cagliari}, \citenamefont {Castorina}, \citenamefont {Bonici},\ and\
  \citenamefont {Bianchi}}]{Cagliari:2023mkq}%
  \BibitemOpen
  \bibfield  {author} {\bibinfo {author} {\bibfnamefont {M.~S.}\ \bibnamefont
  {Cagliari}}, \bibinfo {author} {\bibfnamefont {E.}~\bibnamefont {Castorina}},
  \bibinfo {author} {\bibfnamefont {M.}~\bibnamefont {Bonici}}, \ and\ \bibinfo
  {author} {\bibfnamefont {D.}~\bibnamefont {Bianchi}},\ }\href@noop {} {\
  (\bibinfo {year} {2023})},\ \Eprint {http://arxiv.org/abs/2309.15814}
  {arXiv:2309.15814 [astro-ph.CO]} \BibitemShut {NoStop}%
\bibitem [{\citenamefont {Rezaie}\ \emph {et~al.}(2023)\citenamefont {Rezaie}
  \emph {et~al.}}]{Rezaie:2023lvi}%
  \BibitemOpen
  \bibfield  {author} {\bibinfo {author} {\bibfnamefont {M.}~\bibnamefont
  {Rezaie}} \emph {et~al.},\ }\href@noop {} {\  (\bibinfo {year} {2023})},\
  \Eprint {http://arxiv.org/abs/2307.01753} {arXiv:2307.01753 [astro-ph.CO]}
  \BibitemShut {NoStop}%
\bibitem [{\citenamefont {Amendola}\ \emph {et~al.}(2018)\citenamefont
  {Amendola} \emph {et~al.}}]{Amendola:2016saw}%
  \BibitemOpen
  \bibfield  {author} {\bibinfo {author} {\bibfnamefont {L.}~\bibnamefont
  {Amendola}} \emph {et~al.},\ }\href {\doibase 10.1007/s41114-017-0010-3}
  {\bibfield  {journal} {\bibinfo  {journal} {Living Rev. Rel.}\ }\textbf
  {\bibinfo {volume} {21}},\ \bibinfo {pages} {2} (\bibinfo {year} {2018})},\
  \Eprint {http://arxiv.org/abs/1606.00180} {arXiv:1606.00180 [astro-ph.CO]}
  \BibitemShut {NoStop}%
\bibitem [{\citenamefont {{DESI Collaboration}}\ \emph
  {et~al.}(2016)\citenamefont {{DESI Collaboration}}, \citenamefont
  {{Aghamousa}}, \citenamefont {{Aguilar}}, \citenamefont {{Ahlen}},
  \citenamefont {{Alam}}, \citenamefont {{Allen}}, \citenamefont {{Allende
  Prieto}}, \citenamefont {{Annis}}, \citenamefont {{Bailey}}, \citenamefont
  {{Balland}}, \citenamefont {{Ballester}}, \citenamefont {{Baltay}},
  \citenamefont {{Beaufore}}, \citenamefont {{Bebek}}, \citenamefont {{Beers}},
  \citenamefont {{Bell}}, \citenamefont {{Bernal}}, \citenamefont {{Besuner}},
  \citenamefont {{Beutler}}, \citenamefont {{Blake}}, \citenamefont
  {{Bleuler}}, \citenamefont {{Blomqvist}}, \citenamefont {{Blum}},
  \citenamefont {{Bolton}}, \citenamefont {{Briceno}}, \citenamefont
  {{Brooks}}, \citenamefont {{Brownstein}}, \citenamefont {{Buckley-Geer}},
  \citenamefont {{Burden}}, \citenamefont {{Burtin}}, \citenamefont {{Busca}},
  \citenamefont {{Cahn}}, \citenamefont {{Cai}}, \citenamefont {{Cardiel-Sas}},
  \citenamefont {{Carlberg}}, \citenamefont {{Carton}}, \citenamefont
  {{Casas}}, \citenamefont {{Castander}}, \citenamefont {{Cervantes-Cota}},
  \citenamefont {{Claybaugh}}, \citenamefont {{Close}}, \citenamefont
  {{Coker}}, \citenamefont {{Cole}}, \citenamefont {{Comparat}}, \citenamefont
  {{Cooper}}, \citenamefont {{Cousinou}}, \citenamefont {{Crocce}},
  \citenamefont {{Cuby}}, \citenamefont {{Cunningham}}, \citenamefont
  {{Davis}}, \citenamefont {{Dawson}}, \citenamefont {{de la Macorra}},
  \citenamefont {{De Vicente}}, \citenamefont {{Delubac}}, \citenamefont
  {{Derwent}}, \citenamefont {{Dey}}, \citenamefont {{Dhungana}}, \citenamefont
  {{Ding}}, \citenamefont {{Doel}}, \citenamefont {{Duan}}, \citenamefont
  {{Ealet}}, \citenamefont {{Edelstein}}, \citenamefont {{Eftekharzadeh}},
  \citenamefont {{Eisenstein}}, \citenamefont {{Elliott}}, \citenamefont
  {{Escoffier}}, \citenamefont {{Evatt}}, \citenamefont {{Fagrelius}},
  \citenamefont {{Fan}}, \citenamefont {{Fanning}}, \citenamefont {{Farahi}},
  \citenamefont {{Farihi}}, \citenamefont {{Favole}}, \citenamefont {{Feng}},
  \citenamefont {{Fernandez}}, \citenamefont {{Findlay}}, \citenamefont
  {{Finkbeiner}}, \citenamefont {{Fitzpatrick}}, \citenamefont {{Flaugher}},
  \citenamefont {{Flender}}, \citenamefont {{Font-Ribera}}, \citenamefont
  {{Forero-Romero}}, \citenamefont {{Fosalba}}, \citenamefont {{Frenk}},
  \citenamefont {{Fumagalli}}, \citenamefont {{Gaensicke}}, \citenamefont
  {{Gallo}}, \citenamefont {{Garcia-Bellido}}, \citenamefont {{Gaztanaga}},
  \citenamefont {{Pietro Gentile Fusillo}}, \citenamefont {{Gerard}},
  \citenamefont {{Gershkovich}}, \citenamefont {{Giannantonio}}, \citenamefont
  {{Gillet}}, \citenamefont {{Gonzalez-de-Rivera}}, \citenamefont
  {{Gonzalez-Perez}}, \citenamefont {{Gott}}, \citenamefont {{Graur}},
  \citenamefont {{Gutierrez}}, \citenamefont {{Guy}}, \citenamefont {{Habib}},
  \citenamefont {{Heetderks}}, \citenamefont {{Heetderks}}, \citenamefont
  {{Heitmann}}, \citenamefont {{Hellwing}}, \citenamefont {{Herrera}},
  \citenamefont {{Ho}}, \citenamefont {{Holland}}, \citenamefont {{Honscheid}},
  \citenamefont {{Huff}}, \citenamefont {{Hutchinson}}, \citenamefont
  {{Huterer}}, \citenamefont {{Hwang}}, \citenamefont {{Illa Laguna}},
  \citenamefont {{Ishikawa}}, \citenamefont {{Jacobs}}, \citenamefont
  {{Jeffrey}}, \citenamefont {{Jelinsky}}, \citenamefont {{Jennings}},
  \citenamefont {{Jiang}}, \citenamefont {{Jimenez}}, \citenamefont
  {{Johnson}}, \citenamefont {{Joyce}}, \citenamefont {{Jullo}}, \citenamefont
  {{Juneau}}, \citenamefont {{Kama}}, \citenamefont {{Karcher}}, \citenamefont
  {{Karkar}}, \citenamefont {{Kehoe}}, \citenamefont {{Kennamer}},
  \citenamefont {{Kent}}, \citenamefont {{Kilbinger}}, \citenamefont {{Kim}},
  \citenamefont {{Kirkby}}, \citenamefont {{Kisner}}, \citenamefont
  {{Kitanidis}}, \citenamefont {{Kneib}}, \citenamefont {{Koposov}},
  \citenamefont {{Kovacs}}, \citenamefont {{Koyama}}, \citenamefont {{Kremin}},
  \citenamefont {{Kron}}, \citenamefont {{Kronig}}, \citenamefont
  {{Kueter-Young}}, \citenamefont {{Lacey}}, \citenamefont {{Lafever}},
  \citenamefont {{Lahav}}, \citenamefont {{Lambert}}, \citenamefont
  {{Lampton}}, \citenamefont {{Landriau}}, \citenamefont {{Lang}},
  \citenamefont {{Lauer}}, \citenamefont {{Le Goff}}, \citenamefont {{Le
  Guillou}}, \citenamefont {{Le Van Suu}}, \citenamefont {{Lee}}, \citenamefont
  {{Lee}}, \citenamefont {{Leitner}}, \citenamefont {{Lesser}}, \citenamefont
  {{Levi}}, \citenamefont {{L'Huillier}}, \citenamefont {{Li}}, \citenamefont
  {{Liang}}, \citenamefont {{Lin}}, \citenamefont {{Linder}}, \citenamefont
  {{Loebman}}, \citenamefont {{Luki{\'c}}}, \citenamefont {{Ma}}, \citenamefont
  {{MacCrann}}, \citenamefont {{Magneville}}, \citenamefont {{Makarem}},
  \citenamefont {{Manera}}, \citenamefont {{Manser}}, \citenamefont
  {{Marshall}}, \citenamefont {{Martini}}, \citenamefont {{Massey}},
  \citenamefont {{Matheson}}, \citenamefont {{McCauley}}, \citenamefont
  {{McDonald}}, \citenamefont {{McGreer}}, \citenamefont {{Meisner}},
  \citenamefont {{Metcalfe}}, \citenamefont {{Miller}}, \citenamefont
  {{Miquel}}, \citenamefont {{Moustakas}}, \citenamefont {{Myers}},
  \citenamefont {{Naik}}, \citenamefont {{Newman}}, \citenamefont {{Nichol}},
  \citenamefont {{Nicola}}, \citenamefont {{Nicolati da Costa}}, \citenamefont
  {{Nie}}, \citenamefont {{Niz}}, \citenamefont {{Norberg}}, \citenamefont
  {{Nord}}, \citenamefont {{Norman}}, \citenamefont {{Nugent}}, \citenamefont
  {{O'Brien}}, \citenamefont {{Oh}}, \citenamefont {{Olsen}}, \citenamefont
  {{Padilla}}, \citenamefont {{Padmanabhan}}, \citenamefont {{Padmanabhan}},
  \citenamefont {{Palanque-Delabrouille}}, \citenamefont {{Palmese}},
  \citenamefont {{Pappalardo}}, \citenamefont {{P{\^a}ris}}, \citenamefont
  {{Park}}, \citenamefont {{Patej}}, \citenamefont {{Peacock}}, \citenamefont
  {{Peiris}}, \citenamefont {{Peng}}, \citenamefont {{Percival}}, \citenamefont
  {{Perruchot}}, \citenamefont {{Pieri}}, \citenamefont {{Pogge}},
  \citenamefont {{Pollack}}, \citenamefont {{Poppett}}, \citenamefont
  {{Prada}}, \citenamefont {{Prakash}}, \citenamefont {{Probst}}, \citenamefont
  {{Rabinowitz}}, \citenamefont {{Raichoor}}, \citenamefont {{Ree}},
  \citenamefont {{Refregier}}, \citenamefont {{Regal}}, \citenamefont {{Reid}},
  \citenamefont {{Reil}}, \citenamefont {{Rezaie}}, \citenamefont {{Rockosi}},
  \citenamefont {{Roe}}, \citenamefont {{Ronayette}}, \citenamefont
  {{Roodman}}, \citenamefont {{Ross}}, \citenamefont {{Ross}}, \citenamefont
  {{Rossi}}, \citenamefont {{Rozo}}, \citenamefont {{Ruhlmann-Kleider}},
  \citenamefont {{Rykoff}}, \citenamefont {{Sabiu}}, \citenamefont
  {{Samushia}}, \citenamefont {{Sanchez}}, \citenamefont {{Sanchez}},
  \citenamefont {{Schlegel}}, \citenamefont {{Schneider}}, \citenamefont
  {{Schubnell}}, \citenamefont {{Secroun}}, \citenamefont {{Seljak}},
  \citenamefont {{Seo}}, \citenamefont {{Serrano}}, \citenamefont
  {{Shafieloo}}, \citenamefont {{Shan}}, \citenamefont {{Sharples}},
  \citenamefont {{Sholl}}, \citenamefont {{Shourt}}, \citenamefont {{Silber}},
  \citenamefont {{Silva}}, \citenamefont {{Sirk}}, \citenamefont {{Slosar}},
  \citenamefont {{Smith}}, \citenamefont {{Smoot}}, \citenamefont {{Som}},
  \citenamefont {{Song}}, \citenamefont {{Sprayberry}}, \citenamefont
  {{Staten}}, \citenamefont {{Stefanik}}, \citenamefont {{Tarle}},
  \citenamefont {{Sien Tie}}, \citenamefont {{Tinker}}, \citenamefont
  {{Tojeiro}}, \citenamefont {{Valdes}}, \citenamefont {{Valenzuela}},
  \citenamefont {{Valluri}}, \citenamefont {{Vargas-Magana}}, \citenamefont
  {{Verde}}, \citenamefont {{Walker}}, \citenamefont {{Wang}}, \citenamefont
  {{Wang}}, \citenamefont {{Weaver}}, \citenamefont {{Weaverdyck}},
  \citenamefont {{Wechsler}}, \citenamefont {{Weinberg}}, \citenamefont
  {{White}}, \citenamefont {{Yang}}, \citenamefont {{Yeche}}, \citenamefont
  {{Zhang}}, \citenamefont {{Zhao}}, \citenamefont {{Zheng}}, \citenamefont
  {{Zhou}}, \citenamefont {{Zhou}}, \citenamefont {{Zhu}}, \citenamefont
  {{Zou}},\ and\ \citenamefont {{Zu}}}]{DESI:2016}%
  \BibitemOpen
  \bibfield  {author} {\bibinfo {author} {\bibnamefont {{DESI Collaboration}}},
  \bibinfo {author} {\bibfnamefont {A.}~\bibnamefont {{Aghamousa}}}, \bibinfo
  {author} {\bibfnamefont {J.}~\bibnamefont {{Aguilar}}}, \bibinfo {author}
  {\bibfnamefont {S.}~\bibnamefont {{Ahlen}}}, \bibinfo {author} {\bibfnamefont
  {S.}~\bibnamefont {{Alam}}}, \bibinfo {author} {\bibfnamefont {L.~E.}\
  \bibnamefont {{Allen}}}, \bibinfo {author} {\bibfnamefont {C.}~\bibnamefont
  {{Allende Prieto}}}, \bibinfo {author} {\bibfnamefont {J.}~\bibnamefont
  {{Annis}}}, \bibinfo {author} {\bibfnamefont {S.}~\bibnamefont {{Bailey}}},
  \bibinfo {author} {\bibfnamefont {C.}~\bibnamefont {{Balland}}}, \bibinfo
  {author} {\bibfnamefont {O.}~\bibnamefont {{Ballester}}}, \bibinfo {author}
  {\bibfnamefont {C.}~\bibnamefont {{Baltay}}}, \bibinfo {author}
  {\bibfnamefont {L.}~\bibnamefont {{Beaufore}}}, \bibinfo {author}
  {\bibfnamefont {C.}~\bibnamefont {{Bebek}}}, \bibinfo {author} {\bibfnamefont
  {T.~C.}\ \bibnamefont {{Beers}}}, \bibinfo {author} {\bibfnamefont {E.~F.}\
  \bibnamefont {{Bell}}}, \bibinfo {author} {\bibfnamefont {J.~L.}\
  \bibnamefont {{Bernal}}}, \bibinfo {author} {\bibfnamefont {R.}~\bibnamefont
  {{Besuner}}}, \bibinfo {author} {\bibfnamefont {F.}~\bibnamefont
  {{Beutler}}}, \bibinfo {author} {\bibfnamefont {C.}~\bibnamefont {{Blake}}},
  \bibinfo {author} {\bibfnamefont {H.}~\bibnamefont {{Bleuler}}}, \bibinfo
  {author} {\bibfnamefont {M.}~\bibnamefont {{Blomqvist}}}, \bibinfo {author}
  {\bibfnamefont {R.}~\bibnamefont {{Blum}}}, \bibinfo {author} {\bibfnamefont
  {A.~S.}\ \bibnamefont {{Bolton}}}, \bibinfo {author} {\bibfnamefont
  {C.}~\bibnamefont {{Briceno}}}, \bibinfo {author} {\bibfnamefont
  {D.}~\bibnamefont {{Brooks}}}, \bibinfo {author} {\bibfnamefont {J.~R.}\
  \bibnamefont {{Brownstein}}}, \bibinfo {author} {\bibfnamefont
  {E.}~\bibnamefont {{Buckley-Geer}}}, \bibinfo {author} {\bibfnamefont
  {A.}~\bibnamefont {{Burden}}}, \bibinfo {author} {\bibfnamefont
  {E.}~\bibnamefont {{Burtin}}}, \bibinfo {author} {\bibfnamefont {N.~G.}\
  \bibnamefont {{Busca}}}, \bibinfo {author} {\bibfnamefont {R.~N.}\
  \bibnamefont {{Cahn}}}, \bibinfo {author} {\bibfnamefont {Y.-C.}\
  \bibnamefont {{Cai}}}, \bibinfo {author} {\bibfnamefont {L.}~\bibnamefont
  {{Cardiel-Sas}}}, \bibinfo {author} {\bibfnamefont {R.~G.}\ \bibnamefont
  {{Carlberg}}}, \bibinfo {author} {\bibfnamefont {P.-H.}\ \bibnamefont
  {{Carton}}}, \bibinfo {author} {\bibfnamefont {R.}~\bibnamefont {{Casas}}},
  \bibinfo {author} {\bibfnamefont {F.~J.}\ \bibnamefont {{Castander}}},
  \bibinfo {author} {\bibfnamefont {J.~L.}\ \bibnamefont {{Cervantes-Cota}}},
  \bibinfo {author} {\bibfnamefont {T.~M.}\ \bibnamefont {{Claybaugh}}},
  \bibinfo {author} {\bibfnamefont {M.}~\bibnamefont {{Close}}}, \bibinfo
  {author} {\bibfnamefont {C.~T.}\ \bibnamefont {{Coker}}}, \bibinfo {author}
  {\bibfnamefont {S.}~\bibnamefont {{Cole}}}, \bibinfo {author} {\bibfnamefont
  {J.}~\bibnamefont {{Comparat}}}, \bibinfo {author} {\bibfnamefont {A.~P.}\
  \bibnamefont {{Cooper}}}, \bibinfo {author} {\bibfnamefont {M.~C.}\
  \bibnamefont {{Cousinou}}}, \bibinfo {author} {\bibfnamefont
  {M.}~\bibnamefont {{Crocce}}}, \bibinfo {author} {\bibfnamefont {J.-G.}\
  \bibnamefont {{Cuby}}}, \bibinfo {author} {\bibfnamefont {D.~P.}\
  \bibnamefont {{Cunningham}}}, \bibinfo {author} {\bibfnamefont {T.~M.}\
  \bibnamefont {{Davis}}}, \bibinfo {author} {\bibfnamefont {K.~S.}\
  \bibnamefont {{Dawson}}}, \bibinfo {author} {\bibfnamefont {A.}~\bibnamefont
  {{de la Macorra}}}, \bibinfo {author} {\bibfnamefont {J.}~\bibnamefont {{De
  Vicente}}}, \bibinfo {author} {\bibfnamefont {T.}~\bibnamefont {{Delubac}}},
  \bibinfo {author} {\bibfnamefont {M.}~\bibnamefont {{Derwent}}}, \bibinfo
  {author} {\bibfnamefont {A.}~\bibnamefont {{Dey}}}, \bibinfo {author}
  {\bibfnamefont {G.}~\bibnamefont {{Dhungana}}}, \bibinfo {author}
  {\bibfnamefont {Z.}~\bibnamefont {{Ding}}}, \bibinfo {author} {\bibfnamefont
  {P.}~\bibnamefont {{Doel}}}, \bibinfo {author} {\bibfnamefont {Y.~T.}\
  \bibnamefont {{Duan}}}, \bibinfo {author} {\bibfnamefont {A.}~\bibnamefont
  {{Ealet}}}, \bibinfo {author} {\bibfnamefont {J.}~\bibnamefont
  {{Edelstein}}}, \bibinfo {author} {\bibfnamefont {S.}~\bibnamefont
  {{Eftekharzadeh}}}, \bibinfo {author} {\bibfnamefont {D.~J.}\ \bibnamefont
  {{Eisenstein}}}, \bibinfo {author} {\bibfnamefont {A.}~\bibnamefont
  {{Elliott}}}, \bibinfo {author} {\bibfnamefont {S.}~\bibnamefont
  {{Escoffier}}}, \bibinfo {author} {\bibfnamefont {M.}~\bibnamefont
  {{Evatt}}}, \bibinfo {author} {\bibfnamefont {P.}~\bibnamefont
  {{Fagrelius}}}, \bibinfo {author} {\bibfnamefont {X.}~\bibnamefont {{Fan}}},
  \bibinfo {author} {\bibfnamefont {K.}~\bibnamefont {{Fanning}}}, \bibinfo
  {author} {\bibfnamefont {A.}~\bibnamefont {{Farahi}}}, \bibinfo {author}
  {\bibfnamefont {J.}~\bibnamefont {{Farihi}}}, \bibinfo {author}
  {\bibfnamefont {G.}~\bibnamefont {{Favole}}}, \bibinfo {author}
  {\bibfnamefont {Y.}~\bibnamefont {{Feng}}}, \bibinfo {author} {\bibfnamefont
  {E.}~\bibnamefont {{Fernandez}}}, \bibinfo {author} {\bibfnamefont {J.~R.}\
  \bibnamefont {{Findlay}}}, \bibinfo {author} {\bibfnamefont {D.~P.}\
  \bibnamefont {{Finkbeiner}}}, \bibinfo {author} {\bibfnamefont {M.~J.}\
  \bibnamefont {{Fitzpatrick}}}, \bibinfo {author} {\bibfnamefont
  {B.}~\bibnamefont {{Flaugher}}}, \bibinfo {author} {\bibfnamefont
  {S.}~\bibnamefont {{Flender}}}, \bibinfo {author} {\bibfnamefont
  {A.}~\bibnamefont {{Font-Ribera}}}, \bibinfo {author} {\bibfnamefont {J.~E.}\
  \bibnamefont {{Forero-Romero}}}, \bibinfo {author} {\bibfnamefont
  {P.}~\bibnamefont {{Fosalba}}}, \bibinfo {author} {\bibfnamefont {C.~S.}\
  \bibnamefont {{Frenk}}}, \bibinfo {author} {\bibfnamefont {M.}~\bibnamefont
  {{Fumagalli}}}, \bibinfo {author} {\bibfnamefont {B.~T.}\ \bibnamefont
  {{Gaensicke}}}, \bibinfo {author} {\bibfnamefont {G.}~\bibnamefont
  {{Gallo}}}, \bibinfo {author} {\bibfnamefont {J.}~\bibnamefont
  {{Garcia-Bellido}}}, \bibinfo {author} {\bibfnamefont {E.}~\bibnamefont
  {{Gaztanaga}}}, \bibinfo {author} {\bibfnamefont {N.}~\bibnamefont {{Pietro
  Gentile Fusillo}}}, \bibinfo {author} {\bibfnamefont {T.}~\bibnamefont
  {{Gerard}}}, \bibinfo {author} {\bibfnamefont {I.}~\bibnamefont
  {{Gershkovich}}}, \bibinfo {author} {\bibfnamefont {T.}~\bibnamefont
  {{Giannantonio}}}, \bibinfo {author} {\bibfnamefont {D.}~\bibnamefont
  {{Gillet}}}, \bibinfo {author} {\bibfnamefont {G.}~\bibnamefont
  {{Gonzalez-de-Rivera}}}, \bibinfo {author} {\bibfnamefont {V.}~\bibnamefont
  {{Gonzalez-Perez}}}, \bibinfo {author} {\bibfnamefont {S.}~\bibnamefont
  {{Gott}}}, \bibinfo {author} {\bibfnamefont {O.}~\bibnamefont {{Graur}}},
  \bibinfo {author} {\bibfnamefont {G.}~\bibnamefont {{Gutierrez}}}, \bibinfo
  {author} {\bibfnamefont {J.}~\bibnamefont {{Guy}}}, \bibinfo {author}
  {\bibfnamefont {S.}~\bibnamefont {{Habib}}}, \bibinfo {author} {\bibfnamefont
  {H.}~\bibnamefont {{Heetderks}}}, \bibinfo {author} {\bibfnamefont
  {I.}~\bibnamefont {{Heetderks}}}, \bibinfo {author} {\bibfnamefont
  {K.}~\bibnamefont {{Heitmann}}}, \bibinfo {author} {\bibfnamefont {W.~A.}\
  \bibnamefont {{Hellwing}}}, \bibinfo {author} {\bibfnamefont {D.~A.}\
  \bibnamefont {{Herrera}}}, \bibinfo {author} {\bibfnamefont {S.}~\bibnamefont
  {{Ho}}}, \bibinfo {author} {\bibfnamefont {S.}~\bibnamefont {{Holland}}},
  \bibinfo {author} {\bibfnamefont {K.}~\bibnamefont {{Honscheid}}}, \bibinfo
  {author} {\bibfnamefont {E.}~\bibnamefont {{Huff}}}, \bibinfo {author}
  {\bibfnamefont {T.~A.}\ \bibnamefont {{Hutchinson}}}, \bibinfo {author}
  {\bibfnamefont {D.}~\bibnamefont {{Huterer}}}, \bibinfo {author}
  {\bibfnamefont {H.~S.}\ \bibnamefont {{Hwang}}}, \bibinfo {author}
  {\bibfnamefont {J.~M.}\ \bibnamefont {{Illa Laguna}}}, \bibinfo {author}
  {\bibfnamefont {Y.}~\bibnamefont {{Ishikawa}}}, \bibinfo {author}
  {\bibfnamefont {D.}~\bibnamefont {{Jacobs}}}, \bibinfo {author}
  {\bibfnamefont {N.}~\bibnamefont {{Jeffrey}}}, \bibinfo {author}
  {\bibfnamefont {P.}~\bibnamefont {{Jelinsky}}}, \bibinfo {author}
  {\bibfnamefont {E.}~\bibnamefont {{Jennings}}}, \bibinfo {author}
  {\bibfnamefont {L.}~\bibnamefont {{Jiang}}}, \bibinfo {author} {\bibfnamefont
  {J.}~\bibnamefont {{Jimenez}}}, \bibinfo {author} {\bibfnamefont
  {J.}~\bibnamefont {{Johnson}}}, \bibinfo {author} {\bibfnamefont
  {R.}~\bibnamefont {{Joyce}}}, \bibinfo {author} {\bibfnamefont
  {E.}~\bibnamefont {{Jullo}}}, \bibinfo {author} {\bibfnamefont
  {S.}~\bibnamefont {{Juneau}}}, \bibinfo {author} {\bibfnamefont
  {S.}~\bibnamefont {{Kama}}}, \bibinfo {author} {\bibfnamefont
  {A.}~\bibnamefont {{Karcher}}}, \bibinfo {author} {\bibfnamefont
  {S.}~\bibnamefont {{Karkar}}}, \bibinfo {author} {\bibfnamefont
  {R.}~\bibnamefont {{Kehoe}}}, \bibinfo {author} {\bibfnamefont
  {N.}~\bibnamefont {{Kennamer}}}, \bibinfo {author} {\bibfnamefont
  {S.}~\bibnamefont {{Kent}}}, \bibinfo {author} {\bibfnamefont
  {M.}~\bibnamefont {{Kilbinger}}}, \bibinfo {author} {\bibfnamefont {A.~G.}\
  \bibnamefont {{Kim}}}, \bibinfo {author} {\bibfnamefont {D.}~\bibnamefont
  {{Kirkby}}}, \bibinfo {author} {\bibfnamefont {T.}~\bibnamefont {{Kisner}}},
  \bibinfo {author} {\bibfnamefont {E.}~\bibnamefont {{Kitanidis}}}, \bibinfo
  {author} {\bibfnamefont {J.-P.}\ \bibnamefont {{Kneib}}}, \bibinfo {author}
  {\bibfnamefont {S.}~\bibnamefont {{Koposov}}}, \bibinfo {author}
  {\bibfnamefont {E.}~\bibnamefont {{Kovacs}}}, \bibinfo {author}
  {\bibfnamefont {K.}~\bibnamefont {{Koyama}}}, \bibinfo {author}
  {\bibfnamefont {A.}~\bibnamefont {{Kremin}}}, \bibinfo {author}
  {\bibfnamefont {R.}~\bibnamefont {{Kron}}}, \bibinfo {author} {\bibfnamefont
  {L.}~\bibnamefont {{Kronig}}}, \bibinfo {author} {\bibfnamefont
  {A.}~\bibnamefont {{Kueter-Young}}}, \bibinfo {author} {\bibfnamefont
  {C.~G.}\ \bibnamefont {{Lacey}}}, \bibinfo {author} {\bibfnamefont
  {R.}~\bibnamefont {{Lafever}}}, \bibinfo {author} {\bibfnamefont
  {O.}~\bibnamefont {{Lahav}}}, \bibinfo {author} {\bibfnamefont
  {A.}~\bibnamefont {{Lambert}}}, \bibinfo {author} {\bibfnamefont
  {M.}~\bibnamefont {{Lampton}}}, \bibinfo {author} {\bibfnamefont
  {M.}~\bibnamefont {{Landriau}}}, \bibinfo {author} {\bibfnamefont
  {D.}~\bibnamefont {{Lang}}}, \bibinfo {author} {\bibfnamefont {T.~R.}\
  \bibnamefont {{Lauer}}}, \bibinfo {author} {\bibfnamefont {J.-M.}\
  \bibnamefont {{Le Goff}}}, \bibinfo {author} {\bibfnamefont {L.}~\bibnamefont
  {{Le Guillou}}}, \bibinfo {author} {\bibfnamefont {A.}~\bibnamefont {{Le Van
  Suu}}}, \bibinfo {author} {\bibfnamefont {J.~H.}\ \bibnamefont {{Lee}}},
  \bibinfo {author} {\bibfnamefont {S.-J.}\ \bibnamefont {{Lee}}}, \bibinfo
  {author} {\bibfnamefont {D.}~\bibnamefont {{Leitner}}}, \bibinfo {author}
  {\bibfnamefont {M.}~\bibnamefont {{Lesser}}}, \bibinfo {author}
  {\bibfnamefont {M.~E.}\ \bibnamefont {{Levi}}}, \bibinfo {author}
  {\bibfnamefont {B.}~\bibnamefont {{L'Huillier}}}, \bibinfo {author}
  {\bibfnamefont {B.}~\bibnamefont {{Li}}}, \bibinfo {author} {\bibfnamefont
  {M.}~\bibnamefont {{Liang}}}, \bibinfo {author} {\bibfnamefont
  {H.}~\bibnamefont {{Lin}}}, \bibinfo {author} {\bibfnamefont
  {E.}~\bibnamefont {{Linder}}}, \bibinfo {author} {\bibfnamefont {S.~R.}\
  \bibnamefont {{Loebman}}}, \bibinfo {author} {\bibfnamefont {Z.}~\bibnamefont
  {{Luki{\'c}}}}, \bibinfo {author} {\bibfnamefont {J.}~\bibnamefont {{Ma}}},
  \bibinfo {author} {\bibfnamefont {N.}~\bibnamefont {{MacCrann}}}, \bibinfo
  {author} {\bibfnamefont {C.}~\bibnamefont {{Magneville}}}, \bibinfo {author}
  {\bibfnamefont {L.}~\bibnamefont {{Makarem}}}, \bibinfo {author}
  {\bibfnamefont {M.}~\bibnamefont {{Manera}}}, \bibinfo {author}
  {\bibfnamefont {C.~J.}\ \bibnamefont {{Manser}}}, \bibinfo {author}
  {\bibfnamefont {R.}~\bibnamefont {{Marshall}}}, \bibinfo {author}
  {\bibfnamefont {P.}~\bibnamefont {{Martini}}}, \bibinfo {author}
  {\bibfnamefont {R.}~\bibnamefont {{Massey}}}, \bibinfo {author}
  {\bibfnamefont {T.}~\bibnamefont {{Matheson}}}, \bibinfo {author}
  {\bibfnamefont {J.}~\bibnamefont {{McCauley}}}, \bibinfo {author}
  {\bibfnamefont {P.}~\bibnamefont {{McDonald}}}, \bibinfo {author}
  {\bibfnamefont {I.~D.}\ \bibnamefont {{McGreer}}}, \bibinfo {author}
  {\bibfnamefont {A.}~\bibnamefont {{Meisner}}}, \bibinfo {author}
  {\bibfnamefont {N.}~\bibnamefont {{Metcalfe}}}, \bibinfo {author}
  {\bibfnamefont {T.~N.}\ \bibnamefont {{Miller}}}, \bibinfo {author}
  {\bibfnamefont {R.}~\bibnamefont {{Miquel}}}, \bibinfo {author}
  {\bibfnamefont {J.}~\bibnamefont {{Moustakas}}}, \bibinfo {author}
  {\bibfnamefont {A.}~\bibnamefont {{Myers}}}, \bibinfo {author} {\bibfnamefont
  {M.}~\bibnamefont {{Naik}}}, \bibinfo {author} {\bibfnamefont {J.~A.}\
  \bibnamefont {{Newman}}}, \bibinfo {author} {\bibfnamefont {R.~C.}\
  \bibnamefont {{Nichol}}}, \bibinfo {author} {\bibfnamefont {A.}~\bibnamefont
  {{Nicola}}}, \bibinfo {author} {\bibfnamefont {L.}~\bibnamefont {{Nicolati da
  Costa}}}, \bibinfo {author} {\bibfnamefont {J.}~\bibnamefont {{Nie}}},
  \bibinfo {author} {\bibfnamefont {G.}~\bibnamefont {{Niz}}}, \bibinfo
  {author} {\bibfnamefont {P.}~\bibnamefont {{Norberg}}}, \bibinfo {author}
  {\bibfnamefont {B.}~\bibnamefont {{Nord}}}, \bibinfo {author} {\bibfnamefont
  {D.}~\bibnamefont {{Norman}}}, \bibinfo {author} {\bibfnamefont
  {P.}~\bibnamefont {{Nugent}}}, \bibinfo {author} {\bibfnamefont
  {T.}~\bibnamefont {{O'Brien}}}, \bibinfo {author} {\bibfnamefont
  {M.}~\bibnamefont {{Oh}}}, \bibinfo {author} {\bibfnamefont {K.~A.~G.}\
  \bibnamefont {{Olsen}}}, \bibinfo {author} {\bibfnamefont {C.}~\bibnamefont
  {{Padilla}}}, \bibinfo {author} {\bibfnamefont {H.}~\bibnamefont
  {{Padmanabhan}}}, \bibinfo {author} {\bibfnamefont {N.}~\bibnamefont
  {{Padmanabhan}}}, \bibinfo {author} {\bibfnamefont {N.}~\bibnamefont
  {{Palanque-Delabrouille}}}, \bibinfo {author} {\bibfnamefont
  {A.}~\bibnamefont {{Palmese}}}, \bibinfo {author} {\bibfnamefont
  {D.}~\bibnamefont {{Pappalardo}}}, \bibinfo {author} {\bibfnamefont
  {I.}~\bibnamefont {{P{\^a}ris}}}, \bibinfo {author} {\bibfnamefont
  {C.}~\bibnamefont {{Park}}}, \bibinfo {author} {\bibfnamefont
  {A.}~\bibnamefont {{Patej}}}, \bibinfo {author} {\bibfnamefont {J.~A.}\
  \bibnamefont {{Peacock}}}, \bibinfo {author} {\bibfnamefont {H.~V.}\
  \bibnamefont {{Peiris}}}, \bibinfo {author} {\bibfnamefont {X.}~\bibnamefont
  {{Peng}}}, \bibinfo {author} {\bibfnamefont {W.~J.}\ \bibnamefont
  {{Percival}}}, \bibinfo {author} {\bibfnamefont {S.}~\bibnamefont
  {{Perruchot}}}, \bibinfo {author} {\bibfnamefont {M.~M.}\ \bibnamefont
  {{Pieri}}}, \bibinfo {author} {\bibfnamefont {R.}~\bibnamefont {{Pogge}}},
  \bibinfo {author} {\bibfnamefont {J.~E.}\ \bibnamefont {{Pollack}}}, \bibinfo
  {author} {\bibfnamefont {C.}~\bibnamefont {{Poppett}}}, \bibinfo {author}
  {\bibfnamefont {F.}~\bibnamefont {{Prada}}}, \bibinfo {author} {\bibfnamefont
  {A.}~\bibnamefont {{Prakash}}}, \bibinfo {author} {\bibfnamefont {R.~G.}\
  \bibnamefont {{Probst}}}, \bibinfo {author} {\bibfnamefont {D.}~\bibnamefont
  {{Rabinowitz}}}, \bibinfo {author} {\bibfnamefont {A.}~\bibnamefont
  {{Raichoor}}}, \bibinfo {author} {\bibfnamefont {C.~H.}\ \bibnamefont
  {{Ree}}}, \bibinfo {author} {\bibfnamefont {A.}~\bibnamefont {{Refregier}}},
  \bibinfo {author} {\bibfnamefont {X.}~\bibnamefont {{Regal}}}, \bibinfo
  {author} {\bibfnamefont {B.}~\bibnamefont {{Reid}}}, \bibinfo {author}
  {\bibfnamefont {K.}~\bibnamefont {{Reil}}}, \bibinfo {author} {\bibfnamefont
  {M.}~\bibnamefont {{Rezaie}}}, \bibinfo {author} {\bibfnamefont {C.~M.}\
  \bibnamefont {{Rockosi}}}, \bibinfo {author} {\bibfnamefont {N.}~\bibnamefont
  {{Roe}}}, \bibinfo {author} {\bibfnamefont {S.}~\bibnamefont {{Ronayette}}},
  \bibinfo {author} {\bibfnamefont {A.}~\bibnamefont {{Roodman}}}, \bibinfo
  {author} {\bibfnamefont {A.~J.}\ \bibnamefont {{Ross}}}, \bibinfo {author}
  {\bibfnamefont {N.~P.}\ \bibnamefont {{Ross}}}, \bibinfo {author}
  {\bibfnamefont {G.}~\bibnamefont {{Rossi}}}, \bibinfo {author} {\bibfnamefont
  {E.}~\bibnamefont {{Rozo}}}, \bibinfo {author} {\bibfnamefont
  {V.}~\bibnamefont {{Ruhlmann-Kleider}}}, \bibinfo {author} {\bibfnamefont
  {E.~S.}\ \bibnamefont {{Rykoff}}}, \bibinfo {author} {\bibfnamefont
  {C.}~\bibnamefont {{Sabiu}}}, \bibinfo {author} {\bibfnamefont
  {L.}~\bibnamefont {{Samushia}}}, \bibinfo {author} {\bibfnamefont
  {E.}~\bibnamefont {{Sanchez}}}, \bibinfo {author} {\bibfnamefont
  {J.}~\bibnamefont {{Sanchez}}}, \bibinfo {author} {\bibfnamefont {D.~J.}\
  \bibnamefont {{Schlegel}}}, \bibinfo {author} {\bibfnamefont
  {M.}~\bibnamefont {{Schneider}}}, \bibinfo {author} {\bibfnamefont
  {M.}~\bibnamefont {{Schubnell}}}, \bibinfo {author} {\bibfnamefont
  {A.}~\bibnamefont {{Secroun}}}, \bibinfo {author} {\bibfnamefont
  {U.}~\bibnamefont {{Seljak}}}, \bibinfo {author} {\bibfnamefont {H.-J.}\
  \bibnamefont {{Seo}}}, \bibinfo {author} {\bibfnamefont {S.}~\bibnamefont
  {{Serrano}}}, \bibinfo {author} {\bibfnamefont {A.}~\bibnamefont
  {{Shafieloo}}}, \bibinfo {author} {\bibfnamefont {H.}~\bibnamefont {{Shan}}},
  \bibinfo {author} {\bibfnamefont {R.}~\bibnamefont {{Sharples}}}, \bibinfo
  {author} {\bibfnamefont {M.~J.}\ \bibnamefont {{Sholl}}}, \bibinfo {author}
  {\bibfnamefont {W.~V.}\ \bibnamefont {{Shourt}}}, \bibinfo {author}
  {\bibfnamefont {J.~H.}\ \bibnamefont {{Silber}}}, \bibinfo {author}
  {\bibfnamefont {D.~R.}\ \bibnamefont {{Silva}}}, \bibinfo {author}
  {\bibfnamefont {M.~M.}\ \bibnamefont {{Sirk}}}, \bibinfo {author}
  {\bibfnamefont {A.}~\bibnamefont {{Slosar}}}, \bibinfo {author}
  {\bibfnamefont {A.}~\bibnamefont {{Smith}}}, \bibinfo {author} {\bibfnamefont
  {G.~F.}\ \bibnamefont {{Smoot}}}, \bibinfo {author} {\bibfnamefont
  {D.}~\bibnamefont {{Som}}}, \bibinfo {author} {\bibfnamefont {Y.-S.}\
  \bibnamefont {{Song}}}, \bibinfo {author} {\bibfnamefont {D.}~\bibnamefont
  {{Sprayberry}}}, \bibinfo {author} {\bibfnamefont {R.}~\bibnamefont
  {{Staten}}}, \bibinfo {author} {\bibfnamefont {A.}~\bibnamefont
  {{Stefanik}}}, \bibinfo {author} {\bibfnamefont {G.}~\bibnamefont {{Tarle}}},
  \bibinfo {author} {\bibfnamefont {S.}~\bibnamefont {{Sien Tie}}}, \bibinfo
  {author} {\bibfnamefont {J.~L.}\ \bibnamefont {{Tinker}}}, \bibinfo {author}
  {\bibfnamefont {R.}~\bibnamefont {{Tojeiro}}}, \bibinfo {author}
  {\bibfnamefont {F.}~\bibnamefont {{Valdes}}}, \bibinfo {author}
  {\bibfnamefont {O.}~\bibnamefont {{Valenzuela}}}, \bibinfo {author}
  {\bibfnamefont {M.}~\bibnamefont {{Valluri}}}, \bibinfo {author}
  {\bibfnamefont {M.}~\bibnamefont {{Vargas-Magana}}}, \bibinfo {author}
  {\bibfnamefont {L.}~\bibnamefont {{Verde}}}, \bibinfo {author} {\bibfnamefont
  {A.~R.}\ \bibnamefont {{Walker}}}, \bibinfo {author} {\bibfnamefont
  {J.}~\bibnamefont {{Wang}}}, \bibinfo {author} {\bibfnamefont
  {Y.}~\bibnamefont {{Wang}}}, \bibinfo {author} {\bibfnamefont {B.~A.}\
  \bibnamefont {{Weaver}}}, \bibinfo {author} {\bibfnamefont {C.}~\bibnamefont
  {{Weaverdyck}}}, \bibinfo {author} {\bibfnamefont {R.~H.}\ \bibnamefont
  {{Wechsler}}}, \bibinfo {author} {\bibfnamefont {D.~H.}\ \bibnamefont
  {{Weinberg}}}, \bibinfo {author} {\bibfnamefont {M.}~\bibnamefont {{White}}},
  \bibinfo {author} {\bibfnamefont {Q.}~\bibnamefont {{Yang}}}, \bibinfo
  {author} {\bibfnamefont {C.}~\bibnamefont {{Yeche}}}, \bibinfo {author}
  {\bibfnamefont {T.}~\bibnamefont {{Zhang}}}, \bibinfo {author} {\bibfnamefont
  {G.-B.}\ \bibnamefont {{Zhao}}}, \bibinfo {author} {\bibfnamefont
  {Y.}~\bibnamefont {{Zheng}}}, \bibinfo {author} {\bibfnamefont
  {X.}~\bibnamefont {{Zhou}}}, \bibinfo {author} {\bibfnamefont
  {Z.}~\bibnamefont {{Zhou}}}, \bibinfo {author} {\bibfnamefont
  {Y.}~\bibnamefont {{Zhu}}}, \bibinfo {author} {\bibfnamefont
  {H.}~\bibnamefont {{Zou}}}, \ and\ \bibinfo {author} {\bibfnamefont
  {Y.}~\bibnamefont {{Zu}}},\ }\href {\doibase 10.48550/arXiv.1611.00036}
  {\bibfield  {journal} {\bibinfo  {journal} {arXiv e-prints}\ ,\ \bibinfo
  {eid} {arXiv:1611.00036}} (\bibinfo {year} {2016})},\ \Eprint
  {http://arxiv.org/abs/1611.00036} {arXiv:1611.00036 [astro-ph.IM]}
  \BibitemShut {NoStop}%
\bibitem [{\citenamefont {Dor\'e}\ \emph {et~al.}(2014)\citenamefont {Dor\'e}
  \emph {et~al.}}]{Dore:2014cca}%
  \BibitemOpen
  \bibfield  {author} {\bibinfo {author} {\bibfnamefont {O.}~\bibnamefont
  {Dor\'e}} \emph {et~al.},\ }\href@noop {} {\  (\bibinfo {year} {2014})},\
  \Eprint {http://arxiv.org/abs/1412.4872} {arXiv:1412.4872 [astro-ph.CO]}
  \BibitemShut {NoStop}%
\bibitem [{\citenamefont {Krolewski}\ \emph {et~al.}(2023)\citenamefont
  {Krolewski} \emph {et~al.}}]{DESI:2023duv}%
  \BibitemOpen
  \bibfield  {author} {\bibinfo {author} {\bibfnamefont {A.}~\bibnamefont
  {Krolewski}} \emph {et~al.} (\bibinfo {collaboration} {DESI}),\ }\href@noop
  {} {\  (\bibinfo {year} {2023})},\ \Eprint {http://arxiv.org/abs/2305.07650}
  {arXiv:2305.07650 [astro-ph.CO]} \BibitemShut {NoStop}%
\bibitem [{\citenamefont {{Giri}}\ \emph {et~al.}(2023)\citenamefont {{Giri}},
  \citenamefont {{M{\"u}nchmeyer}},\ and\ \citenamefont {{Smith}}}]{Giri:2023}%
  \BibitemOpen
  \bibfield  {author} {\bibinfo {author} {\bibfnamefont {U.}~\bibnamefont
  {{Giri}}}, \bibinfo {author} {\bibfnamefont {M.}~\bibnamefont
  {{M{\"u}nchmeyer}}}, \ and\ \bibinfo {author} {\bibfnamefont {K.~M.}\
  \bibnamefont {{Smith}}},\ }\href {\doibase 10.48550/arXiv.2305.03070}
  {\bibfield  {journal} {\bibinfo  {journal} {arXiv e-prints}\ ,\ \bibinfo
  {eid} {arXiv:2305.03070}} (\bibinfo {year} {2023})},\ \Eprint
  {http://arxiv.org/abs/2305.03070} {arXiv:2305.03070 [astro-ph.CO]}
  \BibitemShut {NoStop}%
\bibitem [{\citenamefont {Gualdi}\ \emph {et~al.}(2021)\citenamefont {Gualdi},
  \citenamefont {Gil-Mar\'\i{}n},\ and\ \citenamefont
  {Verde}}]{Gualdi:2021yvq}%
  \BibitemOpen
  \bibfield  {author} {\bibinfo {author} {\bibfnamefont {D.}~\bibnamefont
  {Gualdi}}, \bibinfo {author} {\bibfnamefont {H.~e.}\ \bibnamefont
  {Gil-Mar\'\i{}n}}, \ and\ \bibinfo {author} {\bibfnamefont {L.}~\bibnamefont
  {Verde}},\ }\href {\doibase 10.1088/1475-7516/2021/07/008} {\bibfield
  {journal} {\bibinfo  {journal} {JCAP}\ }\textbf {\bibinfo {volume} {07}},\
  \bibinfo {pages} {008} (\bibinfo {year} {2021})},\ \Eprint
  {http://arxiv.org/abs/2104.03976} {arXiv:2104.03976 [astro-ph.CO]}
  \BibitemShut {NoStop}%
\bibitem [{\citenamefont {Andrews}\ \emph {et~al.}(2023)\citenamefont
  {Andrews}, \citenamefont {Jasche}, \citenamefont {Lavaux},\ and\
  \citenamefont {Schmidt}}]{Andrews:2022nvv}%
  \BibitemOpen
  \bibfield  {author} {\bibinfo {author} {\bibfnamefont {A.}~\bibnamefont
  {Andrews}}, \bibinfo {author} {\bibfnamefont {J.}~\bibnamefont {Jasche}},
  \bibinfo {author} {\bibfnamefont {G.}~\bibnamefont {Lavaux}}, \ and\ \bibinfo
  {author} {\bibfnamefont {F.}~\bibnamefont {Schmidt}},\ }\href {\doibase
  10.1093/mnras/stad432} {\bibfield  {journal} {\bibinfo  {journal} {Mon. Not.
  Roy. Astron. Soc.}\ }\textbf {\bibinfo {volume} {520}},\ \bibinfo {pages}
  {5746} (\bibinfo {year} {2023})},\ \Eprint {http://arxiv.org/abs/2203.08838}
  {arXiv:2203.08838 [astro-ph.CO]} \BibitemShut {NoStop}%
\bibitem [{\citenamefont {Gagrani}\ and\ \citenamefont
  {Samushia}(2017)}]{Gagrani:2016rfy}%
  \BibitemOpen
  \bibfield  {author} {\bibinfo {author} {\bibfnamefont {P.}~\bibnamefont
  {Gagrani}}\ and\ \bibinfo {author} {\bibfnamefont {L.}~\bibnamefont
  {Samushia}},\ }\href {\doibase 10.1093/mnras/stx135} {\bibfield  {journal}
  {\bibinfo  {journal} {Mon. Not. Roy. Astron. Soc.}\ }\textbf {\bibinfo
  {volume} {467}},\ \bibinfo {pages} {928} (\bibinfo {year} {2017})},\ \Eprint
  {http://arxiv.org/abs/1610.03488} {arXiv:1610.03488 [astro-ph.CO]}
  \BibitemShut {NoStop}%
\bibitem [{\citenamefont {Byun}\ and\ \citenamefont
  {Krause}(2023)}]{Byun:2022rvn}%
  \BibitemOpen
  \bibfield  {author} {\bibinfo {author} {\bibfnamefont {J.}~\bibnamefont
  {Byun}}\ and\ \bibinfo {author} {\bibfnamefont {E.}~\bibnamefont {Krause}},\
  }\href {\doibase 10.1093/mnras/stac2313} {\bibfield  {journal} {\bibinfo
  {journal} {Mon. Not. Roy. Astron. Soc.}\ }\textbf {\bibinfo {volume} {525}},\
  \bibinfo {pages} {4854} (\bibinfo {year} {2023})},\ \Eprint
  {http://arxiv.org/abs/2205.04579} {arXiv:2205.04579 [astro-ph.CO]}
  \BibitemShut {NoStop}%
\bibitem [{\citenamefont {Tellarini}\ \emph {et~al.}(2016)\citenamefont
  {Tellarini}, \citenamefont {Ross}, \citenamefont {Tasinato},\ and\
  \citenamefont {Wands}}]{Tellarini:2016sgp}%
  \BibitemOpen
  \bibfield  {author} {\bibinfo {author} {\bibfnamefont {M.}~\bibnamefont
  {Tellarini}}, \bibinfo {author} {\bibfnamefont {A.~J.}\ \bibnamefont {Ross}},
  \bibinfo {author} {\bibfnamefont {G.}~\bibnamefont {Tasinato}}, \ and\
  \bibinfo {author} {\bibfnamefont {D.}~\bibnamefont {Wands}},\ }\href
  {\doibase 10.1088/1475-7516/2016/06/014} {\bibfield  {journal} {\bibinfo
  {journal} {JCAP}\ }\textbf {\bibinfo {volume} {06}},\ \bibinfo {pages} {014}
  (\bibinfo {year} {2016})},\ \Eprint {http://arxiv.org/abs/1603.06814}
  {arXiv:1603.06814 [astro-ph.CO]} \BibitemShut {NoStop}%
\bibitem [{\citenamefont {Catelan}\ \emph {et~al.}(2000)\citenamefont
  {Catelan}, \citenamefont {Porciani},\ and\ \citenamefont
  {Kamionkowski}}]{Catelan:2000vn}%
  \BibitemOpen
  \bibfield  {author} {\bibinfo {author} {\bibfnamefont {P.}~\bibnamefont
  {Catelan}}, \bibinfo {author} {\bibfnamefont {C.}~\bibnamefont {Porciani}}, \
  and\ \bibinfo {author} {\bibfnamefont {M.}~\bibnamefont {Kamionkowski}},\
  }\href {\doibase 10.1046/j.1365-8711.2000.04023.x} {\bibfield  {journal}
  {\bibinfo  {journal} {Mon. Not. Roy. Astron. Soc.}\ }\textbf {\bibinfo
  {volume} {318}},\ \bibinfo {pages} {39} (\bibinfo {year} {2000})},\ \Eprint
  {http://arxiv.org/abs/astro-ph/0005544} {arXiv:astro-ph/0005544} \BibitemShut
  {NoStop}%
\bibitem [{\citenamefont {{Baldauf}}\ \emph {et~al.}(2012)\citenamefont
  {{Baldauf}}, \citenamefont {{Seljak}}, \citenamefont {{Desjacques}},\ and\
  \citenamefont {{McDonald}}}]{2012PhRvD..86h3540B}%
  \BibitemOpen
  \bibfield  {author} {\bibinfo {author} {\bibfnamefont {T.}~\bibnamefont
  {{Baldauf}}}, \bibinfo {author} {\bibfnamefont {U.}~\bibnamefont {{Seljak}}},
  \bibinfo {author} {\bibfnamefont {V.}~\bibnamefont {{Desjacques}}}, \ and\
  \bibinfo {author} {\bibfnamefont {P.}~\bibnamefont {{McDonald}}},\ }\href
  {\doibase 10.1103/PhysRevD.86.083540} {\bibfield  {journal} {\bibinfo
  {journal} {\prd}\ }\textbf {\bibinfo {volume} {86}},\ \bibinfo {eid} {083540}
  (\bibinfo {year} {2012})},\ \Eprint {http://arxiv.org/abs/1201.4827}
  {arXiv:1201.4827 [astro-ph.CO]} \BibitemShut {NoStop}%
\bibitem [{\citenamefont {Rizzo}\ \emph {et~al.}(2023)\citenamefont {Rizzo},
  \citenamefont {Moretti}, \citenamefont {Pardede}, \citenamefont {Eggemeier},
  \citenamefont {Oddo}, \citenamefont {Sefusatti}, \citenamefont {Porciani},\
  and\ \citenamefont {Monaco}}]{Rizzo:2022lmh}%
  \BibitemOpen
  \bibfield  {author} {\bibinfo {author} {\bibfnamefont {F.}~\bibnamefont
  {Rizzo}}, \bibinfo {author} {\bibfnamefont {C.}~\bibnamefont {Moretti}},
  \bibinfo {author} {\bibfnamefont {K.}~\bibnamefont {Pardede}}, \bibinfo
  {author} {\bibfnamefont {A.}~\bibnamefont {Eggemeier}}, \bibinfo {author}
  {\bibfnamefont {A.}~\bibnamefont {Oddo}}, \bibinfo {author} {\bibfnamefont
  {E.}~\bibnamefont {Sefusatti}}, \bibinfo {author} {\bibfnamefont
  {C.}~\bibnamefont {Porciani}}, \ and\ \bibinfo {author} {\bibfnamefont
  {P.}~\bibnamefont {Monaco}},\ }\href {\doibase 10.1088/1475-7516/2023/01/031}
  {\bibfield  {journal} {\bibinfo  {journal} {JCAP}\ }\textbf {\bibinfo
  {volume} {01}},\ \bibinfo {pages} {031} (\bibinfo {year} {2023})},\ \Eprint
  {http://arxiv.org/abs/2204.13628} {arXiv:2204.13628 [astro-ph.CO]}
  \BibitemShut {NoStop}%
\bibitem [{\citenamefont {{Dalal}}\ \emph {et~al.}(2008)\citenamefont
  {{Dalal}}, \citenamefont {{Dor{\'e}}}, \citenamefont {{Huterer}},\ and\
  \citenamefont {{Shirokov}}}]{dalal2008}%
  \BibitemOpen
  \bibfield  {author} {\bibinfo {author} {\bibfnamefont {N.}~\bibnamefont
  {{Dalal}}}, \bibinfo {author} {\bibfnamefont {O.}~\bibnamefont {{Dor{\'e}}}},
  \bibinfo {author} {\bibfnamefont {D.}~\bibnamefont {{Huterer}}}, \ and\
  \bibinfo {author} {\bibfnamefont {A.}~\bibnamefont {{Shirokov}}},\ }\href
  {\doibase 10.1103/PhysRevD.77.123514} {\bibfield  {journal} {\bibinfo
  {journal} {\prd}\ }\textbf {\bibinfo {volume} {77}},\ \bibinfo {eid} {123514}
  (\bibinfo {year} {2008})},\ \Eprint {http://arxiv.org/abs/0710.4560}
  {arXiv:0710.4560 [astro-ph]} \BibitemShut {NoStop}%
\bibitem [{\citenamefont {Barreira}(2020)}]{Barreira:2020ekm}%
  \BibitemOpen
  \bibfield  {author} {\bibinfo {author} {\bibfnamefont {A.}~\bibnamefont
  {Barreira}},\ }\href {\doibase 10.1088/1475-7516/2020/12/031} {\bibfield
  {journal} {\bibinfo  {journal} {JCAP}\ }\textbf {\bibinfo {volume} {12}},\
  \bibinfo {pages} {031} (\bibinfo {year} {2020})},\ \Eprint
  {http://arxiv.org/abs/2009.06622} {arXiv:2009.06622 [astro-ph.CO]}
  \BibitemShut {NoStop}%
\bibitem [{\citenamefont {Barreira}(2022)}]{Barreira:2022sey}%
  \BibitemOpen
  \bibfield  {author} {\bibinfo {author} {\bibfnamefont {A.}~\bibnamefont
  {Barreira}},\ }\href {\doibase 10.1088/1475-7516/2022/11/013} {\bibfield
  {journal} {\bibinfo  {journal} {JCAP}\ }\textbf {\bibinfo {volume} {11}},\
  \bibinfo {pages} {013} (\bibinfo {year} {2022})},\ \Eprint
  {http://arxiv.org/abs/2205.05673} {arXiv:2205.05673 [astro-ph.CO]}
  \BibitemShut {NoStop}%
\bibitem [{\citenamefont {Barreira}\ and\ \citenamefont
  {Krause}(2023)}]{Barreira:2023rxn}%
  \BibitemOpen
  \bibfield  {author} {\bibinfo {author} {\bibfnamefont {A.}~\bibnamefont
  {Barreira}}\ and\ \bibinfo {author} {\bibfnamefont {E.}~\bibnamefont
  {Krause}},\ }\href@noop {} {\  (\bibinfo {year} {2023})},\ \Eprint
  {http://arxiv.org/abs/2302.09066} {arXiv:2302.09066 [astro-ph.CO]}
  \BibitemShut {NoStop}%
\bibitem [{\citenamefont {Scoccimarro}(2015)}]{Scoccimarro:2015bla}%
  \BibitemOpen
  \bibfield  {author} {\bibinfo {author} {\bibfnamefont {R.}~\bibnamefont
  {Scoccimarro}},\ }\href {\doibase 10.1103/PhysRevD.92.083532} {\bibfield
  {journal} {\bibinfo  {journal} {Phys. Rev. D}\ }\textbf {\bibinfo {volume}
  {92}},\ \bibinfo {pages} {083532} (\bibinfo {year} {2015})},\ \Eprint
  {http://arxiv.org/abs/1506.02729} {arXiv:1506.02729 [astro-ph.CO]}
  \BibitemShut {NoStop}%
\bibitem [{\citenamefont {Aghanim}\ \emph {et~al.}(2020)\citenamefont {Aghanim}
  \emph {et~al.}}]{Planck:2018vyg}%
  \BibitemOpen
  \bibfield  {author} {\bibinfo {author} {\bibfnamefont {N.}~\bibnamefont
  {Aghanim}} \emph {et~al.} (\bibinfo {collaboration} {Planck}),\ }\href
  {\doibase 10.1051/0004-6361/201833910} {\bibfield  {journal} {\bibinfo
  {journal} {Astron. Astrophys.}\ }\textbf {\bibinfo {volume} {641}},\ \bibinfo
  {pages} {A6} (\bibinfo {year} {2020})},\ \bibinfo {note} {[Erratum:
  Astron.Astrophys. 652, C4 (2021)]},\ \Eprint
  {http://arxiv.org/abs/1807.06209} {arXiv:1807.06209 [astro-ph.CO]}
  \BibitemShut {NoStop}%
\bibitem [{\citenamefont {Lazeyras}\ \emph {et~al.}(2016)\citenamefont
  {Lazeyras}, \citenamefont {Wagner}, \citenamefont {Baldauf},\ and\
  \citenamefont {Schmidt}}]{Lazeyras:2015lgp}%
  \BibitemOpen
  \bibfield  {author} {\bibinfo {author} {\bibfnamefont {T.}~\bibnamefont
  {Lazeyras}}, \bibinfo {author} {\bibfnamefont {C.}~\bibnamefont {Wagner}},
  \bibinfo {author} {\bibfnamefont {T.}~\bibnamefont {Baldauf}}, \ and\
  \bibinfo {author} {\bibfnamefont {F.}~\bibnamefont {Schmidt}},\ }\href
  {\doibase 10.1088/1475-7516/2016/02/018} {\bibfield  {journal} {\bibinfo
  {journal} {JCAP}\ }\textbf {\bibinfo {volume} {02}},\ \bibinfo {pages} {018}
  (\bibinfo {year} {2016})},\ \Eprint {http://arxiv.org/abs/1511.01096}
  {arXiv:1511.01096 [astro-ph.CO]} \BibitemShut {NoStop}%
\bibitem [{\citenamefont {Scoville}\ \emph {et~al.}(2007)\citenamefont
  {Scoville} \emph {et~al.}}]{Scoville:2006vq}%
  \BibitemOpen
  \bibfield  {author} {\bibinfo {author} {\bibfnamefont {N.}~\bibnamefont
  {Scoville}} \emph {et~al.},\ }\href {\doibase 10.1086/516585} {\bibfield
  {journal} {\bibinfo  {journal} {Astrophys. J. Suppl.}\ }\textbf {\bibinfo
  {volume} {172}},\ \bibinfo {pages} {1} (\bibinfo {year} {2007})},\ \Eprint
  {http://arxiv.org/abs/astro-ph/0612305} {arXiv:astro-ph/0612305} \BibitemShut
  {NoStop}%
\bibitem [{\citenamefont {Ilbert}\ \emph {et~al.}(2009)\citenamefont {Ilbert}
  \emph {et~al.}}]{Ilbert:2008hz}%
  \BibitemOpen
  \bibfield  {author} {\bibinfo {author} {\bibfnamefont {O.}~\bibnamefont
  {Ilbert}} \emph {et~al.},\ }\href {\doibase 10.1088/0004-637X/690/2/1236}
  {\bibfield  {journal} {\bibinfo  {journal} {Astrophys. J.}\ }\textbf
  {\bibinfo {volume} {690}},\ \bibinfo {pages} {1236} (\bibinfo {year}
  {2009})},\ \Eprint {http://arxiv.org/abs/0809.2101} {arXiv:0809.2101
  [astro-ph]} \BibitemShut {NoStop}%
\bibitem [{\citenamefont {Tinker}\ \emph {et~al.}(2008)\citenamefont {Tinker},
  \citenamefont {Kravtsov}, \citenamefont {Klypin}, \citenamefont {Abazajian},
  \citenamefont {Warren}, \citenamefont {Yepes}, \citenamefont {Gottlober},\
  and\ \citenamefont {Holz}}]{Tinker:2008ff}%
  \BibitemOpen
  \bibfield  {author} {\bibinfo {author} {\bibfnamefont {J.~L.}\ \bibnamefont
  {Tinker}}, \bibinfo {author} {\bibfnamefont {A.~V.}\ \bibnamefont
  {Kravtsov}}, \bibinfo {author} {\bibfnamefont {A.}~\bibnamefont {Klypin}},
  \bibinfo {author} {\bibfnamefont {K.}~\bibnamefont {Abazajian}}, \bibinfo
  {author} {\bibfnamefont {M.~S.}\ \bibnamefont {Warren}}, \bibinfo {author}
  {\bibfnamefont {G.}~\bibnamefont {Yepes}}, \bibinfo {author} {\bibfnamefont
  {S.}~\bibnamefont {Gottlober}}, \ and\ \bibinfo {author} {\bibfnamefont
  {D.~E.}\ \bibnamefont {Holz}},\ }\href {\doibase 10.1086/591439} {\bibfield
  {journal} {\bibinfo  {journal} {Astrophys. J.}\ }\textbf {\bibinfo {volume}
  {688}},\ \bibinfo {pages} {709} (\bibinfo {year} {2008})},\ \Eprint
  {http://arxiv.org/abs/0803.2706} {arXiv:0803.2706 [astro-ph]} \BibitemShut
  {NoStop}%
\bibitem [{\citenamefont {{Tinker}}\ \emph {et~al.}(2010)\citenamefont
  {{Tinker}}, \citenamefont {{Robertson}}, \citenamefont {{Kravtsov}},
  \citenamefont {{Klypin}}, \citenamefont {{Warren}}, \citenamefont {{Yepes}},\
  and\ \citenamefont {{Gottl{\"o}ber}}}]{2010ApJ...724..878T}%
  \BibitemOpen
  \bibfield  {author} {\bibinfo {author} {\bibfnamefont {J.~L.}\ \bibnamefont
  {{Tinker}}}, \bibinfo {author} {\bibfnamefont {B.~E.}\ \bibnamefont
  {{Robertson}}}, \bibinfo {author} {\bibfnamefont {A.~V.}\ \bibnamefont
  {{Kravtsov}}}, \bibinfo {author} {\bibfnamefont {A.}~\bibnamefont
  {{Klypin}}}, \bibinfo {author} {\bibfnamefont {M.~S.}\ \bibnamefont
  {{Warren}}}, \bibinfo {author} {\bibfnamefont {G.}~\bibnamefont {{Yepes}}}, \
  and\ \bibinfo {author} {\bibfnamefont {S.}~\bibnamefont {{Gottl{\"o}ber}}},\
  }\href {\doibase 10.1088/0004-637X/724/2/878} {\bibfield  {journal} {\bibinfo
   {journal} {\apj}\ }\textbf {\bibinfo {volume} {724}},\ \bibinfo {pages}
  {878} (\bibinfo {year} {2010})},\ \Eprint {http://arxiv.org/abs/1001.3162}
  {arXiv:1001.3162 [astro-ph.CO]} \BibitemShut {NoStop}%
\bibitem [{\citenamefont {Clarkson}\ \emph {et~al.}(2019)\citenamefont
  {Clarkson}, \citenamefont {de~Weerd}, \citenamefont {Jolicoeur},
  \citenamefont {Maartens},\ and\ \citenamefont {Umeh}}]{Clarkson:2018dwn}%
  \BibitemOpen
  \bibfield  {author} {\bibinfo {author} {\bibfnamefont {C.}~\bibnamefont
  {Clarkson}}, \bibinfo {author} {\bibfnamefont {E.~M.}\ \bibnamefont
  {de~Weerd}}, \bibinfo {author} {\bibfnamefont {S.}~\bibnamefont {Jolicoeur}},
  \bibinfo {author} {\bibfnamefont {R.}~\bibnamefont {Maartens}}, \ and\
  \bibinfo {author} {\bibfnamefont {O.}~\bibnamefont {Umeh}},\ }\href {\doibase
  10.1093/mnrasl/slz066} {\bibfield  {journal} {\bibinfo  {journal} {Mon. Not.
  Roy. Astron. Soc.}\ }\textbf {\bibinfo {volume} {486}},\ \bibinfo {pages}
  {L101} (\bibinfo {year} {2019})},\ \Eprint {http://arxiv.org/abs/1812.09512}
  {arXiv:1812.09512 [astro-ph.CO]} \BibitemShut {NoStop}%
\bibitem [{\citenamefont {Jeong}\ and\ \citenamefont
  {Schmidt}(2020)}]{Jeong:2019igb}%
  \BibitemOpen
  \bibfield  {author} {\bibinfo {author} {\bibfnamefont {D.}~\bibnamefont
  {Jeong}}\ and\ \bibinfo {author} {\bibfnamefont {F.}~\bibnamefont
  {Schmidt}},\ }\href {\doibase 10.1103/PhysRevD.102.023530} {\bibfield
  {journal} {\bibinfo  {journal} {Phys. Rev. D}\ }\textbf {\bibinfo {volume}
  {102}},\ \bibinfo {pages} {023530} (\bibinfo {year} {2020})},\ \Eprint
  {http://arxiv.org/abs/1906.05198} {arXiv:1906.05198 [astro-ph.CO]}
  \BibitemShut {NoStop}%
\bibitem [{\citenamefont {Maartens}\ \emph {et~al.}(2020)\citenamefont
  {Maartens}, \citenamefont {Jolicoeur}, \citenamefont {Umeh}, \citenamefont
  {De~Weerd}, \citenamefont {Clarkson},\ and\ \citenamefont
  {Camera}}]{Maartens:2019yhx}%
  \BibitemOpen
  \bibfield  {author} {\bibinfo {author} {\bibfnamefont {R.}~\bibnamefont
  {Maartens}}, \bibinfo {author} {\bibfnamefont {S.}~\bibnamefont {Jolicoeur}},
  \bibinfo {author} {\bibfnamefont {O.}~\bibnamefont {Umeh}}, \bibinfo {author}
  {\bibfnamefont {E.~M.}\ \bibnamefont {De~Weerd}}, \bibinfo {author}
  {\bibfnamefont {C.}~\bibnamefont {Clarkson}}, \ and\ \bibinfo {author}
  {\bibfnamefont {S.}~\bibnamefont {Camera}},\ }\href {\doibase
  10.1088/1475-7516/2020/03/065} {\bibfield  {journal} {\bibinfo  {journal}
  {JCAP}\ }\textbf {\bibinfo {volume} {03}},\ \bibinfo {pages} {065} (\bibinfo
  {year} {2020})},\ \Eprint {http://arxiv.org/abs/1911.02398} {arXiv:1911.02398
  [astro-ph.CO]} \BibitemShut {NoStop}%
\bibitem [{\citenamefont {Jolicoeur}\ \emph {et~al.}(2021)\citenamefont
  {Jolicoeur}, \citenamefont {Maartens}, \citenamefont {De~Weerd},
  \citenamefont {Umeh}, \citenamefont {Clarkson},\ and\ \citenamefont
  {Camera}}]{Jolicoeur:2020eup}%
  \BibitemOpen
  \bibfield  {author} {\bibinfo {author} {\bibfnamefont {S.}~\bibnamefont
  {Jolicoeur}}, \bibinfo {author} {\bibfnamefont {R.}~\bibnamefont {Maartens}},
  \bibinfo {author} {\bibfnamefont {E.~M.}\ \bibnamefont {De~Weerd}}, \bibinfo
  {author} {\bibfnamefont {O.}~\bibnamefont {Umeh}}, \bibinfo {author}
  {\bibfnamefont {C.}~\bibnamefont {Clarkson}}, \ and\ \bibinfo {author}
  {\bibfnamefont {S.}~\bibnamefont {Camera}},\ }\href {\doibase
  10.1088/1475-7516/2021/06/039} {\bibfield  {journal} {\bibinfo  {journal}
  {JCAP}\ }\textbf {\bibinfo {volume} {06}},\ \bibinfo {pages} {039} (\bibinfo
  {year} {2021})},\ \Eprint {http://arxiv.org/abs/2009.06197} {arXiv:2009.06197
  [astro-ph.CO]} \BibitemShut {NoStop}%
\bibitem [{\citenamefont {Noorikuhani}\ and\ \citenamefont
  {Scoccimarro}(2023)}]{Noorikuhani:2022bwc}%
  \BibitemOpen
  \bibfield  {author} {\bibinfo {author} {\bibfnamefont {M.}~\bibnamefont
  {Noorikuhani}}\ and\ \bibinfo {author} {\bibfnamefont {R.}~\bibnamefont
  {Scoccimarro}},\ }\href {\doibase 10.1103/PhysRevD.107.083528} {\bibfield
  {journal} {\bibinfo  {journal} {Phys. Rev. D}\ }\textbf {\bibinfo {volume}
  {107}},\ \bibinfo {pages} {083528} (\bibinfo {year} {2023})},\ \Eprint
  {http://arxiv.org/abs/2207.12383} {arXiv:2207.12383 [astro-ph.CO]}
  \BibitemShut {NoStop}%
\bibitem [{\citenamefont {Foglieni}\ \emph {et~al.}(2023)\citenamefont
  {Foglieni}, \citenamefont {Pantiri}, \citenamefont {Di~Dio},\ and\
  \citenamefont {Castorina}}]{Foglieni:2023xca}%
  \BibitemOpen
  \bibfield  {author} {\bibinfo {author} {\bibfnamefont {M.}~\bibnamefont
  {Foglieni}}, \bibinfo {author} {\bibfnamefont {M.}~\bibnamefont {Pantiri}},
  \bibinfo {author} {\bibfnamefont {E.}~\bibnamefont {Di~Dio}}, \ and\ \bibinfo
  {author} {\bibfnamefont {E.}~\bibnamefont {Castorina}},\ }\href {\doibase
  10.1103/PhysRevLett.131.111201} {\bibfield  {journal} {\bibinfo  {journal}
  {Phys. Rev. Lett.}\ }\textbf {\bibinfo {volume} {131}},\ \bibinfo {pages}
  {111201} (\bibinfo {year} {2023})},\ \Eprint
  {http://arxiv.org/abs/2303.03142} {arXiv:2303.03142 [astro-ph.CO]}
  \BibitemShut {NoStop}%
\bibitem [{\citenamefont {Pardede}\ \emph {et~al.}(2023)\citenamefont
  {Pardede}, \citenamefont {Di~Dio},\ and\ \citenamefont
  {Castorina}}]{Pardede:2023ddq}%
  \BibitemOpen
  \bibfield  {author} {\bibinfo {author} {\bibfnamefont {K.}~\bibnamefont
  {Pardede}}, \bibinfo {author} {\bibfnamefont {E.}~\bibnamefont {Di~Dio}}, \
  and\ \bibinfo {author} {\bibfnamefont {E.}~\bibnamefont {Castorina}},\ }\href
  {\doibase 10.1088/1475-7516/2023/09/030} {\bibfield  {journal} {\bibinfo
  {journal} {JCAP}\ }\textbf {\bibinfo {volume} {09}},\ \bibinfo {pages} {030}
  (\bibinfo {year} {2023})},\ \Eprint {http://arxiv.org/abs/2302.12789}
  {arXiv:2302.12789 [astro-ph.CO]} \BibitemShut {NoStop}%
\bibitem [{\citenamefont {Hartlap}\ \emph {et~al.}(2007)\citenamefont
  {Hartlap}, \citenamefont {Simon},\ and\ \citenamefont
  {Schneider}}]{Hartlap:2006kj}%
  \BibitemOpen
  \bibfield  {author} {\bibinfo {author} {\bibfnamefont {J.}~\bibnamefont
  {Hartlap}}, \bibinfo {author} {\bibfnamefont {P.}~\bibnamefont {Simon}}, \
  and\ \bibinfo {author} {\bibfnamefont {P.}~\bibnamefont {Schneider}},\ }\href
  {\doibase 10.1051/0004-6361:20066170} {\bibfield  {journal} {\bibinfo
  {journal} {Astron. Astrophys.}\ }\textbf {\bibinfo {volume} {464}},\ \bibinfo
  {pages} {399} (\bibinfo {year} {2007})},\ \Eprint
  {http://arxiv.org/abs/astro-ph/0608064} {arXiv:astro-ph/0608064} \BibitemShut
  {NoStop}%
\bibitem [{\citenamefont {Taylor}\ and\ \citenamefont
  {Joachimi}(2014)}]{Taylor:2014ota}%
  \BibitemOpen
  \bibfield  {author} {\bibinfo {author} {\bibfnamefont {A.}~\bibnamefont
  {Taylor}}\ and\ \bibinfo {author} {\bibfnamefont {B.}~\bibnamefont
  {Joachimi}},\ }\href {\doibase 10.1093/mnras/stu996} {\bibfield  {journal}
  {\bibinfo  {journal} {Mon. Not. Roy. Astron. Soc.}\ }\textbf {\bibinfo
  {volume} {442}},\ \bibinfo {pages} {2728} (\bibinfo {year} {2014})},\ \Eprint
  {http://arxiv.org/abs/1402.6983} {arXiv:1402.6983 [astro-ph.CO]} \BibitemShut
  {NoStop}%
\bibitem [{\citenamefont {{LSST Science Collaboration}}\ \emph
  {et~al.}(2009)\citenamefont {{LSST Science Collaboration}}, \citenamefont
  {{Abell}}, \citenamefont {{Allison}}, \citenamefont {{Anderson}},
  \citenamefont {{Andrew}}, \citenamefont {{Angel}}, \citenamefont {{Armus}},
  \citenamefont {{Arnett}}, \citenamefont {{Asztalos}}, \citenamefont
  {{Axelrod}}, \citenamefont {{Bailey}}, \citenamefont {{Ballantyne}},
  \citenamefont {{Bankert}}, \citenamefont {{Barkhouse}}, \citenamefont
  {{Barr}}, \citenamefont {{Barrientos}}, \citenamefont {{Barth}},
  \citenamefont {{Bartlett}}, \citenamefont {{Becker}}, \citenamefont
  {{Becla}}, \citenamefont {{Beers}}, \citenamefont {{Bernstein}},
  \citenamefont {{Biswas}}, \citenamefont {{Blanton}}, \citenamefont {{Bloom}},
  \citenamefont {{Bochanski}}, \citenamefont {{Boeshaar}}, \citenamefont
  {{Borne}}, \citenamefont {{Bradac}}, \citenamefont {{Brandt}}, \citenamefont
  {{Bridge}}, \citenamefont {{Brown}}, \citenamefont {{Brunner}}, \citenamefont
  {{Bullock}}, \citenamefont {{Burgasser}}, \citenamefont {{Burge}},
  \citenamefont {{Burke}}, \citenamefont {{Cargile}}, \citenamefont
  {{Chandrasekharan}}, \citenamefont {{Chartas}}, \citenamefont {{Chesley}},
  \citenamefont {{Chu}}, \citenamefont {{Cinabro}}, \citenamefont {{Claire}},
  \citenamefont {{Claver}}, \citenamefont {{Clowe}}, \citenamefont
  {{Connolly}}, \citenamefont {{Cook}}, \citenamefont {{Cooke}}, \citenamefont
  {{Cooray}}, \citenamefont {{Covey}}, \citenamefont {{Culliton}},
  \citenamefont {{de Jong}}, \citenamefont {{de Vries}}, \citenamefont
  {{Debattista}}, \citenamefont {{Delgado}}, \citenamefont {{Dell'Antonio}},
  \citenamefont {{Dhital}}, \citenamefont {{Di Stefano}}, \citenamefont
  {{Dickinson}}, \citenamefont {{Dilday}}, \citenamefont {{Djorgovski}},
  \citenamefont {{Dobler}}, \citenamefont {{Donalek}}, \citenamefont
  {{Dubois-Felsmann}}, \citenamefont {{Durech}}, \citenamefont {{Eliasdottir}},
  \citenamefont {{Eracleous}}, \citenamefont {{Eyer}}, \citenamefont {{Falco}},
  \citenamefont {{Fan}}, \citenamefont {{Fassnacht}}, \citenamefont
  {{Ferguson}}, \citenamefont {{Fernandez}}, \citenamefont {{Fields}},
  \citenamefont {{Finkbeiner}}, \citenamefont {{Figueroa}}, \citenamefont
  {{Fox}}, \citenamefont {{Francke}}, \citenamefont {{Frank}}, \citenamefont
  {{Frieman}}, \citenamefont {{Fromenteau}}, \citenamefont {{Furqan}},
  \citenamefont {{Galaz}}, \citenamefont {{Gal-Yam}}, \citenamefont
  {{Garnavich}}, \citenamefont {{Gawiser}}, \citenamefont {{Geary}},
  \citenamefont {{Gee}}, \citenamefont {{Gibson}}, \citenamefont {{Gilmore}},
  \citenamefont {{Grace}}, \citenamefont {{Green}}, \citenamefont {{Gressler}},
  \citenamefont {{Grillmair}}, \citenamefont {{Habib}}, \citenamefont
  {{Haggerty}}, \citenamefont {{Hamuy}}, \citenamefont {{Harris}},
  \citenamefont {{Hawley}}, \citenamefont {{Heavens}}, \citenamefont {{Hebb}},
  \citenamefont {{Henry}}, \citenamefont {{Hileman}}, \citenamefont {{Hilton}},
  \citenamefont {{Hoadley}}, \citenamefont {{Holberg}}, \citenamefont
  {{Holman}}, \citenamefont {{Howell}}, \citenamefont {{Infante}},
  \citenamefont {{Ivezic}}, \citenamefont {{Jacoby}}, \citenamefont {{Jain}},
  \citenamefont {{R}}, \citenamefont {{Jedicke}}, \citenamefont {{Jee}},
  \citenamefont {{Garrett Jernigan}}, \citenamefont {{Jha}}, \citenamefont
  {{Johnston}}, \citenamefont {{Jones}}, \citenamefont {{Juric}}, \citenamefont
  {{Kaasalainen}}, \citenamefont {{Styliani}}, \citenamefont {{Kafka}},
  \citenamefont {{Kahn}}, \citenamefont {{Kaib}}, \citenamefont {{Kalirai}},
  \citenamefont {{Kantor}}, \citenamefont {{Kasliwal}}, \citenamefont
  {{Keeton}}, \citenamefont {{Kessler}}, \citenamefont {{Knezevic}},
  \citenamefont {{Kowalski}}, \citenamefont {{Krabbendam}}, \citenamefont
  {{Krughoff}}, \citenamefont {{Kulkarni}}, \citenamefont {{Kuhlman}},
  \citenamefont {{Lacy}}, \citenamefont {{Lepine}}, \citenamefont {{Liang}},
  \citenamefont {{Lien}}, \citenamefont {{Lira}}, \citenamefont {{Long}},
  \citenamefont {{Lorenz}}, \citenamefont {{Lotz}}, \citenamefont {{Lupton}},
  \citenamefont {{Lutz}}, \citenamefont {{Macri}}, \citenamefont {{Mahabal}},
  \citenamefont {{Mandelbaum}}, \citenamefont {{Marshall}}, \citenamefont
  {{May}}, \citenamefont {{McGehee}}, \citenamefont {{Meadows}}, \citenamefont
  {{Meert}}, \citenamefont {{Milani}}, \citenamefont {{Miller}}, \citenamefont
  {{Miller}}, \citenamefont {{Mills}}, \citenamefont {{Minniti}}, \citenamefont
  {{Monet}}, \citenamefont {{Mukadam}}, \citenamefont {{Nakar}}, \citenamefont
  {{Neill}}, \citenamefont {{Newman}}, \citenamefont {{Nikolaev}},
  \citenamefont {{Nordby}}, \citenamefont {{O'Connor}}, \citenamefont
  {{Oguri}}, \citenamefont {{Oliver}}, \citenamefont {{Olivier}}, \citenamefont
  {{Olsen}}, \citenamefont {{Olsen}}, \citenamefont {{Olszewski}},
  \citenamefont {{Oluseyi}}, \citenamefont {{Padilla}}, \citenamefont
  {{Parker}}, \citenamefont {{Pepper}}, \citenamefont {{Peterson}},
  \citenamefont {{Petry}}, \citenamefont {{Pinto}}, \citenamefont {{Pizagno}},
  \citenamefont {{Popescu}}, \citenamefont {{Prsa}}, \citenamefont {{Radcka}},
  \citenamefont {{Raddick}}, \citenamefont {{Rasmussen}}, \citenamefont
  {{Rau}}, \citenamefont {{Rho}}, \citenamefont {{Rhoads}}, \citenamefont
  {{Richards}}, \citenamefont {{Ridgway}}, \citenamefont {{Robertson}},
  \citenamefont {{Roskar}}, \citenamefont {{Saha}}, \citenamefont
  {{Sarajedini}}, \citenamefont {{Scannapieco}}, \citenamefont {{Schalk}},
  \citenamefont {{Schindler}}, \citenamefont {{Schmidt}}, \citenamefont
  {{Schmidt}}, \citenamefont {{Schneider}}, \citenamefont {{Schumacher}},
  \citenamefont {{Scranton}}, \citenamefont {{Sebag}}, \citenamefont
  {{Seppala}}, \citenamefont {{Shemmer}}, \citenamefont {{Simon}},
  \citenamefont {{Sivertz}}, \citenamefont {{Smith}}, \citenamefont {{Allyn
  Smith}}, \citenamefont {{Smith}}, \citenamefont {{Spitz}}, \citenamefont
  {{Stanford}}, \citenamefont {{Stassun}}, \citenamefont {{Strader}},
  \citenamefont {{Strauss}}, \citenamefont {{Stubbs}}, \citenamefont
  {{Sweeney}}, \citenamefont {{Szalay}}, \citenamefont {{Szkody}},
  \citenamefont {{Takada}}, \citenamefont {{Thorman}}, \citenamefont
  {{Trilling}}, \citenamefont {{Trimble}}, \citenamefont {{Tyson}},
  \citenamefont {{Van Berg}}, \citenamefont {{Vanden Berk}}, \citenamefont
  {{VanderPlas}}, \citenamefont {{Verde}}, \citenamefont {{Vrsnak}},
  \citenamefont {{Walkowicz}}, \citenamefont {{Wandelt}}, \citenamefont
  {{Wang}}, \citenamefont {{Wang}}, \citenamefont {{Warner}}, \citenamefont
  {{Wechsler}}, \citenamefont {{West}}, \citenamefont {{Wiecha}}, \citenamefont
  {{Williams}}, \citenamefont {{Willman}}, \citenamefont {{Wittman}},
  \citenamefont {{Wolff}}, \citenamefont {{Wood-Vasey}}, \citenamefont
  {{Wozniak}}, \citenamefont {{Young}}, \citenamefont {{Zentner}},\ and\
  \citenamefont {{Zhan}}}]{2009arXiv0912.0201L}%
  \BibitemOpen
  \bibfield  {author} {\bibinfo {author} {\bibnamefont {{LSST Science
  Collaboration}}}, \bibinfo {author} {\bibfnamefont {P.~A.}\ \bibnamefont
  {{Abell}}}, \bibinfo {author} {\bibfnamefont {J.}~\bibnamefont {{Allison}}},
  \bibinfo {author} {\bibfnamefont {S.~F.}\ \bibnamefont {{Anderson}}},
  \bibinfo {author} {\bibfnamefont {J.~R.}\ \bibnamefont {{Andrew}}}, \bibinfo
  {author} {\bibfnamefont {J.~R.~P.}\ \bibnamefont {{Angel}}}, \bibinfo
  {author} {\bibfnamefont {L.}~\bibnamefont {{Armus}}}, \bibinfo {author}
  {\bibfnamefont {D.}~\bibnamefont {{Arnett}}}, \bibinfo {author}
  {\bibfnamefont {S.~J.}\ \bibnamefont {{Asztalos}}}, \bibinfo {author}
  {\bibfnamefont {T.~S.}\ \bibnamefont {{Axelrod}}}, \bibinfo {author}
  {\bibfnamefont {S.}~\bibnamefont {{Bailey}}}, \bibinfo {author}
  {\bibfnamefont {D.~R.}\ \bibnamefont {{Ballantyne}}}, \bibinfo {author}
  {\bibfnamefont {J.~R.}\ \bibnamefont {{Bankert}}}, \bibinfo {author}
  {\bibfnamefont {W.~A.}\ \bibnamefont {{Barkhouse}}}, \bibinfo {author}
  {\bibfnamefont {J.~D.}\ \bibnamefont {{Barr}}}, \bibinfo {author}
  {\bibfnamefont {L.~F.}\ \bibnamefont {{Barrientos}}}, \bibinfo {author}
  {\bibfnamefont {A.~J.}\ \bibnamefont {{Barth}}}, \bibinfo {author}
  {\bibfnamefont {J.~G.}\ \bibnamefont {{Bartlett}}}, \bibinfo {author}
  {\bibfnamefont {A.~C.}\ \bibnamefont {{Becker}}}, \bibinfo {author}
  {\bibfnamefont {J.}~\bibnamefont {{Becla}}}, \bibinfo {author} {\bibfnamefont
  {T.~C.}\ \bibnamefont {{Beers}}}, \bibinfo {author} {\bibfnamefont {J.~P.}\
  \bibnamefont {{Bernstein}}}, \bibinfo {author} {\bibfnamefont
  {R.}~\bibnamefont {{Biswas}}}, \bibinfo {author} {\bibfnamefont {M.~R.}\
  \bibnamefont {{Blanton}}}, \bibinfo {author} {\bibfnamefont {J.~S.}\
  \bibnamefont {{Bloom}}}, \bibinfo {author} {\bibfnamefont {J.~J.}\
  \bibnamefont {{Bochanski}}}, \bibinfo {author} {\bibfnamefont
  {P.}~\bibnamefont {{Boeshaar}}}, \bibinfo {author} {\bibfnamefont {K.~D.}\
  \bibnamefont {{Borne}}}, \bibinfo {author} {\bibfnamefont {M.}~\bibnamefont
  {{Bradac}}}, \bibinfo {author} {\bibfnamefont {W.~N.}\ \bibnamefont
  {{Brandt}}}, \bibinfo {author} {\bibfnamefont {C.~R.}\ \bibnamefont
  {{Bridge}}}, \bibinfo {author} {\bibfnamefont {M.~E.}\ \bibnamefont
  {{Brown}}}, \bibinfo {author} {\bibfnamefont {R.~J.}\ \bibnamefont
  {{Brunner}}}, \bibinfo {author} {\bibfnamefont {J.~S.}\ \bibnamefont
  {{Bullock}}}, \bibinfo {author} {\bibfnamefont {A.~J.}\ \bibnamefont
  {{Burgasser}}}, \bibinfo {author} {\bibfnamefont {J.~H.}\ \bibnamefont
  {{Burge}}}, \bibinfo {author} {\bibfnamefont {D.~L.}\ \bibnamefont
  {{Burke}}}, \bibinfo {author} {\bibfnamefont {P.~A.}\ \bibnamefont
  {{Cargile}}}, \bibinfo {author} {\bibfnamefont {S.}~\bibnamefont
  {{Chandrasekharan}}}, \bibinfo {author} {\bibfnamefont {G.}~\bibnamefont
  {{Chartas}}}, \bibinfo {author} {\bibfnamefont {S.~R.}\ \bibnamefont
  {{Chesley}}}, \bibinfo {author} {\bibfnamefont {Y.-H.}\ \bibnamefont
  {{Chu}}}, \bibinfo {author} {\bibfnamefont {D.}~\bibnamefont {{Cinabro}}},
  \bibinfo {author} {\bibfnamefont {M.~W.}\ \bibnamefont {{Claire}}}, \bibinfo
  {author} {\bibfnamefont {C.~F.}\ \bibnamefont {{Claver}}}, \bibinfo {author}
  {\bibfnamefont {D.}~\bibnamefont {{Clowe}}}, \bibinfo {author} {\bibfnamefont
  {A.~J.}\ \bibnamefont {{Connolly}}}, \bibinfo {author} {\bibfnamefont
  {K.~H.}\ \bibnamefont {{Cook}}}, \bibinfo {author} {\bibfnamefont
  {J.}~\bibnamefont {{Cooke}}}, \bibinfo {author} {\bibfnamefont
  {A.}~\bibnamefont {{Cooray}}}, \bibinfo {author} {\bibfnamefont {K.~R.}\
  \bibnamefont {{Covey}}}, \bibinfo {author} {\bibfnamefont {C.~S.}\
  \bibnamefont {{Culliton}}}, \bibinfo {author} {\bibfnamefont
  {R.}~\bibnamefont {{de Jong}}}, \bibinfo {author} {\bibfnamefont {W.~H.}\
  \bibnamefont {{de Vries}}}, \bibinfo {author} {\bibfnamefont {V.~P.}\
  \bibnamefont {{Debattista}}}, \bibinfo {author} {\bibfnamefont
  {F.}~\bibnamefont {{Delgado}}}, \bibinfo {author} {\bibfnamefont {I.~P.}\
  \bibnamefont {{Dell'Antonio}}}, \bibinfo {author} {\bibfnamefont
  {S.}~\bibnamefont {{Dhital}}}, \bibinfo {author} {\bibfnamefont
  {R.}~\bibnamefont {{Di Stefano}}}, \bibinfo {author} {\bibfnamefont
  {M.}~\bibnamefont {{Dickinson}}}, \bibinfo {author} {\bibfnamefont
  {B.}~\bibnamefont {{Dilday}}}, \bibinfo {author} {\bibfnamefont {S.~G.}\
  \bibnamefont {{Djorgovski}}}, \bibinfo {author} {\bibfnamefont
  {G.}~\bibnamefont {{Dobler}}}, \bibinfo {author} {\bibfnamefont
  {C.}~\bibnamefont {{Donalek}}}, \bibinfo {author} {\bibfnamefont
  {G.}~\bibnamefont {{Dubois-Felsmann}}}, \bibinfo {author} {\bibfnamefont
  {J.}~\bibnamefont {{Durech}}}, \bibinfo {author} {\bibfnamefont
  {A.}~\bibnamefont {{Eliasdottir}}}, \bibinfo {author} {\bibfnamefont
  {M.}~\bibnamefont {{Eracleous}}}, \bibinfo {author} {\bibfnamefont
  {L.}~\bibnamefont {{Eyer}}}, \bibinfo {author} {\bibfnamefont {E.~E.}\
  \bibnamefont {{Falco}}}, \bibinfo {author} {\bibfnamefont {X.}~\bibnamefont
  {{Fan}}}, \bibinfo {author} {\bibfnamefont {C.~D.}\ \bibnamefont
  {{Fassnacht}}}, \bibinfo {author} {\bibfnamefont {H.~C.}\ \bibnamefont
  {{Ferguson}}}, \bibinfo {author} {\bibfnamefont {Y.~R.}\ \bibnamefont
  {{Fernandez}}}, \bibinfo {author} {\bibfnamefont {B.~D.}\ \bibnamefont
  {{Fields}}}, \bibinfo {author} {\bibfnamefont {D.}~\bibnamefont
  {{Finkbeiner}}}, \bibinfo {author} {\bibfnamefont {E.~E.}\ \bibnamefont
  {{Figueroa}}}, \bibinfo {author} {\bibfnamefont {D.~B.}\ \bibnamefont
  {{Fox}}}, \bibinfo {author} {\bibfnamefont {H.}~\bibnamefont {{Francke}}},
  \bibinfo {author} {\bibfnamefont {J.~S.}\ \bibnamefont {{Frank}}}, \bibinfo
  {author} {\bibfnamefont {J.}~\bibnamefont {{Frieman}}}, \bibinfo {author}
  {\bibfnamefont {S.}~\bibnamefont {{Fromenteau}}}, \bibinfo {author}
  {\bibfnamefont {M.}~\bibnamefont {{Furqan}}}, \bibinfo {author}
  {\bibfnamefont {G.}~\bibnamefont {{Galaz}}}, \bibinfo {author} {\bibfnamefont
  {A.}~\bibnamefont {{Gal-Yam}}}, \bibinfo {author} {\bibfnamefont
  {P.}~\bibnamefont {{Garnavich}}}, \bibinfo {author} {\bibfnamefont
  {E.}~\bibnamefont {{Gawiser}}}, \bibinfo {author} {\bibfnamefont
  {J.}~\bibnamefont {{Geary}}}, \bibinfo {author} {\bibfnamefont
  {P.}~\bibnamefont {{Gee}}}, \bibinfo {author} {\bibfnamefont {R.~R.}\
  \bibnamefont {{Gibson}}}, \bibinfo {author} {\bibfnamefont {K.}~\bibnamefont
  {{Gilmore}}}, \bibinfo {author} {\bibfnamefont {E.~A.}\ \bibnamefont
  {{Grace}}}, \bibinfo {author} {\bibfnamefont {R.~F.}\ \bibnamefont
  {{Green}}}, \bibinfo {author} {\bibfnamefont {W.~J.}\ \bibnamefont
  {{Gressler}}}, \bibinfo {author} {\bibfnamefont {C.~J.}\ \bibnamefont
  {{Grillmair}}}, \bibinfo {author} {\bibfnamefont {S.}~\bibnamefont
  {{Habib}}}, \bibinfo {author} {\bibfnamefont {J.~S.}\ \bibnamefont
  {{Haggerty}}}, \bibinfo {author} {\bibfnamefont {M.}~\bibnamefont {{Hamuy}}},
  \bibinfo {author} {\bibfnamefont {A.~W.}\ \bibnamefont {{Harris}}}, \bibinfo
  {author} {\bibfnamefont {S.~L.}\ \bibnamefont {{Hawley}}}, \bibinfo {author}
  {\bibfnamefont {A.~F.}\ \bibnamefont {{Heavens}}}, \bibinfo {author}
  {\bibfnamefont {L.}~\bibnamefont {{Hebb}}}, \bibinfo {author} {\bibfnamefont
  {T.~J.}\ \bibnamefont {{Henry}}}, \bibinfo {author} {\bibfnamefont
  {E.}~\bibnamefont {{Hileman}}}, \bibinfo {author} {\bibfnamefont {E.~J.}\
  \bibnamefont {{Hilton}}}, \bibinfo {author} {\bibfnamefont {K.}~\bibnamefont
  {{Hoadley}}}, \bibinfo {author} {\bibfnamefont {J.~B.}\ \bibnamefont
  {{Holberg}}}, \bibinfo {author} {\bibfnamefont {M.~J.}\ \bibnamefont
  {{Holman}}}, \bibinfo {author} {\bibfnamefont {S.~B.}\ \bibnamefont
  {{Howell}}}, \bibinfo {author} {\bibfnamefont {L.}~\bibnamefont {{Infante}}},
  \bibinfo {author} {\bibfnamefont {Z.}~\bibnamefont {{Ivezic}}}, \bibinfo
  {author} {\bibfnamefont {S.~H.}\ \bibnamefont {{Jacoby}}}, \bibinfo {author}
  {\bibfnamefont {B.}~\bibnamefont {{Jain}}}, \bibinfo {author} {\bibnamefont
  {{R}}}, \bibinfo {author} {\bibnamefont {{Jedicke}}}, \bibinfo {author}
  {\bibfnamefont {M.~J.}\ \bibnamefont {{Jee}}}, \bibinfo {author}
  {\bibfnamefont {J.}~\bibnamefont {{Garrett Jernigan}}}, \bibinfo {author}
  {\bibfnamefont {S.~W.}\ \bibnamefont {{Jha}}}, \bibinfo {author}
  {\bibfnamefont {K.~V.}\ \bibnamefont {{Johnston}}}, \bibinfo {author}
  {\bibfnamefont {R.~L.}\ \bibnamefont {{Jones}}}, \bibinfo {author}
  {\bibfnamefont {M.}~\bibnamefont {{Juric}}}, \bibinfo {author} {\bibfnamefont
  {M.}~\bibnamefont {{Kaasalainen}}}, \bibinfo {author} {\bibnamefont
  {{Styliani}}}, \bibinfo {author} {\bibnamefont {{Kafka}}}, \bibinfo {author}
  {\bibfnamefont {S.~M.}\ \bibnamefont {{Kahn}}}, \bibinfo {author}
  {\bibfnamefont {N.~A.}\ \bibnamefont {{Kaib}}}, \bibinfo {author}
  {\bibfnamefont {J.}~\bibnamefont {{Kalirai}}}, \bibinfo {author}
  {\bibfnamefont {J.}~\bibnamefont {{Kantor}}}, \bibinfo {author}
  {\bibfnamefont {M.~M.}\ \bibnamefont {{Kasliwal}}}, \bibinfo {author}
  {\bibfnamefont {C.~R.}\ \bibnamefont {{Keeton}}}, \bibinfo {author}
  {\bibfnamefont {R.}~\bibnamefont {{Kessler}}}, \bibinfo {author}
  {\bibfnamefont {Z.}~\bibnamefont {{Knezevic}}}, \bibinfo {author}
  {\bibfnamefont {A.}~\bibnamefont {{Kowalski}}}, \bibinfo {author}
  {\bibfnamefont {V.~L.}\ \bibnamefont {{Krabbendam}}}, \bibinfo {author}
  {\bibfnamefont {K.~S.}\ \bibnamefont {{Krughoff}}}, \bibinfo {author}
  {\bibfnamefont {S.}~\bibnamefont {{Kulkarni}}}, \bibinfo {author}
  {\bibfnamefont {S.}~\bibnamefont {{Kuhlman}}}, \bibinfo {author}
  {\bibfnamefont {M.}~\bibnamefont {{Lacy}}}, \bibinfo {author} {\bibfnamefont
  {S.}~\bibnamefont {{Lepine}}}, \bibinfo {author} {\bibfnamefont
  {M.}~\bibnamefont {{Liang}}}, \bibinfo {author} {\bibfnamefont
  {A.}~\bibnamefont {{Lien}}}, \bibinfo {author} {\bibfnamefont
  {P.}~\bibnamefont {{Lira}}}, \bibinfo {author} {\bibfnamefont {K.~S.}\
  \bibnamefont {{Long}}}, \bibinfo {author} {\bibfnamefont {S.}~\bibnamefont
  {{Lorenz}}}, \bibinfo {author} {\bibfnamefont {J.~M.}\ \bibnamefont
  {{Lotz}}}, \bibinfo {author} {\bibfnamefont {R.~H.}\ \bibnamefont
  {{Lupton}}}, \bibinfo {author} {\bibfnamefont {J.}~\bibnamefont {{Lutz}}},
  \bibinfo {author} {\bibfnamefont {L.~M.}\ \bibnamefont {{Macri}}}, \bibinfo
  {author} {\bibfnamefont {A.~A.}\ \bibnamefont {{Mahabal}}}, \bibinfo {author}
  {\bibfnamefont {R.}~\bibnamefont {{Mandelbaum}}}, \bibinfo {author}
  {\bibfnamefont {P.}~\bibnamefont {{Marshall}}}, \bibinfo {author}
  {\bibfnamefont {M.}~\bibnamefont {{May}}}, \bibinfo {author} {\bibfnamefont
  {P.~M.}\ \bibnamefont {{McGehee}}}, \bibinfo {author} {\bibfnamefont {B.~T.}\
  \bibnamefont {{Meadows}}}, \bibinfo {author} {\bibfnamefont {A.}~\bibnamefont
  {{Meert}}}, \bibinfo {author} {\bibfnamefont {A.}~\bibnamefont {{Milani}}},
  \bibinfo {author} {\bibfnamefont {C.~J.}\ \bibnamefont {{Miller}}}, \bibinfo
  {author} {\bibfnamefont {M.}~\bibnamefont {{Miller}}}, \bibinfo {author}
  {\bibfnamefont {D.}~\bibnamefont {{Mills}}}, \bibinfo {author} {\bibfnamefont
  {D.}~\bibnamefont {{Minniti}}}, \bibinfo {author} {\bibfnamefont
  {D.}~\bibnamefont {{Monet}}}, \bibinfo {author} {\bibfnamefont {A.~S.}\
  \bibnamefont {{Mukadam}}}, \bibinfo {author} {\bibfnamefont {E.}~\bibnamefont
  {{Nakar}}}, \bibinfo {author} {\bibfnamefont {D.~R.}\ \bibnamefont
  {{Neill}}}, \bibinfo {author} {\bibfnamefont {J.~A.}\ \bibnamefont
  {{Newman}}}, \bibinfo {author} {\bibfnamefont {S.}~\bibnamefont
  {{Nikolaev}}}, \bibinfo {author} {\bibfnamefont {M.}~\bibnamefont
  {{Nordby}}}, \bibinfo {author} {\bibfnamefont {P.}~\bibnamefont
  {{O'Connor}}}, \bibinfo {author} {\bibfnamefont {M.}~\bibnamefont {{Oguri}}},
  \bibinfo {author} {\bibfnamefont {J.}~\bibnamefont {{Oliver}}}, \bibinfo
  {author} {\bibfnamefont {S.~S.}\ \bibnamefont {{Olivier}}}, \bibinfo {author}
  {\bibfnamefont {J.~K.}\ \bibnamefont {{Olsen}}}, \bibinfo {author}
  {\bibfnamefont {K.}~\bibnamefont {{Olsen}}}, \bibinfo {author} {\bibfnamefont
  {E.~W.}\ \bibnamefont {{Olszewski}}}, \bibinfo {author} {\bibfnamefont
  {H.}~\bibnamefont {{Oluseyi}}}, \bibinfo {author} {\bibfnamefont {N.~D.}\
  \bibnamefont {{Padilla}}}, \bibinfo {author} {\bibfnamefont {A.}~\bibnamefont
  {{Parker}}}, \bibinfo {author} {\bibfnamefont {J.}~\bibnamefont {{Pepper}}},
  \bibinfo {author} {\bibfnamefont {J.~R.}\ \bibnamefont {{Peterson}}},
  \bibinfo {author} {\bibfnamefont {C.}~\bibnamefont {{Petry}}}, \bibinfo
  {author} {\bibfnamefont {P.~A.}\ \bibnamefont {{Pinto}}}, \bibinfo {author}
  {\bibfnamefont {J.~L.}\ \bibnamefont {{Pizagno}}}, \bibinfo {author}
  {\bibfnamefont {B.}~\bibnamefont {{Popescu}}}, \bibinfo {author}
  {\bibfnamefont {A.}~\bibnamefont {{Prsa}}}, \bibinfo {author} {\bibfnamefont
  {V.}~\bibnamefont {{Radcka}}}, \bibinfo {author} {\bibfnamefont {M.~J.}\
  \bibnamefont {{Raddick}}}, \bibinfo {author} {\bibfnamefont {A.}~\bibnamefont
  {{Rasmussen}}}, \bibinfo {author} {\bibfnamefont {A.}~\bibnamefont {{Rau}}},
  \bibinfo {author} {\bibfnamefont {J.}~\bibnamefont {{Rho}}}, \bibinfo
  {author} {\bibfnamefont {J.~E.}\ \bibnamefont {{Rhoads}}}, \bibinfo {author}
  {\bibfnamefont {G.~T.}\ \bibnamefont {{Richards}}}, \bibinfo {author}
  {\bibfnamefont {S.~T.}\ \bibnamefont {{Ridgway}}}, \bibinfo {author}
  {\bibfnamefont {B.~E.}\ \bibnamefont {{Robertson}}}, \bibinfo {author}
  {\bibfnamefont {R.}~\bibnamefont {{Roskar}}}, \bibinfo {author}
  {\bibfnamefont {A.}~\bibnamefont {{Saha}}}, \bibinfo {author} {\bibfnamefont
  {A.}~\bibnamefont {{Sarajedini}}}, \bibinfo {author} {\bibfnamefont
  {E.}~\bibnamefont {{Scannapieco}}}, \bibinfo {author} {\bibfnamefont
  {T.}~\bibnamefont {{Schalk}}}, \bibinfo {author} {\bibfnamefont
  {R.}~\bibnamefont {{Schindler}}}, \bibinfo {author} {\bibfnamefont
  {S.}~\bibnamefont {{Schmidt}}}, \bibinfo {author} {\bibfnamefont
  {S.}~\bibnamefont {{Schmidt}}}, \bibinfo {author} {\bibfnamefont {D.~P.}\
  \bibnamefont {{Schneider}}}, \bibinfo {author} {\bibfnamefont
  {G.}~\bibnamefont {{Schumacher}}}, \bibinfo {author} {\bibfnamefont
  {R.}~\bibnamefont {{Scranton}}}, \bibinfo {author} {\bibfnamefont
  {J.}~\bibnamefont {{Sebag}}}, \bibinfo {author} {\bibfnamefont {L.~G.}\
  \bibnamefont {{Seppala}}}, \bibinfo {author} {\bibfnamefont {O.}~\bibnamefont
  {{Shemmer}}}, \bibinfo {author} {\bibfnamefont {J.~D.}\ \bibnamefont
  {{Simon}}}, \bibinfo {author} {\bibfnamefont {M.}~\bibnamefont {{Sivertz}}},
  \bibinfo {author} {\bibfnamefont {H.~A.}\ \bibnamefont {{Smith}}}, \bibinfo
  {author} {\bibfnamefont {J.}~\bibnamefont {{Allyn Smith}}}, \bibinfo {author}
  {\bibfnamefont {N.}~\bibnamefont {{Smith}}}, \bibinfo {author} {\bibfnamefont
  {A.~H.}\ \bibnamefont {{Spitz}}}, \bibinfo {author} {\bibfnamefont
  {A.}~\bibnamefont {{Stanford}}}, \bibinfo {author} {\bibfnamefont {K.~G.}\
  \bibnamefont {{Stassun}}}, \bibinfo {author} {\bibfnamefont {J.}~\bibnamefont
  {{Strader}}}, \bibinfo {author} {\bibfnamefont {M.~A.}\ \bibnamefont
  {{Strauss}}}, \bibinfo {author} {\bibfnamefont {C.~W.}\ \bibnamefont
  {{Stubbs}}}, \bibinfo {author} {\bibfnamefont {D.~W.}\ \bibnamefont
  {{Sweeney}}}, \bibinfo {author} {\bibfnamefont {A.}~\bibnamefont {{Szalay}}},
  \bibinfo {author} {\bibfnamefont {P.}~\bibnamefont {{Szkody}}}, \bibinfo
  {author} {\bibfnamefont {M.}~\bibnamefont {{Takada}}}, \bibinfo {author}
  {\bibfnamefont {P.}~\bibnamefont {{Thorman}}}, \bibinfo {author}
  {\bibfnamefont {D.~E.}\ \bibnamefont {{Trilling}}}, \bibinfo {author}
  {\bibfnamefont {V.}~\bibnamefont {{Trimble}}}, \bibinfo {author}
  {\bibfnamefont {A.}~\bibnamefont {{Tyson}}}, \bibinfo {author} {\bibfnamefont
  {R.}~\bibnamefont {{Van Berg}}}, \bibinfo {author} {\bibfnamefont
  {D.}~\bibnamefont {{Vanden Berk}}}, \bibinfo {author} {\bibfnamefont
  {J.}~\bibnamefont {{VanderPlas}}}, \bibinfo {author} {\bibfnamefont
  {L.}~\bibnamefont {{Verde}}}, \bibinfo {author} {\bibfnamefont
  {B.}~\bibnamefont {{Vrsnak}}}, \bibinfo {author} {\bibfnamefont {L.~M.}\
  \bibnamefont {{Walkowicz}}}, \bibinfo {author} {\bibfnamefont {B.~D.}\
  \bibnamefont {{Wandelt}}}, \bibinfo {author} {\bibfnamefont {S.}~\bibnamefont
  {{Wang}}}, \bibinfo {author} {\bibfnamefont {Y.}~\bibnamefont {{Wang}}},
  \bibinfo {author} {\bibfnamefont {M.}~\bibnamefont {{Warner}}}, \bibinfo
  {author} {\bibfnamefont {R.~H.}\ \bibnamefont {{Wechsler}}}, \bibinfo
  {author} {\bibfnamefont {A.~A.}\ \bibnamefont {{West}}}, \bibinfo {author}
  {\bibfnamefont {O.}~\bibnamefont {{Wiecha}}}, \bibinfo {author}
  {\bibfnamefont {B.~F.}\ \bibnamefont {{Williams}}}, \bibinfo {author}
  {\bibfnamefont {B.}~\bibnamefont {{Willman}}}, \bibinfo {author}
  {\bibfnamefont {D.}~\bibnamefont {{Wittman}}}, \bibinfo {author}
  {\bibfnamefont {S.~C.}\ \bibnamefont {{Wolff}}}, \bibinfo {author}
  {\bibfnamefont {W.~M.}\ \bibnamefont {{Wood-Vasey}}}, \bibinfo {author}
  {\bibfnamefont {P.}~\bibnamefont {{Wozniak}}}, \bibinfo {author}
  {\bibfnamefont {P.}~\bibnamefont {{Young}}}, \bibinfo {author} {\bibfnamefont
  {A.}~\bibnamefont {{Zentner}}}, \ and\ \bibinfo {author} {\bibfnamefont
  {H.}~\bibnamefont {{Zhan}}},\ }\href {\doibase 10.48550/arXiv.0912.0201}
  {\bibfield  {journal} {\bibinfo  {journal} {arXiv e-prints}\ ,\ \bibinfo
  {eid} {arXiv:0912.0201}} (\bibinfo {year} {2009})},\ \Eprint
  {http://arxiv.org/abs/0912.0201} {arXiv:0912.0201 [astro-ph.IM]} \BibitemShut
  {NoStop}%
\bibitem [{\citenamefont {Heavens}\ \emph {et~al.}(2000)\citenamefont
  {Heavens}, \citenamefont {Jimenez},\ and\ \citenamefont
  {Lahav}}]{Heavens:1999am}%
  \BibitemOpen
  \bibfield  {author} {\bibinfo {author} {\bibfnamefont {A.}~\bibnamefont
  {Heavens}}, \bibinfo {author} {\bibfnamefont {R.}~\bibnamefont {Jimenez}}, \
  and\ \bibinfo {author} {\bibfnamefont {O.}~\bibnamefont {Lahav}},\ }\href
  {\doibase 10.1046/j.1365-8711.2000.03692.x} {\bibfield  {journal} {\bibinfo
  {journal} {Mon. Not. Roy. Astron. Soc.}\ }\textbf {\bibinfo {volume} {317}},\
  \bibinfo {pages} {965} (\bibinfo {year} {2000})},\ \Eprint
  {http://arxiv.org/abs/astro-ph/9911102} {arXiv:astro-ph/9911102} \BibitemShut
  {NoStop}%
\bibitem [{\citenamefont {Alsing}\ and\ \citenamefont
  {Wandelt}(2018)}]{Alsing:2017var}%
  \BibitemOpen
  \bibfield  {author} {\bibinfo {author} {\bibfnamefont {J.}~\bibnamefont
  {Alsing}}\ and\ \bibinfo {author} {\bibfnamefont {B.}~\bibnamefont
  {Wandelt}},\ }\href {\doibase 10.1093/mnrasl/sly029} {\bibfield  {journal}
  {\bibinfo  {journal} {Mon. Not. Roy. Astron. Soc.}\ }\textbf {\bibinfo
  {volume} {476}},\ \bibinfo {pages} {L60} (\bibinfo {year} {2018})},\ \Eprint
  {http://arxiv.org/abs/1712.00012} {arXiv:1712.00012 [astro-ph.CO]}
  \BibitemShut {NoStop}%
\bibitem [{\citenamefont {Gualdi}\ \emph {et~al.}(2018)\citenamefont {Gualdi},
  \citenamefont {Manera}, \citenamefont {Joachimi},\ and\ \citenamefont
  {Lahav}}]{Gualdi:2017iey}%
  \BibitemOpen
  \bibfield  {author} {\bibinfo {author} {\bibfnamefont {D.}~\bibnamefont
  {Gualdi}}, \bibinfo {author} {\bibfnamefont {M.}~\bibnamefont {Manera}},
  \bibinfo {author} {\bibfnamefont {B.}~\bibnamefont {Joachimi}}, \ and\
  \bibinfo {author} {\bibfnamefont {O.}~\bibnamefont {Lahav}},\ }\href
  {\doibase 10.1093/mnras/sty261} {\bibfield  {journal} {\bibinfo  {journal}
  {Mon. Not. Roy. Astron. Soc.}\ }\textbf {\bibinfo {volume} {476}},\ \bibinfo
  {pages} {4045} (\bibinfo {year} {2018})},\ \Eprint
  {http://arxiv.org/abs/1709.03600} {arXiv:1709.03600 [astro-ph.CO]}
  \BibitemShut {NoStop}%
\bibitem [{\citenamefont {Biagetti}\ \emph {et~al.}(2022)\citenamefont
  {Biagetti}, \citenamefont {Castiblanco}, \citenamefont {Nore\~na},\ and\
  \citenamefont {Sefusatti}}]{Biagetti:2021tua}%
  \BibitemOpen
  \bibfield  {author} {\bibinfo {author} {\bibfnamefont {M.}~\bibnamefont
  {Biagetti}}, \bibinfo {author} {\bibfnamefont {L.}~\bibnamefont
  {Castiblanco}}, \bibinfo {author} {\bibfnamefont {J.}~\bibnamefont
  {Nore\~na}}, \ and\ \bibinfo {author} {\bibfnamefont {E.}~\bibnamefont
  {Sefusatti}},\ }\href {\doibase 10.1088/1475-7516/2022/09/009} {\bibfield
  {journal} {\bibinfo  {journal} {JCAP}\ }\textbf {\bibinfo {volume} {09}},\
  \bibinfo {pages} {009} (\bibinfo {year} {2022})},\ \Eprint
  {http://arxiv.org/abs/2111.05887} {arXiv:2111.05887 [astro-ph.CO]}
  \BibitemShut {NoStop}%
\bibitem [{\citenamefont {Fl\"oss}\ \emph {et~al.}(2023)\citenamefont
  {Fl\"oss}, \citenamefont {Biagetti},\ and\ \citenamefont
  {Meerburg}}]{Floss:2022wkq}%
  \BibitemOpen
  \bibfield  {author} {\bibinfo {author} {\bibfnamefont {T.}~\bibnamefont
  {Fl\"oss}}, \bibinfo {author} {\bibfnamefont {M.}~\bibnamefont {Biagetti}}, \
  and\ \bibinfo {author} {\bibfnamefont {P.~D.}\ \bibnamefont {Meerburg}},\
  }\href {\doibase 10.1103/PhysRevD.107.023528} {\bibfield  {journal} {\bibinfo
   {journal} {Phys. Rev. D}\ }\textbf {\bibinfo {volume} {107}},\ \bibinfo
  {pages} {023528} (\bibinfo {year} {2023})},\ \Eprint
  {http://arxiv.org/abs/2206.10458} {arXiv:2206.10458 [astro-ph.CO]}
  \BibitemShut {NoStop}%
\end{thebibliography}%

\end{document}